%% file: pathint_propagator_JHEP.tex
\documentclass[11pt,a4paper]{article}
\usepackage{jheppub}

\usepackage[utf8]{inputenc}
\usepackage{floatpag}
\usepackage{amsthm}
\usepackage{enumerate}
\usepackage{amssymb}
\usepackage{amsmath}
\usepackage{mathtools}
\usepackage{braket}
\usepackage{hyperref}
\usepackage{xcolor}
\definecolor{rosa}{HTML}{E63946}
\definecolor{azul}{HTML}{1D3557}
\hypersetup{
	colorlinks   = true, 
	urlcolor     = azul, 
	linkcolor    = azul, 
	citecolor   = rosa 
}

\usepackage{graphicx}
\usepackage{epstopdf}
\usepackage{color} 
\graphicspath{{./img/}}
\usepackage{caption}
\usepackage{subcaption}
\usepackage[capitalise]{cleveref}

\newcommand{\avp}{\langle p^+ \rangle}
\newcommand{\breakp}[2]{\nonumber \\ & \hskip#1cm #2}
\newcommand{\deltabar}{{\mathchar '26\mkern -10mu\delta}}

\title{\boldmath Scalar propagator in a background gluon field beyond the eikonal approximation}

\author[a,b]{Pedro Agostini}

\affiliation[a]{Departamento de F\'{\i}sica de Part\'{\i}culas and IGFAE, Universidade de Santiago de Compostela, 15782 Santiago de Compostela,Galicia--Spain}
\affiliation[b]{Theoretical Physics Division, National Centre for Nuclear Research, Pasteura 7, Warsaw 02-093, Poland}

\emailAdd{pedro.agostini@usc.es}

\abstract{We investigate the path integral representation of the scalar propagator in a background gluon field, extending beyond the eikonal approximation by considering all gauge field components and incorporating its $x^-$ dependence. Utilizing the worldline formalism, we integrate the Schwinger proper time to express the scalar propagator in light-cone coordinates, facilitating a direct comparison with known results in the literature. The derived propagator captures the change of longitudinal momentum of the projectile within the medium. In the high-energy limit, our result simplifies to the effective gluon propagator employed in the BDMPS-Z formalism. Hence, we propose that our outcome serves as a foundational point for investigating corrections to the BDMPS-Z spectrum arising from the longitudinal momentum transfer of the radiated gluon with the medium, as well as for studying collisional energy loss phenomena. Lastly, by employing an expansion around the classical saddle point solution, we systematically derive an eikonal expansion in inverse powers of the boost parameter, encompassing corrections related to longitudinal momentum transfer and interactions of the projectile with the transverse component of the field.}

\begin{document}
	
\maketitle

\flushbottom	
	
\section{Introduction}
\label{sec:intro}

	The propagation of fast-moving particles in a classical background field, denoted by $A^\mu(x)$, generated by an external source, is a crucial element in describing high-energy scattering processes. In high-energy Quantum Chromodynamics (QCD), the classical field is generated by a dense system that can come from two main scenarios: (i) a nucleus (cold nuclear matter) in proton-nucleus ($p$A) collisions or in nuclear deep inelastic scattering (DIS), and (ii) a hot medium produced after a heavy-ion collision.
	
	In high-energy $p$A collisions, the projectile is typically treated as a dilute system composed of collinear partons, which behave as quasi-free particles. This is possible because the scale for partonic interactions is significantly larger than the interaction time due to Lorentz dilation, allowing us to describe the collision in terms of the partonic structure of the projectile. However, the situation is different for the target, where gluons occupy a substantial portion of the phase space at high energies. This is a consequence of the rapid growth of gluon density at small values of $x$ (higher energies) and higher values of the nucleus mass number $A$. Nevertheless, this growth is expected to be tamed by non-linear effects, leading to a \emph{saturation regime} \cite{GRIBOV19831,MUELLER1986427}. In this regime, the target can be effectively described by a classical field, and the Color Glass Condensate (CGC) emerges as a semi-classical effective field theory (EFT) \cite{McLerran_1994,McLerran_1994b,McLerran_1994c,kovchegov_levin_2012,Blaizot_2017} to describe this phenomenon.	
	
	On the other hand, numerous studies have been conducted to investigate final state interactions in heavy-ion collisions, with a key focus on understanding the significant suppression observed in large transverse momentum hadron spectra when compared to proton-proton collisions. This phenomenon, known as \emph{jet quenching}, is attributed to the energy loss experienced by the initially produced hard parton. As the hard parton traverses through the dense hot medium formed after the collision, which can be described by a classical field, it emits gluons, leading to a partial depletion of its energy. The theory that describes this process of radiative parton energy loss is referred to as the  Baier, Dokshitzer, Mueller, Peigné, Schiff, and Zakharov (BDMPS-Z) formalism \cite{Baier:1996kr,Baier:1996sk,Baier:1998kq,Zakharov:1996fv,Zakharov:1997uu}.
	
	In the study of initial stage effects using the Color Glass Condensate (CGC) effective field theory and final stage effects using the BDMPS-Z formalism, the propagation of partons in a classical background field is of utmost interest. Both approaches rely on the \emph{eikonal approximation}, which exploits the high rapidity difference between the projectile and the target. Under this approximation, the field equations lead to a Weizsäcker-Williams-like field, where only the longitudinal component is significant and independent of $x^-$. Consequently, the longitudinal momentum of the partons remains conserved during the interaction due to the $x^-$ independence of the field. The Lorentz contraction causes the color charge density of the nucleus to be concentrated around the light-cone $x^+=0$, giving rise to the so-called shockwave approximation. In this scenario, the interaction between the projectile and the nucleus is described by the Wilson line along the nucleus's longitudinal direction, with only the longitudinal component of the field contributing. However, in the BDMPS-Z formalism, the shockwave approximation is relaxed to account for a finite extent of the medium. As a result, the propagation of the projectile within the medium becomes more intricate, characterized by transverse Brownian motion inside the medium together with a color rotation. Finally, the spin structure of the partonic current become negligible in the eikonal limit since they are suppressed by a inverse power of longitudinal momentum.
	
	Despite the eikonal approximation being well-justified and having yielded numerous successful phenomenological results in high-energy collisions, there is a growing interest in relaxing its strict application. This interest stems from the realization that while the eikonal approximation simplifies the structure of scattering amplitudes, it may also obscure important phenomena present in the scattering processes, as we will discuss further below. Furthermore, with the upcoming Electron-Ion Collider (EIC) at BNL, which will explore DIS at high but moderate energies, a regime where low-$x$ physics is still relevant but non-eikonal effects may become significant, there has been a notable surge in the scientific community's interest in studying the implications of non-eikonal corrections.
	
	Various methods exist for relaxing the eikonal approximation. One such approach involves relaxing the assumption of infinite boost. In the context of the CGC framework, pioneering studies in this direction were conducted in \cite{Altinoluk_2014,Altinoluk_2016}. These investigations, similar to the analysis presented in this manuscript, focused on computing corrections to the finite width of the target by expanding the field at next-to-leading order in the saddle-point approximation. Subsequently, the study was extended to include improvements for the quark propagator by incorporating the transverse component of the background field \cite{Altinoluk_2021} and accounting for its $x^-$ dependence \cite{Altinoluk_2022}. This approach has been employed to compute various observables beyond the eikonal approximation \cite{Altinoluk:2022jkk,Agostini_2019,Agostini:2022ctk,Agostini:2019hkj,Agostini:2022oge}.
	
	On the other hand, a notable challenge in QCD, known as the \emph{proton spin puzzle} \cite{EuropeanMuon:1987isl}, arises from the discrepancy between theoretical predictions and experimental results showing that quarks carry a smaller fraction of the proton's spin than expected. The eikonal observables, being spin-independent, might not account for the "missing spin," which may reside in the low-$x$ region of the proton's wave function. Consequently, there has been a growing interest in the low-$x$ physics community \cite{Chirilli_2021,Hatta:2016aoc,Chirilli:2018kkw,Jalilian-Marian:2018iui,Jalilian-Marian:2019kaf,Altinoluk:2023qfr,Bhattacharya:2022vvo,Boussarie:2020vzf,Boussarie:2020fpb,Li:2020uhl,Li:2021zaw} to explore the inclusion of (anti)quark exchanges in the background field of the target, leading to spin-dependent observables. In a series of papers \cite{Kovchegov_2016,Kovchegov:2016weo,Kovchegov:2016zex,Kovchegov:2017jxc,Kovchegov:2017lsr,Kovchegov:2018znm,Kovchegov:2018zeq,Kovchegov:2020hgb,Borden:2023ugd}, it has been investigated the effects of relaxing the eikonal approximation on the helicity and orbital angular momentum parton distributions at low-$x$.
	Furthermore, recent works \cite{Cougoulic:2019aja,Cougoulic:2020tbc,Cougoulic:2022gbk} have taken a step further by extending the non-linear evolution equations and the McLerran-Venugopalan (MV) model at next-to-eikonal order. In a related development, an effective Hamiltonian approach incorporating both quark and gluon fields within the background field has been presented in \cite{Li:2023tlw}. Additionally, some notable progress has been made in understanding the proton's spin structure using the worldline formalism \cite{Strassler:1992zr,Schubert:2001he}, as presented in \cite{Tarasov:2020cwl, Tarasov:2021yll}.
	
	Finally, studies on the corrections to the eikonal approximation have also been relevant in the BDMPS-Z formalism. Calculations of parton energy loss using the eikonal approximation suffer from significant uncertainties when extrapolated to the full kinematic range and neglect collisional energy loss, those deficiencies were analyzed in \cite{Zapp:2012nw,Feal:2018bru,Feal:2018sml,Caron-Huot:2010qjx,Andres:2020vxs}. Moreover, in \cite{Sadofyev:2021ohn,Andres:2022ndd}, the eikonal approximation is relaxed by incorporating a transverse flow model in the nuclear matter due to a finite transverse component of the gauge field, resulting in an anisotropic jet broadening distribution.
	
	The aim of this manuscript is to expand upon the existing literature by relaxing the eikonal approximation in high-energy scattering amplitudes. To achieve this, we comprehensively analyze all components of the gauge field, including $A^+$ and ${\bf A}$, which are omitted in the eikonal approximation, along with the dynamical $x^-$ dependence of the field. The novelty of our approach lies in presenting the result as a path integral in light-cone coordinates, enabling us to compute the scattering amplitude at arbitrary powers of the inverse of the energy. While the spin of the projectile is currently disregarded, we intend to incorporate it in future investigations. Therefore, our focus centers on studying the scalar propagator in the presence of a classical background field, which is the primary object in the calculation of scattering amplitudes when the spin is neglected. The path integral representation of the scalar propagator provides an initial framework for evaluating non-eikonal corrections at arbitrary orders and serves as the starting point for analyzing more complex quantities, such as the quark or gluon propagator.

	The results presented within this manuscript are derived through a methodology analogous to that of \cite{Laenen:2008gt, Fabbrichesi:1993kz} and the references therein. This derivation is grounded in the framework of the worldline formalism, to be introduced in \cref{sec:theoreticalback}. Notably, a distinguishing feature of our approach lies in the utilization of light-cone coordinates. Specifically, the retarded scalar propagator obtained in this work, which is explained in detail in \cref{sec:scalar_prop_LC}, can be expressed as follows:
	\begin{align}\label{eq:ret_scalar_prop_config2}
		\Delta_R(x,y) &= 
		\Theta(x^+-y^+)
		\int \mathcal{D}p^+ (\tau^+)
		\frac{\Theta(\avp)}{2 \avp }
		\int_{\vec{y}}^{\vec{x}} 
		\mathcal{D}^3\vec{z} (\tau^+)
		\nonumber \\ & \hskip0cm \times
		\exp\left\{ i \int_{y^+}^{x^+} d\tau^+ \left[ - \frac{m^2}{2 \avp} + \frac{\avp}{2} \dot{\bf z}^2 - p^+ \dot{z}^- \right] \right\} 
		\mathcal{U}_{[x^+,y^+]}[z^+,\vec{z}]
		,
	\end{align}
	with the Wilson line given by
	\begin{align}
		\mathcal{U}_{[x^+,y^+]}[z^+(\tau^+),\vec{z}(\tau^+)] = \mathcal{P}_+ 
		\exp \left\{
		-ig \int_{y^+}^{x^+} d\tau^+ \left[
		A^-(z^+, \vec{z}) \frac{p^+}{\avp} + \vec{A}\big(z^+,\vec{z}\big) \cdot \dot{\vec{z}}
		\right]
		\right\},
	\end{align}
	where $\vec{A} \equiv (A^+, {\bf A})$.
	
	The expression in \cref{eq:ret_scalar_prop_config2} describes the propagation of a particle with mass $m$, following a trajectory $\vec{z}(\tau^+)$ within the medium, where $\vec{z} \equiv (z^-, {\bf z})$. As the particle undergoes multiple transverse scatterings inside the medium, its transverse path ${\bf z}(\tau^+)$ resembles a Brownian motion. On the other hand, due to the dynamical $z^-$ dependence of the gauge field, the medium imparts longitudinal momentum to the projectile, resulting in a non-constant longitudinal momentum $p^+(\tau^+)$ along the path. The quantity $\avp$ represents the mean longitudinal momentum of the particle within the medium. Consequently, the particle follows a finite and non-constant longitudinal trajectory $z^-(\tau^+)$ within the medium. In the case where the field is time ($z^-$) independent and solely longitudinal ($\vec{A}=0$), \cref{eq:ret_scalar_prop_config2} simplifies to the effective propagator utilized in the BDMPS-Z formalism.
	
	While \cref{eq:ret_scalar_prop_config2} is formulated for a generic field, specifying a model for the gauge field enables the calculation of corrections to the BDMPS-Z spectrum arising from the longitudinal momentum exchange between the radiated gluon and the non-eikonal field. Furthermore, by considering the non-constancy of the projectile's longitudinal momentum, this propagator facilitates the investigation of collisional energy loss, assuming a vanishing parton's spin. Lastly, an eikonal expansion in powers of inverse energy can be systematically obtained by expanding \cref{eq:ret_scalar_prop_config2} around its saddle point solution.

	This manuscript is organized as follows: In \cref{sec:theoreticalback}, we introduce the well established result of the Feynman propagator for a scalar field in the presence of a background field using the worldline formalism. In \cref{sec:scalar_prop_LC}, we present the Feynman scalar propagator expressed in terms of light-cone coordinates, establishing a clearer connection with known expressions used in the existing literature. Furthermore, we compare our result with the effective scalar propagator utilized in the jet quenching formalism. Additionally, we compute the propagator when the medium has a finite extent, which is relevant for the study of $p$A collisions, as well as DIS. Moving forward, in \cref{sec:eikonal_expansion}, we perform an expansion of the path integral around its classical saddle point solution and introduce a systematic approach to calculate eikonal corrections to the scalar propagator. Lastly, in \cref{sec:conclusions}, we provide a summary of our findings and present prospects for future research.

\section{The scalar propagator in a background non-Abelian field}
\label{sec:sec2}

	Although the scalar propagator does not directly influence QCD observables, it plays a crucial role in processes where the spin of quarks is negligible, as observed in ultra high-energy collisions. Moreover, its derivation lays the groundwork for computing more complex objects, such as the quark propagator, which involves the incorporation of the so-called spin factor \cite{Brink:1976sc,Fradkin:1991ci} or gluon propagator. In this section, we will comprehensively examine the path integral representation of the scalar propagator within a classical field using the worldline formalism and subsequently express it in light-cone coordinates. This latter simplification will enable us to establish a more lucid connection between the worldline formalism and the conventional notation commonly employed in small-$x$ physics.
	
\subsection{The scalar propagator in the worldline formalism}
\label{sec:theoreticalback}

	Let us consider a scalar field $\phi(x)$ with mass $m$ interacting with a non-abelian classical field $A^\mu(x)=T_{\mathcal{R}}^a A_a^\mu(x)$, where $T_{\mathcal{R}}^a$ are the generators of the gauge group in the representation $\mathcal{R}$. The Lagrangian describing this theory is given by
	\begin{equation}\label{key}
		\mathcal{L} = - \frac{1}{4} F_{\mu \nu}F^{\mu \nu} + |D_\mu \phi|^2-m^2|\phi|^2,
	\end{equation}
	where $F_{\mu \nu} \equiv \partial_\mu A_\nu - \partial_\nu A_\mu +i g [A_\mu,A_\nu]$ is the field strength tensor, and $D_\mu \equiv \partial_\mu + ig A_\mu$ represents the covariant derivative.
	
	The Feynman propagator of the scalar field in a 4-dimensional space-time\footnote{In this manuscript we use the metric $g^{\mu \nu}={\rm diag}(1,-1,-1,-1)$.}, $\Delta_F(x,y)$, is given by the Klein-Gordon equation in presence of a gauge field:
	\begin{equation}\label{key}
		(D^\mu D_\mu + m^2) \Delta_F(x,y) = - i \delta^{(4)}(x-y).
	\end{equation}
	
	To represent this equation in integral form, we introduce the operator $\hat{H} = D^\mu D_\mu + m^2$, which acts on the Hilbert space of square integrable functions. The Green's function operator is defined as
	\begin{equation}\label{key}
		\hat{\Delta}_F = -i \hat{H}^{-1}.
	\end{equation}
	Therefore, the inverse of the operator $i\hat{H}$ can be expressed using the Schwinger representation:
	\begin{equation}\label{key}
		\hat{\Delta}_F = \int_0^{\infty} dT e^{-iT(\hat{H}-i\epsilon)},
	\end{equation}
	where the inclusion of $i\epsilon$ ensures the convergence of the integral. Here, $e^{-iT\hat{H}}$ is an unitary operator satisfying the Schrödinger equation, allowing us to interpret $\hat{H}$ as the Hamiltonian of a quantum system, and $T$ as an internal time coordinate known as the Schwinger proper time.
	
	By introducing the complete set of single-particle states $\ket{x}$ in the Hilbert space, we can express the scalar propagator as the matrix element
	\begin{equation}\label{key}
		\Delta_F(x,y) = \braket{x | \hat{\Delta}_F | y}.
	\end{equation}
	Consequently, the propagator describes the evolution of a scalar particle from the space-time point $y$ at proper time $t=0$ to a point $x$ at $t=T$, accounting for all possible interactions with the background field $A^\mu$ across different values of $T$:
	\begin{equation}\label{eq:prop1}
		\Delta_F(x,y) = \int_0^{\infty} dT \braket{x | e^{-iT(\hat{H}-i\epsilon)} | y}.
	\end{equation}
	
	Analogously to non-relativistic quantum mechanics, we can represent the propagator as a path integral by introducing the complete set of momentum states $\ket{p}$ satisfying
	\begin{equation}\label{key}
		\braket{x | p} = e^{i x \cdot p}, \qquad \int d^4x \ket{x}\bra{x}=\int \frac{d^4p}{(2\pi)^4} \ket{p}\bra{p}=1,
	\end{equation}
	as well as slicing the Schwinger proper time into $N$ steps of length $\Delta t = \frac{T}{N}$ such that
	\begin{equation}\label{key}
		e^{-i(\hat{H}-i \epsilon)T} = e^{-i(\hat{H}-i \epsilon)\sum_{n=1}^{N} \Delta t} = \prod_{n=1}^{N} e^{-i(\hat{H}-i \epsilon) \Delta t + \mathcal{O}(\Delta t^2)}.
	\end{equation}
	
	Thus, inserting $N$ and $N-1$ spectral decomposition in terms of momentum and coordinate eigenstates, respectively, in \cref{eq:prop1} we can write the scalar propagator as
	\begin{align}\label{key}
		\Delta_F(x,y) &=\int_0^{\infty} dT \braket{x | e^{-iT(\hat{H}-i\epsilon)} | y} 
		\nonumber \\ & =
		\int_0^{\infty} dT
		\lim_{N \to \infty} \int \prod_{k=1}^{N-1} d^4 z_k \int \prod_{n=1}^N \frac{d^4p_n}{(2\pi)^4}
		\braket{z_n | e^{-i \Delta t (\hat{H}-i\epsilon)} | p_n} \braket{p_n |  z_{n-1}}
		\nonumber \\ & =
		\int_0^{\infty} dT
		\lim_{N \to \infty} \int \prod_{k=1}^{N-1} d^4 z_k \int \prod_{n=1}^N \frac{d^4p_n}{(2\pi)^4}
		\breakp{1}{\times}
		\mathcal{P}
		\exp\left\{ i \sum_{l=1}^{N} \Delta t \left[  \left(  p^\mu_l + g A^\mu(z_l) \right)^2 - m_\epsilon^2 +i p_l \cdot \frac{z_l-z_{l-1}}{\Delta t} \right] \right\}
		,
	\end{align}
	where we have defined $z_0 \equiv y$, $z_N \equiv x$ and $m_\epsilon^2 \equiv m^2-i\epsilon$. The path ordering operator $\mathcal{P}$ is introduced to account to the fact that $A_\mu(z_l)$ does not commute at different values of $z$, or equivalently, of the proper time. We have also used the fact that matrix element of the Weyl ordered Hamiltonian is $\braket{x|\hat{H}(\hat{x},\hat{p}) | p} = H(x,p) \braket{x|p}$, where $H(x,p)$ is a c-number obtained from replacing the operators with their corresponding variables, such that
	\begin{equation}\label{key}
		\braket{x | e^{-i(\hat{H}-i \epsilon) \Delta t} | p} = e^{i\left[ (p_\mu+gA_\mu(x))^2-m^2+i \epsilon \right] \Delta t + \mathcal{O}(\Delta t^2)} e^{i x \cdot p}.
	\end{equation}
	
	Finally, taking the continuous $N\to \infty$ limit, the propagator can be written in its phase space path integral representation:
	\begin{align}\label{eq:prop_momentum}
		\Delta_F(x,y) &= \int_0^\infty dT e^{-i m_\epsilon^2 T}
		\breakp{1}{\times}
		\int_{z(0)=y}^{z(T)=x} 
		\mathcal{D}^4 z(t) \int \mathcal{D}^4 p(t) 
		\
		e^{i \int_0^T dt (p^2 +p \cdot \dot{z})}
		\mathcal{P} e^{-ig \int_0^T dt A^\mu(z) \dot{z}_\mu},
	\end{align}
	where we have shifted the 4-momentum, defined $\dot{z}^\mu = dz^\mu/dt$ and introduced the path integral measure:
	\begin{equation}
		\mathcal{D}^4 z(t) \equiv \lim_{N \to \infty} \prod_{n=1}^{N-1} d^4 z_n, \qquad \mathcal{D}^4 p(t) \equiv \lim_{N \to \infty} \prod_{n=1}^{N} d^4 z_n \frac{d^4p_n}{(2\pi)^4},
	\end{equation}
	which accounts for all possible configurations of the trajectories $z(t)$ and $p(t)$. We also have omitted the dependence of $p^\mu(t)$ and $z^\mu(t)$ on the path parameter $t$ in the integrand of \cref{eq:prop_momentum}. From now on, unless necessary for clarification, we will not write the dependence on the curve parameter explicitly.
	
	Alternatively, the 4-momenta in \cref{eq:prop_momentum} can be integrated out as they are analytic continuation of gaussian integrals. The result after integrating over $p$ reads
	\begin{align}\label{key}
		\Delta_F(x,y) &= \int_0^\infty dT \lim_{N \to \infty} \left( \frac{i}{4 \pi \Delta t} \right)^{2N} \int 	\prod_{k=1}^{N-1} d^4 z_k
		\breakp{1}{\times} 
		\mathcal{P} \exp\left\{ - i \sum_{l=1}^{N} \Delta t \left[ \frac{(z_l-z_{l-1})^2}{4 \Delta t^2} + m_\epsilon^2 +g 	A(z_l) \cdot \frac{z_l-z_{l-1}}{\Delta t} \right] \right\}.
	\end{align}
	Taking the continuous limit, we obtain the configuration space path integral representation of the scalar propagator in a background field in terms of the worldline $z(t)$:
	\begin{equation}\label{eq:worldline_prop}
		\Delta_F(x,y) = \int_0^\infty dT e^{-i m_\epsilon^2 T}\int_{z(0)=y}^{z(T)=x} \mathcal{D}^4 z  
		\
		e^{  -i \int_0^T dt \frac{\dot{z}^2}{4} }
		\
		\mathcal{P} e^{-ig \int_0^T dt A^\mu(z) \dot{z}_\mu},
	\end{equation}
	where now, as usual in the path integral formalism, the path integral measure when the momentum is integrated out is normalized by a factor $\left(- 4 \pi i \Delta t \right)^{-2N}$ which regulates the integral in its discrete representation. 
	
	\Cref{eq:prop_momentum,eq:worldline_prop} are the worldline path integral representation of the propagator of a scalar particle coupled to a classical background field. This approach is the basis of the worldline formalism \cite{Strassler:1992zr,Schubert:2001he} where one replaces the Quantum Field Theory (QFT) path integral with a path integral over the worldline of a single particle, coupled to a background field.\footnote{It is noteworthy that the field employed within this manuscript is held as a fixed background. In order to extend the formalism to encompass a dynamical field, we direct the reader's attention to \cite{Affleck:1981bma,Feal:2022iyn,Feal:2022ufw}, as well as the references therein.} This simplifies the calculation of QFT amplitudes by reducing the problem to a one-dimensional problem in the worldline parameter. The main advantage of \cref{eq:prop_momentum,eq:worldline_prop} in scattering processes with respect to the usual perturbative approach is that instead of summing over all possible Feynman diagrams that contribute to the process we just represent the scattering amplitude as a resummed and exponentiated quantity.
	
\begin{figure}[h!]
		\centering
		\includegraphics[scale=0.4]{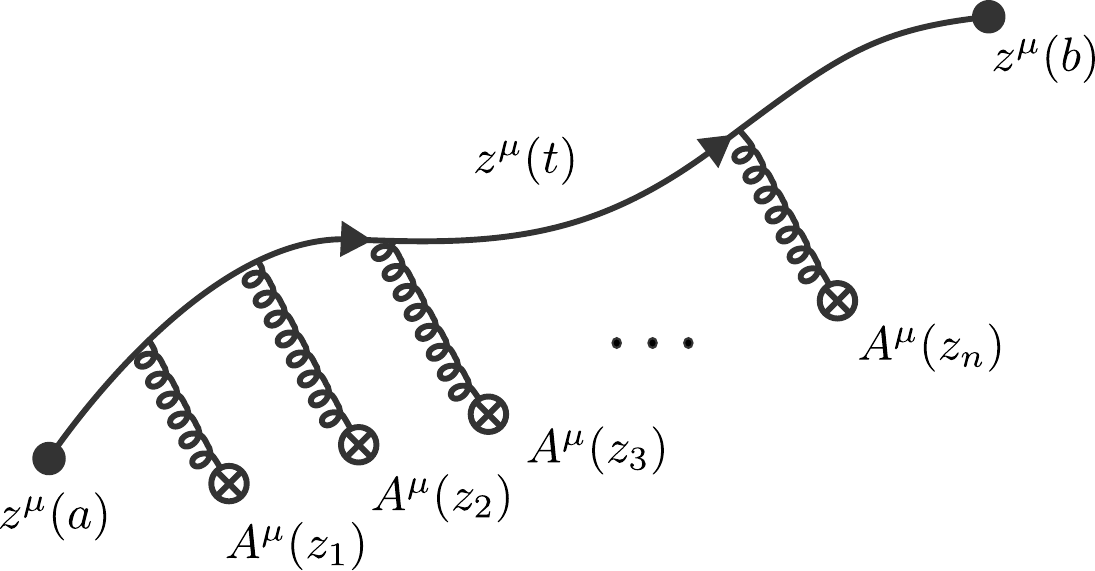}
		\caption{Picture of the $n^{\rm th}$ term in the expansion of the Wilson line $U_{[b,a]}[z]$ over a path $z^\mu(t)$ ($t\in[a,b]$). The worldline interacts with gluons generated by the medium at the coordinates $z(t_i)$ (abbreviated as $z_i^+$).}
		\label{fig:wilson_line}
	\end{figure}

	The worldline path integral is written in terms of the particle's position and momentum as a function of the worldline parameter. By integrating over all possible paths of the particle in the presence of the background field, we can compute the scattering amplitude for the process of interest. Moreover, we can read from \cref{eq:worldline_prop} that the part of the propagator that depends on the background gauge field is given by the Wilson line along the path $z(t)$:
	\begin{equation}\label{key}
		\mathcal{U}_{[T,0]}[z(t)]= 
		\mathcal{P} \exp\left\{-ig \int_0^T dt A^\mu(z) \dot{z}_\mu \right\} = 
		\mathcal{P} \exp\left\{ -ig \int_y^x dz_\mu A^\mu(z) \right\},
	\end{equation}
	which represents the phase accumulated by the particle's wave function as it travels through the medium. Indeed, by expanding the path ordered exponential, and using the shorthand notation $z_i \equiv z(t_i)$:
	\begin{align}\label{key}
		\mathcal{U}_{[b,a]}[z] &= \sum_{n=0}^{\infty} (-ig)^n 
		\int_a^b dt_n \int_a^{t_n}dt_{n-1} \cdots \int_a^{t_{2}} dt_1 
		\breakp{1}{\times}
		\dot{z}^{\mu_n}(t_n) A_{\mu_n}(z_n) \dot{z}^{\mu_{n-1}}(t_{n-1}) A_{\mu_{n-1}}(z_{n-1}) \cdots \dot{z}^{\mu_1}(t_1) A_{\mu_1}(z_1),
	\end{align}
	we see that the $n^{\rm th}$ order term in the expansion represents the interaction of the worldline with $n$ gluons at position $z(t_i)$ ($i=1,\dots,n$), see \cref{fig:wilson_line}. Therefore the exponential is a resummation of all possible interactions of the path with the gluons emitted by the external source. When considering a Wilson line in the fundamental (adjoint) representation, it can also describe a fast-moving quark (gluon) interacting with the medium. In such cases, we assume that the longitudinal momentum transfer arising from the particle's interaction with the background gluons is negligible, which corresponds to the eikonal approximation discussed in \cref{sec:eikonal_expansion}.
	
\subsection{The scalar propagator in light-cone coordinates}
\label{sec:scalar_prop_LC}

	In the preceding section, we derived the path integral representation of the scalar propagator using Lorentz-invariant equations. Now, we take a different approach by expressing both the worldline and gauge field in terms of light-cone (LC) coordinates. LC coordinates are particularly advantageous in high-energy scattering processes where longitudinal momenta play a dominant role. By separating the 4-momentum components into longitudinal and transverse parts, we can identify the leading terms in the scattering amplitude. In LC coordinates, we define $x^\mu=(x^+, x^-, {\bf x})$, where $x^{\pm} = (x^0 \pm x^3)/\sqrt{2}$ and ${\bf x}=(x^1,x^2)$. The dot product in these coordinates is given by $x \cdot y = x^+ y^- + x^- y^+ -{\bf x} \cdot {\bf y}$.\footnote{In this manuscript, we adopt the convention that the dot product, when applied to transverse coordinates, is Euclidean: ${\bf x} \cdot {\bf y} = {\bf x}^i {\bf y}^i \ne {\bf x}^i {\bf y}_i$.}
	
	Using LC coordinates, we can rewrite the scalar propagator given in \cref{eq:prop_momentum} as follows:	
	\begin{align}\label{eq:aux3}
		\Delta_F(x,y) &= \int_0^\infty dT \int_{z(0)=y}^{z(1)=x} \mathcal{D}^4 z \int \mathcal{D}^4 p 
		\breakp{-1}{\times}
		\exp\left\{ i \int_0^1 d\tau \left[ p^-(2 T p^+ - \dot{z}^+) - T ({\bf p}^2 + m_\epsilon^2) - p^+ \dot{z}^-  + {\bf p} \cdot \dot{\bf z} \right] \right\}  \mathcal{U}_{[1,0]}[z(\tau)],
	\end{align}
	where we have made the change of variables $p^\mu \to -p^\mu$ for convenience, and we re-parametrized the worldline using a dimensionless parameter. Specifically, we performed the reparametrization\footnote{We should note that the Wilson line remains invariant under reparametrizations. Thus, we adopt the convention that the boundaries specified in its sub-index have the same dimensions as the path parameter. In other words,
	\begin{align}
		\mathcal{U}_{[T,0]}[z(t)] &= \mathcal{P}_t \exp\left\{ -ig \int_0^T dt \frac{dz^\mu(t)}{dt} A_\mu[z(t)] \right\}
		= \mathcal{P}_\tau \exp\left\{ -ig \int_0^1 d\tau \frac{dz^\mu(\tau)}{d\tau} A_\mu[z(\tau)] \right\} 
		\breakp{0}{=}
		\mathcal{U}_{[1,0]}[z(\tau)].
		\end{align}} 
	$\tau = t/T$, with $\tau \in [0, 1]$. 
	
	As the argument of the exponential in \cref{eq:aux3} is linear in $p^-$, the $p^-$ path integral yields a Dirac delta function that fixes $\dot{z}^+(\tau)=2 T p^+(\tau)$. This constraint determines the trajectory\footnote{In this manuscript, we have chosen to fix the $z^+$ trajectory by integrating over $p^-$. However, it is important to note that the argument of the exponential in \cref{eq:aux3} is symmetric in the $+$ and $-$ components. Hence, an alternative approach would be to fix the $z^-$ trajectory by integrating over $p^+$. The reason for adopting our current convention is that we later assume the particle is boosted in the right direction, allowing us to interpret $p^+$ and $p^-$ as the longitudinal momentum and light-cone energy of the particle, respectively, while also regarding $z^+$ as the light-cone time.} $z^+(\tau)$, which we refer to as the {\it light-cone time}:
	\begin{equation}\label{eq:constraint}
		z^+(\tau) = y^+ + 2 T \int_0^\tau d\tau' p^+(\tau'),
	\end{equation}
	where we have used the fact that $z^+(0)=y^+$. Furthermore, since $z^+(1)=x^+$ is fixed, \cref{eq:constraint} also introduces the constraint $x^+-y^+ = 2 T \avp$, where
	\begin{align}\label{eq:avp_t}
		\avp = \int_0^1 d\tau p^+(\tau),
	\end{align}
	is the average longitudinal momentum over the particle path. Hence, the Schwinger proper time can be written in terms of the longitudinal separation and $\avp$ as $T = (x^+-y^+)/2 \avp$. 
	
	Indeed, in \cref{app:scalar_prop}, we explicitly perform the $p^-$ and $z^+$ path integrals, which leads to the following result:
	\begin{align}\label{eq:aux_1b}
		\Delta_F(x,y) &= \int_0^\infty dT \int_{\vec{z}(0)=\vec{y}}^{\vec{z}(1)=\vec{x}} \mathcal{D}^3 \vec{z}(\tau) \int \mathcal{D}^3 \vec{p}(\tau) 
		\
		\delta\left( 2 \avp T - (x^+-y^+) \right)
		\nonumber \\ & \hskip2cm \times
		\exp\left\{ - i \int_0^1 d\tau \left[ \hat{p}^- \dot{z}^+ + \vec{p} \cdot \dot{\vec{z}} \right] \right\}  \mathcal{U}_{[1,0]}[z^+(\tau), \vec{z}(\tau)],
	\end{align}
	where we have defined the on-shell light-cone energy $\hat{p}^-=({\bf p}^2+m_\epsilon^2)/2p^+$, $\vec{z} \equiv (z^-, {\bf z})$, $\vec{p} \equiv (p^+,{\bf p})$ and the 3-dimensional dot product as $\vec{p} \cdot \vec{z} = p^+ z^- - {\bf p} \cdot {\bf z}$. We emphasize that in \cref{eq:aux_1b}, the LC time path $z^+(\tau)$ is fixed by \cref{eq:constraint} and, therefore, $\dot{z}^+(\tau) = 2 T p^+(\tau)$.
	
	Thanks to the Dirac delta, when the average longitudinal momentum is finite ($\avp \ne 0$), the integral over $T$ becomes straightforward.
	On the other hand, configurations with $\avp = 0$ correspond to cases where the particle does not propagate in the $z^+$ direction, i.e., $x^+=y^+$. Generally, when using the propagator as an external line in a physical process, we can justify the assumption $x^+ \ne y^+$ since at least one of the propagator legs extends to infinity once we amputate it. However, when the propagator is an internal line, as in loop calculations within the medium, there exists a region in the phase space of integration where $x^+=y^+$ (or equivalently, $\avp=0$). In such cases, the expressions presented in this manuscript are incomplete.
	
	To proceed with our analysis, we adopt the convention followed in \cite{Chirilli_2021,Chirilli:2018kkw,Altinoluk_2021,Altinoluk_2022} and examine the propagator in the case where $x^+ \ne y^+$. Thus, we can perform the integration over $T$, and the scalar propagator becomes:
	\begin{align}\label{eq:aux_2b}
		\Delta_F(x,y) &= \int_{\vec{z}(0)=\vec{y}}^{\vec{z}(1)=\vec{x}} \mathcal{D}^3 \vec{z}(\tau) \int \mathcal{D}^3 \vec{p}(\tau) 
		\
		\frac{\Theta(x^+ - y^+) \Theta(\avp) - \Theta(y^+ - x^+) \Theta(-\avp)}{2 \avp }
		\nonumber \\ & \hskip1cm \times
		\exp\left\{- i \int_0^1 d\tau \left[\frac{{\bf p}^2 + m_\epsilon^2}{2 \avp} (x^+ - y^+) + \vec{p} \cdot \dot{\vec{z}} \right] \right\}  \mathcal{U}_{[1,0]}[z^+(\tau), \vec{z}(\tau)].
	\end{align}
	
	To express the path parameter in units of time, as is customary in high-energy QCD, we perform the reparametrization of the dimensionless proper time $\tau$ as follows:\footnote{Although it might seem tempting to define the light-cone time, $z^+$, as the path parameter, as is often done in the high-energy limit of QCD, the reparametrization $\tau \to z^+(\tau) = y^+ + (x^+-y^+) \int_0^\tau d\tau' \frac{p^+(\tau')}{\avp}$ is not well-defined. This is because $\dot{z}^+(\tau) \propto p^+(\tau)$, and the resulting transformation is not a diffeomorphism; the sign of $\dot{z}^+(\tau)$ may change in the domain $[y^+, x^+]$.}
	$\tau \to \tau^+(\tau) = y^+ + \tau (x^+-y^+)$, with $\tau^+(0)=y^+$ and $\tau^+(1)=x^+$. Under this reparametrization, the path-ordered operator in the definition of the Wilson line becomes $\mathcal{P}_\tau \to \Theta(x^+-y^+) \mathcal{P}_+ + \Theta(y^+-x^+) \bar{\mathcal{P}}_+$, where $\mathcal{P}_+$ represents the path-ordered operator along the direction $\tau^+$, and $\bar{\mathcal{P}}_+$ denotes the anti path-ordered operator. Thus, \cref{eq:aux_2b} takes the form:
	\begin{align}\label{eq:aux_3b}
		\Delta_F(x,y) &= \int_{\vec{z}(y^+)=\vec{y}}^{\vec{z}(x^+)=\vec{x}} \mathcal{D}^3 \vec{z}(\tau^+) \int \mathcal{D}^3 \vec{p}(\tau^+) 
		\
		\frac{1}{2 \avp } \exp\left\{- i \int_{y^+}^{x^+} d\tau^+ \left[\hat{p}^- \dot{z}^+ + \vec{p} \cdot \dot{\vec{z}} \right] \right\}
		\breakp{0}{\times}
		\left[ \Theta(\avp)\Theta(x^+-y^+) \mathcal{P}_+ - \Theta(-\avp) \Theta(y^+ - x^+) \bar{\mathcal{P}}_+ \right]
		\breakp{0}{\times}
		\exp\left\{ -i g \int_{y^+}^{x^+} d\tau^+ \left[ A^-\big(z^+,\vec{z}\big) \dot{z}^+ + \vec{A}\big(z^+,\vec{z}\big) \cdot \dot{\vec{z}}\right] \right\},
	\end{align}
	where the LC time, in terms of $\tau^+$, is given by
	\begin{equation}\label{eq:xip}
		z^+(\tau^+) = y^+ + \int_{y^+}^{\tau^+} \frac{p^+(\bar{\tau}^+)}{\avp} d\bar{\tau}^+=x^+-\int_{\tau^+}^{x^+} \frac{p^+(\bar{\tau}^+)}{\avp} d\bar{\tau}^+,
	\end{equation}
	with  $\dot{z}^+(\tau^+) = p^+(\tau^+)/\avp$, $z^+(y^+)=y^+$, $z^+(x^+)=x^+$, and the average longitudinal momentum by
	\begin{align}\label{eq:avp}
		\avp = \frac{1}{x^+-y^+} \int_{y^+}^{x^+} d\tau^+ p^+(\tau^+).
	\end{align}
	
	Since the reparametrized variable $\tau^+$ has a well sign defined interval, $d\tau^+$, we refer to it as the {\it light-cone proper time}. On the other hand, we have $dz^+ = d\tau^+ p^+(\tau^+)/\avp $, where $d\tau^+/\avp > 0$. Hence, we encounter two scenarios: if $p^+(\tau^+) > 0$, the particle travels forward in light-cone time, while if $p^+(\tau^+) < 0$, it travels backward in light-cone time. Furthermore, in the case where $p^+(\tau^+)$ remains constant along the path, as is commonly assumed in high-energy QCD, we find that the light-cone time and proper time coincide, $z^+ = \tau^+$.
	
	Finally, we use the fact that a given a path $z^\mu(\lambda)$, the hermitian of the Wilson line is 
	\begin{align}
		\mathcal{U}^\dagger_{[b,a]} = \bar{\mathcal{P}} \exp \left\{ ig\int_{a}^{b} d\lambda A \cdot \dot{z} \right\}.
	\end{align}
	Thus, we can rewrite \cref{eq:aux_3b} as
	\begin{align}\label{eq:scalar_prop_PS}
		\Delta_F(x,y) &=
		\int
		\frac{\mathcal{D}^3\vec{p}(\tau^+)}{2 \avp }
		\int_{\vec{y}}^{\vec{x}}
		\mathcal{D}^3\vec{z} (\tau^+)
		\
		\exp \Bigg\{ -i \int_{y^+}^{x^+} d\tau^+ \left[ \hat{p}^- \dot{z}^+ + \vec{p} \cdot \dot{\vec{z}}\right] \Bigg\}
		\breakp{-1}{\times}
		\left[ \Theta(\avp)\Theta(x^+-y^+) \mathcal{U}_{[x^+,y^+]}[z^+,\vec{z}] - \Theta(y^+-x^+) \Theta(-\avp) \mathcal{U}^\dagger_{[y^+,x^+]}[z^+,\vec{z}]\right],
	\end{align}
	where
	\begin{align}
		\mathcal{U}_{[x^+,y^+]}[z^+(\tau^+),\vec{z}(\tau^+)] &
		\breakp{-1}{=}
		\mathcal{P}_+ 
		\exp \left\{
		-ig \int_{y^+}^{x^+} d\tau^+ \left[
		A^-(z^+, \vec{z}) \frac{p^+}{\avp} + \vec{A}\big(z^+,\vec{z}\big) \cdot \dot{\vec{z}}
		\right]
		\right\}.
	\end{align}
	
	Analogously to \cref{eq:worldline_prop}, we can solve the Gaussian path integral in ${\bf p}$ with the help of \cref{eq:pathint_gauss}. By doing that, we obtain the scalar propagator in the \emph{configuration space path integral representation}:
	\begin{align}\label{eq:scalar_prop_config}
		\Delta_F(x,y) &= 
		\int
		\frac{\mathcal{D}p^+ (\tau^+)}{2 \avp }
		\int_{\vec{y}}^{\vec{x}} 
		\mathcal{D}^3\vec{z} (\tau^+)
		\
		\exp\left\{ i \int_{y^+}^{x^+} d\tau^+ \left[ - \frac{m_\epsilon^2}{2 \avp} + \frac{\avp}{2} \dot{\bf z}^2 - p^+ \dot{z}^- \right] \right\}
		\breakp{-1}{\times}
		\left[ \Theta(\avp)\Theta(x^+-y^+) \mathcal{U}_{[x^+,y^+]}[z^+,\vec{z}] - \Theta(y^+-x^+) \Theta(-\avp) \mathcal{U}^\dagger_{[y^+,x^+]}[z^+,\vec{z}]\right]
		.
	\end{align}
	
	\Cref{eq:scalar_prop_PS,eq:scalar_prop_config} are the main results of this section and serve as the starting point for the subsequent analysis. In comparison to the worldline representation given by \cref{eq:prop_momentum,eq:worldline_prop}, this new representation offers two key advantages: (i) the integration over the Schwinger proper time $T$ has been performed, simplifying the computation of the propagator, and (ii) the explicit dependence on the longitudinal momentum $p^+$ is revealed, making it well-suited for studying high-energy scattering processes. However, it is important to note that this representation is not applicable when $x^+ = y^+$.
	
	In summary, \cref{eq:scalar_prop_config} describes the propagation of a scalar particle from $y^\mu$ to $x^\mu$ in the presence of a background medium. As a result of multiple transverse scatterings, the particle follows a Brownian path, denoted as ${\bf z}(\tau^+)$, in the transverse plane. Moreover, the presence of a $z^-$ dependence in the gauge field causes the particle's longitudinal momentum inside the medium, denoted as $p^+(\tau^+)$, to vary along its trajectory, leading to a non-trivial $z^-(\tau^+)$ path. Additionally, due to longitudinal interactions with the medium, the LC time $z^+(\tau^+)$ of the particle may not necessarily increase monotonically and can even exhibit backward motion when the longitudinal momentum transfer with the medium is negative.
	
	In general, we are interested in processes where $x^+ > y^+$ so that only the retarded component of the Feynman propagator contributes. For this reason, in order to simplify our analysis in the subsequent section we will study the retarded propagator:
	\begin{align}\label{eq:ret_scalar_prop_config}
		\Delta_R(x,y) &= 
		\Theta(x^+-y^+)
		\int \mathcal{D}p^+ (\tau^+)
		\frac{\Theta(\avp)}{2 \avp }
		\int_{\vec{y}}^{\vec{x}} 
		\mathcal{D}^3\vec{z} (\tau^+)
		\nonumber \\ & \hskip0cm \times
		\exp\left\{ i \int_{y^+}^{x^+} d\tau^+ \left[ - \frac{m_\epsilon^2}{2 \avp} + \frac{\avp}{2} \dot{\bf z}^2 - p^+ \dot{z}^- \right] \right\} 
		\mathcal{U}_{[x^+,y^+]}[z^+,\vec{z}]
		.
	\end{align}
	
\subsubsection{Connection with the BDMPS-Z effective propagator}
\label{sec:static_field}

	In high-energy scattering processes, the right-moving projectile possesses a significantly large longitudinal momentum, surpassing all other scales involved in the process. Consequently, the interactions with the medium are primarily governed by the exchange of soft gluons. In the particle's reference frame, the background field sources are boosted towards the left direction, resulting in the formation of a Weizsäcker-Williams field with the following expression: $\hat{A}^\mu(q) = \delta^{\mu-} \delta(q^+) a(q^-,{\bf q})$, where $a$ represents the distribution of the $q^-,{\bf q}$ modes of the gauge field.	As a result, the projectile interacts exclusively with the $-$-component of the field, which remains independent of the coordinate $z^-$:\footnote{In coordinate space, this is seen as a high boost in the $z^-$ direction so that $A^\mu(z) \to \Lambda^\mu_\nu A^\nu(\Lambda^{-1} z)$, where $\Lambda \in SO(1,3)$ is a Lorentz transformation \cite{Altinoluk_2016}. Thus, the $z^-$ dependence of the field is suppressed at high rapidity, $\omega$, boosts, $\Lambda^{-1} z = (e^{\omega} z^+, e^{-\omega} z^-, {\bf z})$, and the longitudinal component of the field is enhanced, $\Lambda^\mu_\nu A^\nu = (e^{-\omega}A^+, e^{\omega} A^-, {\bf A})$.}
	\begin{align}
		A^\mu(z) = \int_{q} e^{i q \cdot z} \tilde{A}^\mu (q) = \delta^{\mu-} \int_{q^-,{\bf q}} e^{i q^- z^+ - i {\bf q} \cdot {\bf z}} a(q^-,{\bf q}) \equiv \delta^{\mu -} A^-(z^+,{\bf z}).
	\end{align}
	
	In this case, the Wilson line is $z^-$ independent and is given by
	\begin{align}
		\mathcal{U}_{[x^+,y^+]}[z^+(\tau^+),\vec{z}(\tau^+)] = \mathcal{P}_+ 
		\exp \left\{
		-ig \int_{y^+}^{x^+} d\tau^+ \left[
		A^-(z^+, {\bf z}) \frac{p^+}{\avp}
		\right]
		\right\}.
	\end{align}
	From \cref{eq:ret_scalar_prop_config}, it can be observed that the $z^-$ dependence solely arises from the kinetic term $p^+ \dot{z}^-$, allowing us to solve the $z^-$ and $p^+$ path integrals. The $z^-$ path integral determines $p^+(\tau^+) = {\rm cte} \equiv k^+$, thereby establishing a correspondence between the light-cone proper time and the light-cone time, such that $\tau^+ = z^+$. By employing \cref{eq:pathint_dirac,eq:functional_dirac}, we can express \cref{eq:ret_scalar_prop_config} as follows:
	\begin{align}\label{eq:aux14}
		\Delta_R(x,y) &= 
		\Theta(x^+-y^+)
		\int \frac{d k^+}{2 \pi}
		\frac{\Theta(k^+)}{2 k^+} e^{-i k^+ (x^--y^-)} 
		\breakp{0}{\times}
		\int_{{\bf y}}^{{\bf x}} 
		\mathcal{D}^2 {\bf z}(z^+)
		\
		\mathcal{P}_+ 
		\exp\left\{ i \int_{y^+}^{x^+} dz^+ \left[ \frac{k^+}{2} \dot{\bf z}^2 - g A^-(z^+,{\bf z}) \right] \right\},
	\end{align}
	where we have neglected the mass term since the longitudinal momentum is scaled by a large parameter, $k^+ \gg m$.
		
	\Cref{eq:aux14}, represents the well-known result of the (retarded) in-medium scalar propagator when considering that the momentum transfer of the particle with the medium is purely transverse and only accounting for the longitudinal component of the field. The transverse path integral
	\begin{align}\label{eq:BDMPS_prop}
		\mathcal{G}_{k^+}(x^+,{\bf x}; y^+, {\bf y}) &=
		\int_{{\bf y}}^{{\bf x}} 
		\mathcal{D}^2 {\bf z}(z^+)
		\
		\mathcal{P}_+ 
		\exp\left\{ i \int_{y^+}^{x^+} dz^+ \left[ \frac{k^+}{2} \dot{\bf z}^2 - g A^-(z^+,{\bf z}) \right] \right\},
	\end{align}
	 is the propagator of a Schrödinger equation for a non-relativistic particle of mass $k^+$ moving in a time-dependent potential $A^-(z^+, {\bf z})$. \Cref{eq:BDMPS_prop} acts as an effective gluon propagator in the high-energy limit and is one of the key ingredients in the calculation of the BDMPS-Z spectrum, which describes the energy spectrum of a radiated gluon due to the interaction of the parton projectile with the medium \cite{Mehtar-Tani:2006vpj}.
	
	Thus, \cref{eq:ret_scalar_prop_config} can be used to compute corrections to the effective gluon propagator, by including longitudinal momentum transfer with the medium as well as interactions with the $A^+$ and ${\bf A}$ components of the field. With a model for the gauge field $A^\mu(z)$, one can compute non-eikonal corrections to the BDMPS-Z spectrum resulting from the interaction of the radiated gluon with the non-eikonal gauge field.
	
	Furthermore, in the BDMPS-Z formalism, the parent parton is typically treated as an eikonal object that traverses a straight transverse trajectory within the medium, experiencing energy loss solely due to gluon radiation. Considering \cref{eq:ret_scalar_prop_config} as the quark propagator when the parton spin can be neglected, it offers opportunities to investigate the energy loss of the projectile parton resulting from longitudinal scatterings with the dynamic medium. This extension allows for the exploration of collisional energy loss effects beyond the eikonal approximation. To achieve this, a model for the gauge field with an explicit dependence on $z^-$ is required. Under certain conditions, the $z^-$ and $p^+$ path integrals can be exactly solved, leading to an energy spectrum for elastic projectile-medium scatterings. However, we acknowledge that a comprehensive non-eikonal treatment should incorporate the quark spin, which we plan to explore in a future project.

\subsubsection{The propagator in the case of a finite medium}
\label{sec:field_close_region}
	
	To finalize this section, we introduce the case in which the particle travels outside the medium, which is relevant for the analysis of proton-nucleus ($p$A) collisions\footnote{The case presented in this section is also relevant for DIS, where the virtual photon may split into a quark-antiquark pair outside the medium.}. In the hybrid formalism, this situation is represented as a projectile parton propagating from $z^+ = -\infty$ to $z^+ = \infty$, interacting with a classical field generated by the cold nuclear matter. Following standard field theory conventions, we assume that the gauge field $A^\mu(z)$ rapidly decreases, meaning it vanishes faster than any power of $z^+$ as $z^+ \to \pm \infty$. To account for this behavior, we introduce longitudinal points $z_i^+$ and $z_f^+$, where $z_f^+ > z_i^+$, such that the background field becomes negligible for $z^+ < z_i^+$ and $z^+ > z_f^+$. In practical terms, $z_i^+ \sim -R^+$ and $z_f^+ \sim R^+$, where $R^+$ is the longitudinal light-cone radius of the nucleus. Consequently, the parton only interacts with the medium within the region $z^+ \in [z_i^+, z_f^+]$.
	
	In order to simplify our analysis, we choose to work in the light-cone gauge $A^+ = 0$, which leads to a simpler solution for the $z^-$ path integral. However, it is important to note that the following analysis can be extended to a generic gauge if needed. From \cref{eq:ret_scalar_prop_config}, we observe that if $A^\mu(z)$ is nonzero only in the region $z_i^+ < z^+ < z_f^+$, the quantity $p^+(\tau^+)$ remains constant for $\tau^+ < \tau_i^+$ or $\tau^+ > \tau_f^+$. Here, the boundaries of the medium in the LC proper time are defined by $z^+(\tau_i^+) = z_i^+$ and $z^+(\tau_f^+) = z_f^+$. Consequently, the configuration for $p^+(\tau^+)$ is given by
	\begin{align}
		p^+(\tau^+) = 
		\begin{cases}
			k_i^+ & \text{if $y^+ < \tau^+ < \tau_i^+$}\\
			p^+_{\Delta}(\tau^+) & \text{if $\tau_i^+ < \tau^+ < \tau_f^+$}\\
			k_f^+ & \text{if $\tau_f^+ < \tau^+ < x^+$}
		\end{cases}
		,
	\end{align}
	where we define $p^+_\Delta(\tau^+)$ as the longitudinal momentum of the particle inside the medium, which is not constant due to multiple longitudinal scatterings with the nuclear matter. Additionally, we use the shorthand notation $\vec{k}_i \equiv \vec{p}(y^+)$ and $\vec{k}_f \equiv \vec{p}(x^+)$. From \cref{eq:avp}, it can be observed that in this case, the average longitudinal momentum is given by
	\begin{align}\label{eq:inmed_avgp}
		\avp = k_i^+ \frac{\tau^+_i-y^+}{x^+-y^+} + \langle p_\Delta^+ \rangle \frac{\tau^+_f-\tau^+_i}{x^+-y^+} + k_f^+ \frac{x^+-\tau^+_f}{x^+-y^+},
	\end{align}
	where we have defined the average in-medium longitudinal momentum as
	\begin{align}
		\langle p_\Delta^+ \rangle = \frac{1}{\tau^+_f-\tau^+_i} \int_{\tau^+_i}^{\tau^+_f} p_\Delta^+(\tau^+) d\tau^+.
	\end{align}
	
\begin{figure}[h!]
		\centering
		\includegraphics[scale=0.5]{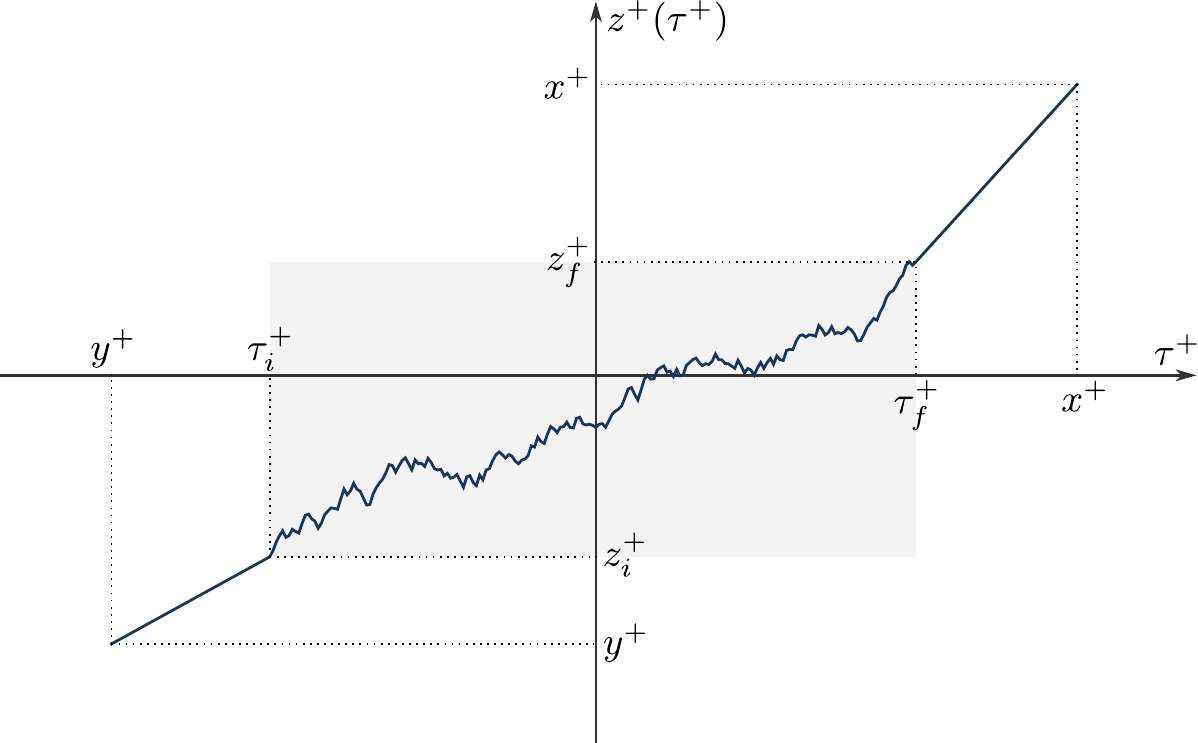}
		\caption{The light cone time, $z^+(\tau^+)$, when the background field is finite only in $z^+ \in [z_i^+,z_f^+]$ at a given configuration of the in-medium longitudinal momentum $p^+_\Delta(\tau^+)$. The shaded region represents the area in which the field is non-zero.}
		\label{fig:xip_vs_zp}
	\end{figure}

	From \cref{eq:xip}, we can observe that the relation between $z^+$ and $\tau^+$ is linear outside the medium, as illustrated in \cref{fig:xip_vs_zp}, where the slopes are given by the following relation:
	\begin{align}\label{eq:geometric_rel2}
		\frac{z_i^+-y^+}{\tau^+_i-y^+} = \frac{k_i^+}{\avp}, \qquad
		\frac{x^+-z_f^+}{x^+-\tau^+_f} = \frac{k_f^+}{\avp}.
	\end{align}
	These equations allow us to relate $z^+$ and $\tau^+$ in the boundary of the medium.
	
	In this case, we can easily solve \cref{eq:ret_scalar_prop_config} with the help of \cref{eq:pathint_dirac,eq:functional_dirac} and using the relationship given in \cref{eq:geometric_rel2}. The result is the following:
	\begin{align}\label{eq:aux9}
		\Delta_R(x,y) &=\Theta(x^+-y^+) 
		\int_{\vec{k}_i, \vec{k}_f} 
		\Theta(k_i^+) \Theta(k_f^+)
		e^{-i \hat{k}_f \cdot {x} + i \hat{k}_i \cdot {y}}
		\int_{\vec{z}_i,\vec{z}_f}
		e^{i \hat{k}_f \cdot {z}_f-i \hat{k}_i \cdot {z}_i}
		\int\mathcal{D}p_\Delta^+ (\tau^+)
		\breakp{-1}{\times}
		\frac{\Theta(\avp)}{2 \avp}
		\int_{\vec{z}(\tau^+_i)=\vec{z}_i}^{\vec{z}(\tau^+_f)=\vec{z}_f} \mathcal{D}^3\vec{z} (\tau^+)
		\
		e^{i \int_{\tau^+_i}^{\tau^+_f} d\tau^+ \left[-\frac{m_\epsilon^2}{2 \avp} + \frac{\avp}{2} \dot{\bf z}^2 -p^+ \dot{z}^- \right]}
		\mathcal{U}_{[\tau^+_f,\tau^+_i]}[z^+,\vec{z}],
	\end{align}
	where $\hat{k} \cdot {x} = \hat{k}^- x^+ + \vec{k} \cdot \vec{x}$. \Cref{eq:aux9} is more generic than \cref{eq:ret_scalar_prop_config}, and for this reason, in the analysis of the next section, we will employ \cref{eq:aux9} rather than \cref{eq:ret_scalar_prop_config}. Furthermore, due to the presence of the "LC energy" term $\hat{k}_f^- x^+ - \hat{k}_i^- y^+$ in the exponent, \cref{eq:aux9} facilitates a straightforward amputation of the propagator using an LSZ-like approach, as detailed in \cref{app:LSZ}.

\section{Eikonal expansion of the scalar propagator}
\label{sec:eikonal_expansion}

	In \cref{sec:intro}, we introduced the importance of the eikonal approximation in high-energy QCD, as well as its limitations. In this section, we will review the eikonal approximation, which corresponds to the saddle-point approximation of \cref{eq:scalar_prop_config,eq:scalar_prop_PS,eq:ret_scalar_prop_config}, where the paths follow their classical trajectory, and the interaction of the particle with the field is manifested by a change of phase in the projectile wave function. We will also present a systematic expansion in terms of powers of the inverse of the energy. Before delving into our analysis, let us introduce the eikonal approximation in the context of scattering theory.
	
	The eikonal approximation assumes that in a scattering process, the energy of the collision is much higher than the momentum transfer, as in the Gribov-Regge limit. In this scenario, the interaction of the particle with the background field is achieved by the exchange of soft gluons. In terms of Lorentz invariant variables, this means that the center-of-mass energy squared, $s$, of the collisions is much higher than the momentum transfer squared, $t$:
	\begin{equation}\label{eq:eikonal_mandelstam}
		s \gg |t| \ {\rm at \ fixed \ }t .
	\end{equation}
	
	Working in the target rest frame, the right-moving projectile ($p_i$) and target ($k_i$) momenta before the scattering are (in Minkowski coordinates)
	\begin{equation}\label{key}
		p_i = (\gamma m, \gamma \beta m, {\bf 0}), \qquad k_i = (M, 0, {\bf 0}),
	\end{equation}
	where $\gamma = \cosh \omega \approx e^\omega/2$ is the Lorentz gamma, $\omega$ is the rapidity of the projectile, $\beta=\sqrt{1-\gamma^{-2}}$, $m$ is the mass of the projectile, and $M$ is the mass of the target. After the scattering, the projectile exchanges momentum $Q^\mu=p_f^\mu - p_i^\mu$ with the target such that
	\begin{equation}\label{key}
		s \approx 2 m M \gamma, \qquad t=Q^2.
	\end{equation}
	
	In LC coordinates, the initial projectile longitudinal momentum is given by $p_i^+=\gamma m = s/2M$. Thus, assuming that the momentum exchange is much smaller than the initial momentum of the projectile, we have the following relations at high energy:
	\begin{align}\label{key}
		P^+ = \frac{p_i^++p_f^+}{2} \approx \frac{s}{2M} \to \infty,
		\qquad
		Q^- =-M \frac{t}{s} \to 0,
		\qquad
		Q^+ = \frac{t}{\sqrt{2}M}+Q^- \ll P^+,
	\end{align}
	where we have written the eikonal limit given by \cref{eq:eikonal_mandelstam}. Since $P^+ \sim s$ and $Q^+ \sim t$, in the following sections, we are going to study the corrections to the eikonal approximation as an expansion in powers of $1/P^+$ and $Q^+/P^+$. We work in the rest frame of the target and use the fact that the projectile is boosted to the right direction so that the longitudinal momentum $P^+ \sim m e^{\omega}$ is scaled by a large parameter. Because the eikonal approximation is better described in momentum space, it is convenient to write the scalar propagator in terms of its Fourier modes:
	\begin{equation}\label{key}
		\tilde{\Delta}_R(x^+,\vec{p}_f;y^+,\vec{p}_i) = \int_{\vec{x}, \vec{y}} e^{i \vec{p}_f \cdot \vec{x} - i \vec{p}_i \cdot 	\vec{y}} 
		\
		\Delta_R(x,y),
	\end{equation}
	where we just Fourier transform the $x^-$ and ${\bf x}$ coordinates for convenience. We note that since we are working in a mixed representation, we also have to take into account that, apart from the LC momentum $P^+$, LC time intervals, $\Delta x^+ \to e^{\omega} \Delta x^+$, are also going to be parametrically large under projectile boosts due to Lorentz dilation. Moreover, we use the representation derived in \cref{sec:field_close_region}, which we remember was derived in the LC gauge $A^+=0$, because it is more general, and in many practical problems, we are interested in the case in which the particle also propagates outside the medium. The Fourier transform of \cref{eq:aux9} is thus given by
	\begin{align}\label{eq:propagator_momentum}
		\tilde{\Delta}_R(x^+,\vec{p}_f;y^+,\vec{p}_i) &=
		\Theta(p_i^+) \Theta(p_f^+)
		\Theta(x^+-y^+)
		e^{-i \hat{p}_f^- {x}^+ + i \hat{p}_i^- {y}^+} \Delta_{\rm m}\left(\frac{\vec{p}_f+\vec{p}_i}{2},\vec{p}_f-\vec{p}_i\right),
	\end{align}
	where we have defined the in-medium scalar propagator as\footnote{We note that $\hat{Q}^- = \hat{p}_f^- - \hat{p}_i^-$ and $\hat{P}^- = \frac{\hat{p}_f^- + \hat{p}_i^-}{2}$.}
	\begin{align}\label{eq:aux_4}
		\Delta_{\rm m}(\vec{P},\vec{Q}) &=
		\int_{\vec{B},\vec{\Delta}}
		e^{i {\Delta} \cdot \hat{P}+i {B} \cdot \hat{Q}}
		\int\mathcal{D}p_\Delta^+ (\tau^+)
		\frac{\Theta(\avp)}{2 \avp}
		\nonumber \\ & \hskip0cm \times
		\int_{\vec{B}-\vec{\Delta}/2}^{\vec{B}+\vec{\Delta}/2} \mathcal{D}^3\vec{z} (\tau^+)
		\
		e^{i \int_{\tau^+_i}^{\tau^+_f} d\tau^+ \left[-\frac{m_\epsilon^2}{2 \avp} + \frac{\avp}{2} \dot{\bf z}^2 -p^+_\Delta \dot{z}^- \right]}
		\mathcal{U}_{[\tau^+_f,\tau^+_i]}[z^+,\vec{z}],
	\end{align}
	and
	\begin{align}
		P^\mu = \frac{p_i^\mu+p_f^\mu}{2}, \qquad Q^\mu=p_f^\mu-p_i^\mu,
		\\
		B^\mu = \frac{z_i^\mu+z_f^\mu}{2}, \qquad \Delta^\mu=z_f^\mu-z_i^\mu,
		\label{eq:B_def}
	\end{align}
	where $P$ is the average 4-momentum of the projectile and $Q$ the 4-momentum transfer by the medium.
	
	Since the eikonal approximation neglects the longitudinal momentum transfer in the scattering, it is convenient to write the path integral in terms of the local longitudinal momentum transfer $q^+(\tau^+)=\dot{p}^+_\Delta (\tau_f^+-\tau_i^+)$, which can be achieved by performing the change of variables:
	\begin{align}\label{eq:momentum_medium}
		p^+_\Delta(\tau^+) = 
		P^+ + \int_{\tau_i^+}^{\tau_f^+} \frac{d \bar{\tau}^+ q^+(\bar{\tau}^+)}{\tau_f^+-\tau_i^+} \frac{\epsilon(\tau^+-\bar{\tau}^+)}{2}.
	\end{align}
	where $\epsilon(x)=\Theta(x)-\Theta(-x)$ is the sign function. We note that $q^+(\tau^+)$ is the longitudinal momentum transfer from the nuclear matter to the particle at a given proper time $\tau^+$.
		
	In \cref{app:local_momentum}, we derive the expression for \cref{eq:aux_4} written in terms of $q^+(\tau^+)$ using the discrete representation of the path integral. The continuous limit, after performing this change of variables, is given by
	\begin{align}\label{eq:medium_propagtor}
		\Delta_{\rm m}(\vec{P},\vec{Q}) &=
		e^{ i \hat{Q}^- B^+ + i \hat{P}^- \Delta^+}
		\int_{{\bf B}, {\bf \Delta}} e^{-i {\bf B} \cdot {\bf Q}-i {\bf \Delta} \cdot {\bf P}} 
		\breakp{0}{\times}
		\int^{q^+(\tau_f^+) = 0} \mathcal{D}q^+ (\tau^+)
		\deltabar\left( \langle q^+ \rangle - Q^+ \right)
		\frac{\Theta(\avp)}{2 \avp }
		\int \mathcal{D} z^- (\tau^+)
		\int_{{{\bf B} - {\bf \Delta}/2}}^{{\bf B} + {\bf \Delta}/2} \mathcal{D}^2 {\bf z} (\tau^+)
		\breakp{0}{\times}
		\exp \Bigg\{ 
		i \int_{\tau^+_i}^{\tau^+_f} d\tau^+ \left[-\frac{m_\epsilon^2}{2 \avp} + \frac{\avp}{2} \dot{\bf z}^2 + \frac{q^+ z^-}{\tau_f^+-\tau_i^+} 	 \right]
		\Bigg\}		
		\mathcal{U}_{[\tau^+_f,\tau^+_i]}[z^+,\vec{z}],
	\end{align}
	where we have introduced $\deltabar(x) \equiv 2 \pi \delta(x)$ and the Dirac delta comes from the fact that the change of variables given in \cref{eq:momentum_medium} introduces the constraint that the average longitudinal momentum transfer is equal to the total longitudinal momentum transfer:
	\begin{align}
		\langle q^+ \rangle \equiv \int_{\tau_i^+}^{\tau_f^+} \frac{d \tau^+}{\tau_f^+-\tau_i^+} q^+(\tau^+) = Q^+.
	\end{align}
	
	Finally, the average longitudinal momentum, written in terms of the local momentum transfer, can be computed by using \cref{eq:inmed_avgp,eq:momentum_medium}, and reads
	\begin{align}\label{eq:avg_momentum_q}
		\avp = P^+ &+ \frac{Q^+}{2} \frac{(x^+-\tau_f^+)-(\tau_i^+-y^+)}{x^+-y^+} 
		\breakp{0}{+}
		\frac{1}{x^+-y^+} \int_{\tau_i^+}^{\tau_f^+} d\tau^+ \frac{q^+(\tau^+)}{2} \frac{(\tau_f^+-\tau^+)-(\tau^+-\tau_i^+)}{\tau_f^+-\tau_i^+}.
	\end{align}

\subsection{The eikonal approximation}
	
	In the eikonal $P^+ \to \infty$ limit, the path integral's action becomes dominated by the free Lagrangian:
	\begin{align}\label{eq:aux11}
		\mathcal{L}_0 = 
		-\frac{m_\epsilon^2}{2 \avp} + \frac{\avp}{2} \dot{\bf z}^2 + \frac{q^+ z^-}{\Delta \tau^+},
	\end{align}
	where in this section, we introduce $\Delta \tau^+ = \tau_f^+ - \tau_i^+$ and $B^+_\tau = (\tau_i^++\tau_f^+)/2$.
	This implies that the particle interacts weakly with the gauge field along its trajectory, and one can approximate its trajectory by the classical one. Thus, the eikonal approximation is equivalent to the saddle-point approximation in the functional approach. The effect of the field, as we are going to see below, is just to introduce an Aharonov-Bohm-like phase in the particle's wave function through the eikonal Wilson line, to be defined below.
	
	Thus, by solving the Euler-Lagrange equations of \cref{eq:aux11}, we obtain a set of equations for the classical trajectories:
	\begin{align}
		z_{\rm cl}^- = - \left( \frac{m^2}{2 (P^+)^2} + \frac{\dot{\bf z}^2}{2}\right) (\tau^+-B_\tau^+), \qquad
		\ddot{\bf z}_{\rm cl} = 0, \qquad
		q_{\rm cl}^+ = 0.
	\end{align}
	As one would expect, the particle's classical trajectory is given by a straight line where the longitudinal momentum is constant $p_{\rm cl}^+(\tau^+) = P^+$. We can express the slope of the transverse trajectory either in momentum space or in coordinate space (by making use of the boundary conditions ${\bf z}(\tau_i^+) = {\bf z}_i$ and ${\bf z}(\tau_f^+) = {\bf z}_f$). For the discussion performed in this chapter, it is convenient to express the slope in terms of momentum coordinates which is possible by making the action stationary with respect to the external points. In order to make the final result explicitly translational invariant, we expand the transverse trajectory around the impact parameter ${\bf B}$ so that:
	\begin{align}
		z_{\rm cl}^- = - \frac{m^2+{\bf P}^2}{2 (P^+)^2} (\tau^+-B_\tau^+), \qquad
		{\bf z}_{\rm cl} = {\bf B} + \frac{{\bf P}}{P^+}(\tau^+- B_\tau^+), \qquad
		z_{\rm cl}^+ = \tau^+.
	\end{align}
	
	In the functional approach presented in this manuscript, the eikonal approximation consists of two steps. First, we use the saddle-point approximation to evaluate the field around the classical straight-line path: $A^\mu(z) \approx A^\mu(z_{\rm cl})$. Second, we use the small-angle approximation \cite{Altinoluk_2014} in the $P^+ \to \infty$ limit to neglect the slope of the $\vec{z}_{\rm cl}$ trajectory, so that $A^\mu(z) \approx A^\mu(\tau^+, 0, {\bf B})$. Hence, using this approximation for the gauge field in \cref{eq:medium_propagtor}, we can solve the ${\bf z}$, $z^-$ and $q^+$ path integrals with the help of \cref{eq:functional_cuadratica,eq:pathint_dirac2,eq:pathint_dirac3}, and we obtain:
	\begin{align}
		\Delta_{\rm m}(\vec{P},\vec{Q})	& =
		e^{ i \hat{Q}^- B^+ + i \hat{P}^- \Delta^+}
		\deltabar(Q^+)
		\frac{\Theta(P^+)}{2 P^+ }
		\frac{P^+ }{2\pi i \Delta^+}
		\int_{{\bf B}, {\bf \Delta}} e^{-i {\bf B} \cdot {\bf Q}-i {\bf \Delta} \cdot {\bf P} + i \frac{P^+}{2\Delta^+} {\bf \Delta}^2} 
		U_{[z_f^+,z_i^+]}(0,{\bf B})
		\nonumber \\ & = 
		e^{i \frac{{\bf P} \cdot {\bf Q}}{P^+}B^+ +i \frac{{\bf Q}^2}{8 P^+}\Delta^+}
		\deltabar (Q^+ )
		\frac{\Theta(P^+)}{2 P^+ }
		\int_{{\bf B}} e^{-i {\bf B} \cdot {\bf Q}} 
		\
		U_{[z_f^+,z_i^+]}(0,{\bf B}),,
	\end{align}
	where we have used the fact that
	\begin{align}
		\hat{Q}^- B^+ + \hat{P}^- \Delta^+ - \frac{{\bf P}^2+m^2}{2P^+}\Delta^+ = \frac{{\bf P} \cdot {\bf Q}}{P^+}B^+ + \frac{{\bf Q}^2}{8 P^+}\Delta^+,
	\end{align}
	when $Q^+=0$. Neglecting the phases, that are subleading in the $P^+ \to \infty$ limit, we obtain the eikonal in-medium propagator:
	\begin{align}\label{eq:eikonal_propagator}
		\Delta_{\rm m}(\vec{P},\vec{Q}) = \deltabar (Q^+)
		\frac{\Theta(P^+)}{2 P^+ }
		\int_{{\bf B}} e^{-i {\bf B} \cdot {\bf Q}} 
		\
		U_{[z_f^+,z_i^+]}(0,{\bf B}).
	\end{align}
	
	Note that this propagator is invariant under transverse Galilean transformations, i.e., it only depends on the transverse momentum transfer ${\bf Q}$ but not on ${\bf P}$. This implies that, when written in coordinate space, it is going to be diagonal in the transverse coordinates. It describes the propagation of a fast-moving scalar particle in a classical background field where the recoil of the medium is neglected. The only effect of the interaction with the medium is a color rotation in the projectile's wave function that is described by the {\it eikonal Wilson line}:\footnote{Note that the transverse component of the background field does not contribute to the eikonal Wilson line since it is suppressed by $\dot{\bf z}\sim 1/P^+$.}
	\begin{align}\label{eq:eikonal_WL}
		U_{[x^+,y^+]}(\vec{z}) = \mathcal{P}_+ e^{-ig \int_{y^+}^{x^+} dz^+ A^-(z^+,\vec{z})}. 
	\end{align}
	Strictly speaking, the eikonal Wilson line is defined with $z^-=0$ since, in the eikonal limit, the particle only probe the field in this region. However, we define it with $z^- \ne 0$ for future convenience.

\subsection{The eikonal expansion}

	By employing the saddle point approximation, $A^\mu(z) \approx A^\mu(z_{\rm cl})$, and the small-angle approximation, $P^+ \to \infty$, we can obtain the eikonal scalar propagator. However, in this section, we aim to go beyond these approximations and systematically expand the scalar propagator in terms of $1/P^+$ by computing the finite boost corrections. To achieve this, we adopt a strategy similar to the one employed in \cite{Altinoluk_2014}, which can be outlined as follows:
	\begin{enumerate}
		\item We relax the saddle point approximation by Taylor expanding the background field around the classical trajectory. To achieve this, we perform the following change of variables:
		\begin{align}\label{eq:transverse_path}
			z^-=B^-+v^-, \qquad {\bf z} =  {\bf z}_{\rm cl} + {\bf v}, \qquad {\bf z}_{\rm cl} \equiv {\bf B} + \frac{\bf \Delta \ }{\Delta\tau^+} (\tau^+-B_\tau^+),
		\end{align}
		where we represent the classical transverse trajectory slope in coordinate space for convenience, and $B^-$ is the conjugate of $Q^+$ and can be identified as the one defined in \cref{eq:B_def}. We adopt the convention of expanding around $z^- = B^-$ instead of $z_{\rm cl}^-$ because it simplifies the analysis, and at the end of the calculation, we revert to the convention of \cite{Altinoluk_2021} and expand around $B^-=0$. The Taylor expansion of the background field is then given by
		\begin{align}
			A^\mu(z^+,z^-,{\bf z}) = \exp \big\{v^- \partial^+ + {\bf v}^i \cdot \partial_{{\bf z}_{\rm cl}^i} \big\} A^\mu(z^+,B^-,{\bf z}_{\rm cl}).
		\end{align}
		
		\item We perform the small angle limit by assuming that the transverse slope is small and we Taylor expand again the background field around ${\bf z}_{\rm cl} = {\bf B}$:
		\begin{align}\label{eq:field_taylor}
			A^\mu(z^+,z^-,{\bf z}) = \exp \Bigg\{ 
			v^- \partial^++
			\left[{\bf v}^i + \frac{\bf \Delta \ }{\Delta\tau^+} (\tau^+-B_\tau^+) \right] \cdot \partial_{{\bf B}^i} \Bigg\} A^\mu(z^+,B^-,{\bf B}).
		\end{align}		
	\end{enumerate}
	
	The main advantage of the change of variables given in \cref{eq:transverse_path} is that it allows as to write the scalar propagator \cref{eq:medium_propagtor} as
	\begin{align}\label{eq:propagator_expansion1}
		\Delta_{\rm m}(\vec{P},\vec{Q}) &
		= 
		e^{ i \hat{Q}^- B^+ + i \hat{P}^- \Delta^+}
		\int_{\vec{B}} e^{i \vec{B} \cdot \vec{Q}} 
		\int_{\bf \Delta} e^{-i {\bf \Delta} \cdot {\bf P} + i \frac{\avp}{2 \Delta\tau^+} {\bf \Delta}^2 - i \frac{m_\epsilon^2}{2 \avp} \Delta\tau^+}
		\breakp{0}{\times}
		\int^{q^+(\tau_f^+) = 0} \mathcal{D}q^+(\tau^+)
		\frac{\Theta(\avp)}{2 \avp }
		\int \mathcal{D} v^- (\tau^+)
		e^{i \int_{\tau_i^+}^{\tau_f^+} d\tau^+ \frac{v^- q^+}{\Delta \tau^+}}
		\breakp{0}{\times}
		\int_{{\bf v}(\tau_i^+) = 0}^{{\bf v}(\tau_f^+) = 0} \mathcal{D}^2 {\bf v}(\tau^+)
		\ e^{i \int_{\tau_i^+}^{\tau_f^+} d\tau^+ \frac{\avp}{2} \dot{\bf v}^2}
		\mathcal{U}_{[\tau_f^+,\tau_i^+]}\big[z^+,B^- + v^-, {\bf z}_{\rm cl} + {\bf v}\big],
	\end{align}
	so that the transverse path integral has periodical boundary conditions: ${\bf v}(\tau_i^+) = {\bf v}(\tau_f^+) = 0$. In order to get \cref{eq:propagator_expansion1}, we have written the Dirac delta that constrains the mean longitudinal momentum transfer as
	\begin{align}
		\deltabar\left( \langle q^+ \rangle - Q^+ \right) = \int_{B^-} e^{i B^- (Q^+-\langle q^+ \rangle)},
	\end{align}
	thus, the second term in the exponent cancels out after the change of variables.
	
	We can solve \cref{eq:propagator_expansion1} by expanding the Wilson line in terms of ${\bf v}$, ${\bf \Delta}$, and $v^-$. As $\avp = \mathcal{O}(P^+)$, it is clear from \cref{eq:propagator_expansion1} that the (path) integral in (${\bf v}$) ${\bf \Delta}$ behaves like a Gaussian-like (path) integral. Consequently, terms of order (${\bf v}^{2n}$) ${\bf \Delta}^{2n}$ in the Wilson line expansion will contribute corrections of $\mathcal{O}(1/(P^+)^{n})$. On the other hand, as we shall see in the subsequent discussion, terms of order $(v^-)^n$ will introduce an $n^{\rm th}$ order derivative in the longitudinal momentum transfer, resulting in a correction of order\footnote{In fact, corrections arising from the $v^-$ expansion yield powers of $1/P^+(x^+-y^+)$ that, as previously discussed, scale with the boost rapidity as $e^{- 2 \omega}$.} $\mathcal{O}(1/(P^+)^{2n})$. 
	
	In order to make the notation cleaner, let us define
	\begin{align}
		\tilde{A}_{z^+}^\mu \equiv - i g A^{\mu} (z^+,B^-,{\bf B}).
	\end{align}
	
	It is also convenient to introduce the {\it insertion} tensor, $\mathcal{I}$, which represents the part in the exponent of the Wilson line that vanishes in the eikonal limit and can be expressed as $\mathcal{I} = -ig[ A^-(z^+, \vec{z}) z^+ - {\bf A} (z^+, \vec{z}) \cdot \dot{\bf z} ] - \tilde{A}_{z^+}^- \dot{z}^+$. Utilizing \cref{eq:field_taylor}, it can be further written as
	\begin{align}\label{eq:insertion}
		&\mathcal{I}_{\tau^+} \left[v^-, {\bf v}, {\bf \Delta}  \right] 
		\breakp{0}{\equiv}
		\exp\left\{ v^- \partial^+ + \left[ {\bf v}^i + \frac{{\bf \Delta}^i}{\Delta\tau^+} (\tau^+-B_\tau^+) \right] {\partial}_{{\bf B}^i} \right\}  
		\left[
		\tilde{A}_{z^+}^- \dot{z}^+
		-
		\tilde{\bf A}_{z^+} \cdot \left( \dot{\bf v} + \frac{\bf \Delta}{\Delta\tau^+} \right)
		\right] 
		-\tilde{A}_{z^+}^- \dot{z}^+
		\breakp{0}{=}
		-
		\tilde{\bf A}_{z^+} \cdot \left( \dot{\bf v} + \frac{\bf \Delta}{\Delta\tau^+} \right)
		+
		\left( v^- \partial^+ + \left[ {\bf v}^i + \frac{{\bf \Delta}^i}{\Delta\tau^+} (\tau^+-B_\tau^+) \right] {\partial}_{{\bf B}^i} \right) 
		\tilde{A}_{z^+}^- \dot{z}^+
		+ \cdots.
	\end{align}
	
	Thus, the Wilson line can be expanded in terms of the insertions as
	\begin{align}
		&\mathcal{U}_{[\tau_f^+,\tau_i^+]}\big[z^+,B^-+v^-, {\bf z}_{\rm cl} + {\bf v}\big] = 
		\mathcal{P}_+
		\exp \Bigg\{ 
		\int_{\tau_i^+}^{\tau_f^+} d\tau^+
		\Big(\mathcal{I}_{\tau^+} \left[v^-, {\bf v}, {\bf \Delta}  \right]+ \tilde{A}_{z^+}^- \dot{z}^+ \Big)
		\Bigg\}
		\breakp{0}{=}
		\mathcal{P}_+ e^{\int_{\tau_i^+}^{\tau_f^+} d\tau^+ \tilde{A}_{z^+}^- \dot{z}^+ }
		\breakp{2}{+}
		\mathcal{P}_+
		\int_{\tau_i^+}^{\tau_f^+} d\tau_1^+ e^{\int_{\tau_1^+}^{\tau_f^+} d\tau^+ \tilde{A}_{z^+}^- \dot{z}^+} 
		\mathcal{I}_{\tau_1^+} \left[v^-, {\bf v}, {\bf \Delta}  \right] 
		e^{\int_{\tau_i^+}^{\tau_1^+} d\tau^+ \tilde{A}_{z^+}^- \dot{z}^+} + \cdots,
	\end{align}
	where the $m^{\rm th}$ order of the expansion is obtained by the path-ordered product of $m$ insertions. The expansion of the Wilson line can be written in a generalized way as
	\begin{align}\label{eq:aux12}
		&\mathcal{U}_{[\tau_f^+,\tau_i^+]}\big[z^+,B^-+v^-, {\bf z}_{\rm cl} + {\bf v}\big] 
		\nonumber \\  & \hskip1cm =
		\sum_{m=0}^{\infty} 
		\Bigg(
		\prod_{n=m}^{1} \mathcal{P}_+ \int_{\tau_i^+}^{\tau_{n+1}^+} d\tau_n^+ 
		e^{\int_{\tau_n^+}^{\tau_{n+1}^+} d\tau^+ \tilde{A}_{z^+}^- \dot{z}^+} 
		\mathcal{I}_{\tau_n^+} \left[v^-, {\bf v}, {\bf \Delta}  \right]
		\Bigg)
		e^{\int_{\tau_i^+}^{\tau_{1}^+} d\tau^+ \tilde{A}_{z^+}^- \dot{z}^+},
	\end{align}
	where $\tau_{m+1}^+ \equiv \tau_f^+$ and $\tau_0^+ \equiv \tau_i^+$.
	
	Given the expansions in \cref{eq:insertion,eq:aux12}, the next step is to solve the integrals over ${\bf \Delta}$, ${\bf v}(\tau^+)$, and $v^-(\tau^+)$. The first one is the most straightforward, as for an infinitely differentiable function $f({\bf \Delta})$, the Gaussian integral is given by
	\begin{equation}\label{eq:gauss_integral}
		\int_{\bf \Delta} e^{-i {\bf \Delta} \cdot {\bf P}-\frac{\avp}{2i \Delta \tau^+} {\bf \Delta}^2- i \frac{m_\epsilon^2}{2 \avp} \Delta \tau^+} f({\bf \Delta}) = 
		\frac{2\pi i \Delta \tau^+}{\avp} f(i \partial_{\bf P}) e^{-i \frac{{\bf P}^2+m_\epsilon^2}{2 \avp} \Delta \tau^+}.
	\end{equation}
	
	In the case of the path integral in ${\bf v}(\tau^+)$, we are interested in solving a path integral of the following type:
	\begin{align}\label{key}
		&\int_{{\bf v}(\tau_i^+) = 0}^{{\bf v}(\tau_f^+) = 0} \mathcal{D}^2{\bf v}(\tau^+) 
		\
		e^{i \int_{\tau_i^+}^{\tau_f^+} d\tau^+ \frac{\avp}{2} \dot{\bf v}^2} F[{\bf v}]
		\breakp{3}{=}
		\int_{{\bf v}(\tau_i^+) = 0}^{{\bf v}(\tau_f^+) = 0} \mathcal{D}^2{\bf v}(\tau^+) 
		\
		e^{- \frac{1}{2} \int_{\tau_i^+}^{\tau_f^+} d\tau^+ {\bf v}\left( i \avp \frac{d^2}{(d\tau^+)^2} \right){\bf v}} F[{\bf v}],
	\end{align}
	where $F[{\bf v}]$ is a generic functional that can be Taylor expanded, and in the last step, we have integrated by parts. This integral can be solved analogously to \cref{eq:gauss_integral} by defining the generating functional:
	\begin{align}\label{key}
		&Z[{\bf J}] = \int_{{\bf v}(\tau_i^+) = 0}^{{\bf v}(\tau_f^+) = 0}  \mathcal{D}^2{\bf v}(\tau^+) 
		e^{\int_{\tau_i^+}^{\tau_f^+} d\tau^+ \left[i \frac{\avp}{2} \dot{\bf v}^2 + {\bf J}(\tau^+) \cdot {\bf v} \right]}
		\breakp{3}{=}
		Z[0] \exp\Bigg\{ \frac{1}{2} \int_{\tau_1^+ \tau_2^+} {\bf J}^i(\tau_1^+) G^{ij}(\tau_1^+,\tau_2^+) {\bf J}^j(\tau_2^+)\Bigg\},
	\end{align}
	where $\tau_1^+,\tau_2^+ \in [\tau_i^+,\tau_f^+]$. The generating functional allows us to write the solution of the Gaussian path integral
	as 
	\begin{align}\label{eq:result_gauss_path_int}
		\int_{{\bf v}(\tau_i^+) = 0}^{{\bf v}(\tau_f^+) = 0} \mathcal{D}^2{\bf v}(\tau^+) 
		\
		e^{i \int_{\tau_i^+}^{\tau_f^+} d\tau^+ \frac{\avp}{2} \dot{\bf v}^2} F[{\bf v}] = F\left[ \frac{\delta \ }{\delta {\bf J}} \right] Z[{\bf J}] \bigg|_{{\bf J}=0}.
	\end{align}
	
	Here, $G^{ij}(\tau_1^+,\tau_2^+)$ is the inverse of the Gaussian coefficient and is the Green's function of the following one dimensional Poisson equation:
	\begin{equation}\label{key}
		i \avp \frac{d^2}{(d\tau_1^+)^2} G^{ij}(\tau_1^+,\tau_2^+) = \delta^{ij} \delta(\tau_1^+-\tau_2^+),
	\end{equation}
	with the Dirichlet boundary conditions $G^{ij}(\tau_f^+,\tau^+)=G^{ij}(\tau^+,\tau_i^+)=0$ for every $\tau^+ \in [\tau_i^+,\tau_f^+]$. The solution for this equation is well known and reads
	\begin{align}\label{eq:2point}
		&G^{ij}(\tau_1^+,\tau_2^+)
		\breakp{0.1}{=}
		\frac{i \delta^{ij}}{\avp} \left[
		\frac{(\tau_f^+-\tau_1^+)(\tau_2^+-\tau_i^+)}{\tau_f^+-\tau_i^+} \Theta(\tau_1^+-\tau_2^+)
		+ \frac{(\tau_f^+-\tau_2^+)(\tau_1^+-\tau_i^+)}{\tau_f^+-\tau_i^+} \Theta(\tau_2^+-\tau_1^+)
		\right].
	\end{align}
	
	On the other hand, the "free" generating functional is given by
	\begin{equation}\label{key}
		Z[0] = \int_{{\bf v}(\tau_i^+) = 0}^{{\bf v}(\tau_f^+) = 0} \mathcal{D}^2{\bf v}(\tau^+)  \ e^{i \int_{\tau_i^+}^{\tau_f^+} \frac{\avp}{2} \dot{\bf v}^2} = \frac{\avp}{2 \pi i \Delta \tau^+}.
	\end{equation}	
	
	Since the generating functional is Gaussian, it implies that any odd power of the Taylor expansion of $F[{\bf v}]$ will vanish, and even powers higher than 2 can be computed just by means of the Wick theorem, i.e., by summing over all possible permutations of the products of the 2-point function given in \cref{eq:2point}. Moreover, because of the $\dot{\bf v}$ term multiplying the transverse field in \cref{eq:insertion}, apart from the 2-point function, we are also going to need its time derivatives whenever a transverse field is inserted in the expansion of the Wilson line. Thus, we define the four building blocks for the expansion over ${\bf v}$ as follows:
	\begin{align}
		G^{ij}(\tau_1^+,\tau_2^+) &\equiv \langle {\bf v}^i(\tau_1^+) {\bf v}^j(\tau_2^+) \rangle,
		\label{eq:2poinder0}
		\\
		G^{ij}(\tau_1^+,\tau_2^+)^\bullet &\equiv \langle {\bf v}^i(\tau_1^+) \dot{\bf v}^j(\tau_2^+) \rangle 
		\breakp{0}{=}
		i\frac{\delta^{ij}}{\avp} \left[
		\frac{\tau_f^+-\tau_1^+}{\tau_f^+-\tau_i^+} \Theta(\tau_1^+-\tau_2^+)
		- \frac{\tau_1^+-\tau_i^+}{\tau_f^+-\tau_i^+} \Theta(\tau_2^+-\tau_1^+)
		\right],
		\label{eq:2poinder1}
		\\
		\prescript{\bullet}{}G^{ij}(\tau_1^+,\tau_2^+) &\equiv \langle \dot{\bf v}^i(\tau_1^+) {\bf v}^j(\tau_2^+) \rangle = G^{ij}(\tau_2^+,\tau_1^+)^\bullet,
		\label{eq:2poinder2}
		\\
		\prescript{\bullet}{}G^{ij}(\tau_1^+,\tau_2^+)^\bullet &\equiv \langle \dot{\bf v}^i(\tau_1^+) \dot{\bf v}^j(\tau_2^+) \rangle = i
		\frac{\delta^{ij}}{\avp} \left[\delta(\tau_1^+-\tau_2^+)-\frac{1}{\tau_f^+-\tau_i^+}
		\right],
		\label{eq:2poinder3}
	\end{align}
	where we have defined
	\begin{align}
		\langle F[{\bf v}] \rangle = Z[0]^{-1}  \int_{{\bf v}(\tau_i^+) = 0}^{{\bf v}(\tau_f^+) = 0} \mathcal{D}^2{\bf v}(\tau^+) \ e^{i \int_{\tau_i^+}^{\tau_f^+} \frac{\avp}{2} \dot{\bf v}^2} F[{\bf v}].
	\end{align}
	
	Finally, the last path integral that we have to deal with is the one over $v^-(\tau^+)$. This one can be easily solved by noting that
	\begin{align}\label{eq:aux13}
		&\int \mathcal{D} v^-(\tau^+) e^{i \int_{\tau_i^+}^{\tau_f^+} d\tau^+ \frac{v^- q^+}{\Delta\tau^+}} v^-(\tau_1^+) \cdots v^-(\tau_n^+) 
		\breakp{3}{=}
		(\Delta\tau^+)^n \frac{\delta \ }{i \delta q^+(\tau_1^+)} \cdots \frac{\delta \ }{i \delta q^+(\tau_n^+)} 
		\int \mathcal{D} v^- e^{i \int_{\tau_i^+}^{\tau_f^+} d\tau^+ \frac{v^- q^+}{\Delta\tau^+}}
		\breakp{3}{\equiv} 
		\frac{\delta \ }{i \delta q^+(\tau_1^+)} \cdots \frac{\delta \ }{i \delta q^+(\tau_n^+)} \deltabar[q^+(\tau^+)],
	\end{align}
	where the functional derivative of the Dirac delta, given in the last step, is defined in \cref{app:pathint_dirac_der}. It implies that the result of the integration over $n$ insertions $v^-(\tau_1^+) \cdots v^-(\tau_n^+) $ is to set $q^+(\tau^+)=0$ in between $\tau^+\in(\tau_{m}^+,\tau_{m+1}^+)$ ($m=0,\dots,n$). Hence, for a generic functional $G[v^-]$ that can be Taylor expanded, we obtain the following result:
	\begin{align}\label{eq:result_dirac_int}
		\int \mathcal{D} v^-e^{i \int_{\tau_i^+}^{\tau_f^+} d\tau^+ \frac{v^- q^+}{\Delta\tau^+}} G[v^-] = G\left[ \frac{\delta \ }{i \delta q^+} \right] \deltabar[q^+(\tau^+)].
	\end{align}
	
	Before deriving the final expression, let us examine the impact of the $v^-$ insertions in \cref{eq:propagator_expansion1,eq:aux12}. We assume that there are $n$ insertions of $v^-(\tau^+)$ at light-cone proper times $\tau_1^+,\dots,\tau_n^+$. Due to \cref{eq:aux13}, the longitudinal momentum transfer vanishes within the intervals $\tau^+ \in (\tau_{m}^+,\tau_{m+1}^+)$. We introduce the shorthand notation $q_m^+ \equiv q^+(\tau_m^+)$, which allows us to express it as $q^+(\tau^+) = \sum_{m=1}^n q_m^+ \delta([\tau^+ - \tau_m^+]/[\tau_f^+ - \tau_i^+])$. Hence, the in-medium longitudinal momentum, \cref{eq:momentum_medium}, is given by
	\begin{align}
		p_\Delta^+(\tau^+) = P^+ + \sum_{m=1}^{n} \frac{q_m^+}{2} \epsilon(\tau^+-\tau_m^+),
	\end{align}
	i.e., it is constant within the region $\tau^+ \in (\tau_{m}^+,\tau_{m+1}^+)$ and at each longitudinal point $\tau_m^+$ it receives a longitudinal "kick" $q_m^+/2$ due to the interaction with the medium. Moreover, using the constraint $\sum_{m=1}^{n} q_m^+ = Q^+$, we see that the longitudinal momentum in the boundary of the medium is given by: $p_\Delta^+(\tau_i^+) = P^+-Q^+/2 = p_i^+$ and $p_\Delta^+(\tau_f^+)=P^++Q^+/2=p_f^+$. 
	
\begin{figure}[h!]
		\centering
		\includegraphics[scale=0.7]{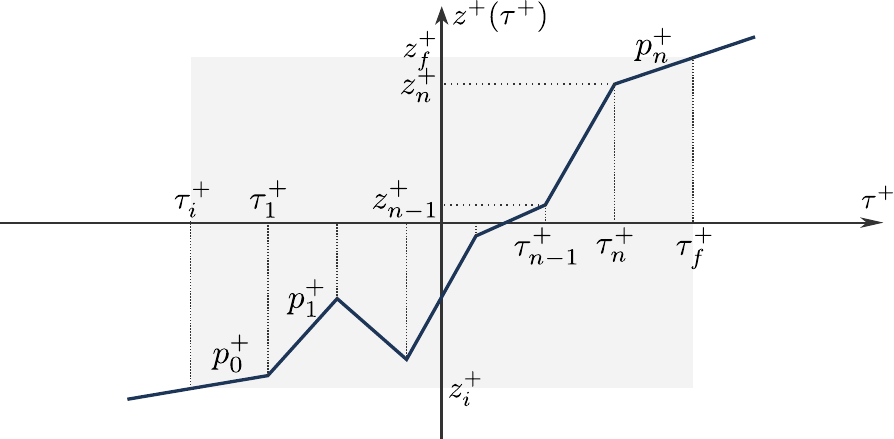}
		\caption{Dependence of the LC time $z^+(\tau^+)$ with the LC proper time $\tau^+$ after $n$ insertions in $v^-(\tau_m^+)$. Each insertion fixes the momentum to be constant within a region $[\tau_m^+,\tau_{m+1}^+]$ so that, because of \cref{eq:xip}, the dependence of $z^+$ with $\tau^+$ is linear in this region and the slope is given by $p_m^+/\avp$.}
		\label{fig:xip_vs_zp_discrete}
	\end{figure}

	Analogous to the discussion performed in \cref{sec:field_close_region}, when the longitudinal momentum is constant in a region $(\tau_{m}^+,\tau_{m+1}^+)$, $z^+(\tau^+)$ is going to be linear in this region, as can be read from \cref{eq:xip}. Thus, as illustrated in \cref{fig:xip_vs_zp_discrete}, we can relate $z_m^+ \equiv z^+(\tau_m^+)$ and $\tau_m^+$ as follows: 
	\begin{align}\label{eq:geometric_relation}
		\frac{z_{m+1}^+-z_m^+}{\tau_{m+1}^+-\tau_{m}^+}=\frac{p_m^+}{\avp}, \qquad \tau_m^+=\tau_i^+ + \sum_{l=0}^{m} \frac{\avp}{p_l^+}(z_{l+1}^+-z_{l}^+),
	\end{align}
	where we have introduced $p_m^+ \equiv p_\Delta^+ (\tau_m^+)$.
	
	Finally, since $z^+(\tau^+)$ is linear in between two insertions, the transformation $\tau^+ \to z^+(\tau^+)$ is a diffeomorphism and we can write the path ordered exponential appearing in \cref{eq:aux12} in terms of the eikonal Wilson lines:
	\begin{align}\label{eq:aux22}
		&\mathcal{P}^+ \exp \left\{ \int_{\tau_{m}^+}^{\tau_{m+1}^+} d\tau^+ \tilde{A}_{z^+}^- \dot{z}^+ \right\} 
		\breakp{0}{=} 
		\Theta(z_{m+1}^+-z_{m}^+) \Theta(p_m^+) U_{[z_{m+1}^+,z_{m}^+]}(\vec{ B})
		+ \Theta(z_{m}^+-z_{m+1}^+) \Theta(-p_m^+) U^\dagger_{[z_{m}^+,z_{m+1}^+]}(\vec{ B}).
	\end{align}
	
	Inserting \cref{eq:aux22} into \cref{eq:aux12} we obtain the expansion of the Wilson line in terms of its eikonal homologous:
	\begin{align}\label{eq:aux15}
		&\mathcal{U}_{[\tau_f^+,\tau_i^+]}\big[z^+,v^-+B^-, {\bf z}_{\rm cl} + {\bf v}\big] 
		\breakp{1}{=}
		\sum_{m=0}^{\infty}
		\Bigg(
		\prod_{n=m}^{1} \int_{\tau_i^+}^{\tau_{n+1}^+} d\tau_n^+ 
		\Bigg[
		\Theta(z_{n+1}^+-z_{n}^+) \Theta(p_{n}^+)  U_{[z_{n+1}^+,z_{n}^+]}(\vec{B})
		\breakp{2}{+}
		\Theta(z_{n}^+-z_{n+1}^+) \Theta(-p_{n}^+) U^\dagger_{[z_{n+1}^+,z_{n}^+]}(\vec{B})
		\Bigg]
		\mathcal{I}_{\tau_n^+} \left[v^-, {\bf v}, {\bf \Delta}  \right]
		\Bigg)U_{[z_{1}^+,z_{i}^+]}(\vec{B}),
	\end{align}
	where $\tau_{m+1}^+ = \tau_f^+$ and $z_{m+1}^+ = z_f^+$.
	
	All in all, inserting the results obtained in \cref{eq:gauss_integral,eq:result_gauss_path_int,eq:result_dirac_int} into \cref{eq:propagator_expansion1}, and the expansion of the Wilson lines in terms of the eikonal ones given in \cref{eq:aux15}, we obtain the final expression for the eikonal expansion:
	\begin{align}\label{eq:eikonal_expansion}
		&\Delta_{\rm m}(\vec{P},\vec{Q})
		= e^{ i \hat{Q}^- B^+ + i \hat{P}^- \Delta^+} 
		\int_{\vec{B}} e^{i \vec{B} \cdot \vec{Q}} 
		\int^{q^+(\tau_f^+)=0} \mathcal{D}q^+
		\frac{\Theta(\avp)}{2 \avp }
		\sum_{m=0}^{\infty} 
		\Bigg(
		\prod_{n=m}^{1} \int_{\tau_i^+}^{\tau_{n+1}^+} d\tau_n^+ 
		\breakp{1}{\times} 
		\Bigg[
		U_{[z_{n+1}^+,z_{n}^+]}(\vec{B}) \Theta(z_{n+1}^+-z_{n}^+) \Theta(p_{n}^+) + U^\dagger_{[z_{n+1}^+,z_{n}^+]}(\vec{B}) \Theta(z_{n}^+-z_{n+1}^+) \Theta(-p_{n}^+)
		\Bigg]
		\breakp{1}{\times} 
		\mathcal{I}_{\tau_n^+} \left[\frac{\delta \ }{i \delta q^+}, \frac{\delta \ }{\delta {\bf J}},i \partial_{\bf P}\right] 
		\Bigg)
		U_{[z_{1}^+,z_{i}^+]}(\vec{B}) \
		e^{-i \frac{{\bf P}^2+m_\epsilon^2}{2 \avp} \Delta \tau^+} \frac{Z[{\bf J}]}{Z[0]} \Bigg|_{{\bf J}=0} \deltabar[q^+(\tau^+)].
	\end{align}
	\Cref{eq:eikonal_expansion} is the most important result of this section and can be utilized to compute the eikonal corrections to the scalar propagator at a specific order in the boost parameter $e^{-\omega}$. We note that although we explicitly write the path integral in $q^+$, its solution is straightforward due to the functional derivatives of the Dirac delta. On the other hand, $p_m^+$ and $\avp$ are given in terms of $q^+$ by \cref{eq:avg_momentum_q,eq:momentum_medium}, respectively. We do not explicitly write this dependence to keep the expression concise. Moreover, the relation of $\tau_n^+$ and $z_n^+$ between two longitudinal insertions is given by \cref{eq:geometric_relation}.
	
	To summarize, the eikonal expansion given by \cref{eq:eikonal_expansion} together with \cref{eq:insertion} has two different sources. On one hand, we have corrections due to the longitudinal momentum transfer from the medium to the projectile. This correction arises from the $z^-$ dependence of the field, which is taken into account by an expansion in terms of $\delta^n/\delta^n q^+$. As illustrated in \cref{sec:first_order_correction}, where we computed the first-order correction to \cref{eq:eikonal_expansion}, the source of this correction comes from the fluctuations of $\avp$ around its classical value $P^+$. One can observe from \cref{eq:avg_momentum_q} that these fluctuations, which appear through the functional derivative of $q^+$, are $\mathcal{O}(1/[P^+(x^+-y^+)]) = \mathcal{O}(e^{-2\omega})$ and are therefore double-suppressed in the eikonal limit. Thus, in order to obtain a correction at order $(e^{-\omega})^n$, it is sufficient to expand \cref{eq:eikonal_expansion} up to order $(\delta/\delta q^+)^{n/2}$.
	
	On the other hand, there are corrections due to a finite width of the medium, $z_f^+ - z_i^+$, as well as interactions with the transverse component of the field, which result in a transverse motion of the particle inside the medium. This makes the propagator non-diagonal in the transverse coordinates. In the functional approach, these corrections arise from fluctuations around the classical transverse trajectory of the particle and are accounted for by an expansion in terms of $\delta/\delta {\bf J}$ and $\partial_{\bf P}$. In this case, since the only dependence of ${\bf P}$ and ${\bf J}$ comes from Gaussian functions whose coefficient is $\mathcal{O}(1/P^+)$, in order to obtain a correction of order $(e^{-\omega})^n$, it is sufficient to perform the transverse expansion of \cref{eq:eikonal_expansion} up to order $2n$.
	
	Finally, we should also perform the expansion in $B^-$. It is easy to read from the Fourier exponent of \cref{eq:eikonal_expansion} that an expansion of the eikonal Wilson lines around $B^-=0$ will result in an expansion in derivatives of $\deltabar(Q^+)$. As we explain in \cref{app:aux1}, each derivative of $\deltabar(Q^+)$ will lead to a suppression of $\mathcal{O}(e^{-\omega})$ once we write the propagator in coordinate space. Therefore, in order to obtain a correction of $\mathcal{O}((e^{-\omega})^n)$ we should expand the eikonal Wilson lines up to order $(B^-)^n$.
	
	Although \cref{eq:eikonal_expansion} provides a systematic expansion that can be expanded at any order in $e^{-\omega}$, evaluating it at high orders can become cumbersome. In such cases, it can be advantageous to introduce "non-eikonal" Feynman rules to simplify the calculation in terms of a diagrammatic approach.
	
	In order to illustrate this approach, in \cref{sec:first_order_correction}, we have computed the scalar propagator at first order in the eikonal expansion. This result has also been independently computed recently in \cite{Altinoluk_2022}, providing a valuable cross-check for the approach presented in this manuscript. At first order, the result we obtain for the retarded scalar propagator in momentum space is as follows:	
	\begin{align}\label{}
		\Delta_{\rm m}(\vec{P},\vec{Q}) &=
		\deltabar(Q^+)
		\frac{\Theta(P^+)}{2 P^+} 
		\int_{\bf{B}} e^{-i \bf{B} \cdot \bf{Q}}
		\Bigg\{
		\left(1 + i \frac{{\bf P} \cdot {\bf Q}}{P^+}B^+\right){U}_{[z_f^+,z_i^+]}(0,{\bf B})
		\nonumber \\ &  \hskip0cm
		-
		\int_{z_i^+}^{z_f^+} dz^+ {U}_{[z_f^+,z^+]}(0,{\bf B})
		\Bigg[
		\frac{{\bf P}^i}{P^+} \overleftrightarrow{D}_{{\bf B}^i}
		+ \frac{i}{2 P^+} 
		\overleftarrow{D}_{{\bf B}^i}	\overrightarrow{D}_{{\bf B}^i}
		\Bigg] {U}_{[z^+, z_i^+]}(0,{\bf B})
		\Bigg\}
		\nonumber \\ &  \hskip0cm
		-i\deltabar'(Q^+)
		\frac{\Theta(P^+)}{2 P^+}
		\int_{\bf{B}} e^{-i \bf{B} \cdot \bf{Q}} \partial^+ {U}_{[z_f^+,z_i^+]}(0,{\bf B})
		,
	\end{align}
	where $D_{{\bf B}^i}$ is the transverse covariant derivative and is defined in \cref{eq:aux26}. We note that in the eikonal limit, $P^+ \to  \infty$, we recover the result given in \cref{eq:eikonal_propagator}.

\section{Conclusions}
\label{sec:conclusions}

	The investigation of sub eikonal corrections to high-energy scattering amplitudes has gained significant importance, particularly in light of the kinematics that will be probed at the upcoming Electron-Ion Collider. These corrections provide a unique opportunity to explore novel effects that remain concealed within the eikonal limit. In this manuscript, we have comprehensively reviewed the scalar propagator in the presence of a non-Abelian classical gauge field, generated by a nuclear medium, utilizing the worldline formalism. The path integral was expressed in light-cone coordinates by integrating the Schwinger proper time, as shown in \cref{eq:scalar_prop_PS,eq:scalar_prop_config}. This representation becomes more useful at high energy since the dependence on the longitudinal momentum of the projectile is explicit, facilitating the identification of leading terms in the propagator.
	
	The interpretation of the propagator in light-cone coordinates reveals that the projectile exchanges both longitudinal and transverse momentum while interacting with the medium. The gauge field of the medium encompasses components $A^+$ and ${\bf A}$, in contrast to the eikonal case. The longitudinal momentum transfer with the medium arises from the $z^-$ dependence of the gauge field. Interestingly, by neglecting this dependence and considering only the $A^-$ component of the field, we recover the effective gluon propagator used in the derivation of the BDMPS-Z spectrum.
	
	While the results presented in this work are formal, they open the door for further investigations and practical applications. Once a model for the gauge field is established, the path integral can be solved for $z^-$ and $p^+$, enabling the computation of (i) corrections to the BDMPS-Z spectrum by incorporating the longitudinal momentum transfer to the effective gluon propagator, and (ii) the computation of collisional energy loss due to the longitudinal interaction of the projectile parton with the medium.
	
	In \cref{sec:eikonal_expansion}, we have developed an expansion around the saddle point solution of the path integral, which corresponds to a power expansion in $e^{- \omega}$, where $\omega$ represents the rapidity of the projectile. While the first order correction has been recently computed in \cite{Altinoluk_2022,Chirilli:2018kkw}, higher-order non-eikonal corrections are crucial for processes where the collision energy is moderate.
	
	As the results presented in this manuscript are derived for a scalar particle and a generic gauge field, there are several avenues for improvement and extension. First, when the eikonal approximation is relaxed, the spin of partons becomes relevant. To account for this, a complete analysis of scattering amplitudes with quarks should be conducted, replacing the scalar propagator with the quark propagator. In the worldline formalism, this can be achieved by incorporating the spin factor \cite{Brink:1976sc,Fradkin:1991ci}. Moreover, the study of the non-eikonal gluon propagator, which is relevant for numerous observables, is another important extension. Hence, we plan to extend our analysis in order to account for both cases.
	
	Next, this work focuses on the interaction of a quantum scalar particle with a classical field. To fully explore non-eikonal effects, we need to include a quark interacting with a classical gluon field, as well as a gluon interacting with a classical quark field. While some progress has been made in this direction at the next-to-eikonal accuracy in \cite{Li:2023tlw}, it is worth investigating the possibility of obtaining an exponentiated path integral result when the background field consists of gluons and quarks.
	
	Furthermore, the result presented here is only valid when the particle propagates in the $z^+$ direction, i.e., $x^+ \ne y^+$. For certain applications, it may be necessary to include the term proportional to $\delta(x^+ - y^+)$. This can be explored by studying \cref{eq:aux_1b} before performing the integration over the Schwinger proper time.
	
	In conclusion, this manuscript lays the groundwork for studying non-eikonal corrections to scattering amplitudes in the presence of a background non-Abelian field. The developments presented here open up opportunities for future research and can be extended to a variety of phenomenological applications.

\section*{Acknowledgments}

	We are very grateful to Xabier Feal, Andrey Tarasov, and Raju
	Venugopalan for reading this manuscript and their valuable comments. We thank Tolga Altinoluk and Guillaume Beuf for insightful discussions that have contributed to the development of this manuscript. Special thanks go to N\'estor Armesto and Fabio Dominguez for their invaluable revision and constructive feedback on the manuscript, as well as for their inspiring and fruitful discussions. This work has been supported by Conseller\'ia de Cultura, Educac\'ion e Universidade of Xunta de Galicia (Spain) under the grant ED481B-2022-050. The author has received financial support from Xunta de Galicia (Centro singular de investigaci\'on de Galicia accreditation 2019-2022, ref. ED421G-2019/05), by European Union ERDF, by the ”Mar\'{\i}a de Maeztu” Units of Excellence program MDM2016-0692, and by the Spanish Research State Agency under project PID2020-119632GB-I00. This work has been performed in the framework of the European Research Council project ERC-2018-ADG-835105 YoctoLHC and the MSCA RISE 823947 ”Heavy ion collisions: collectivity and precision in saturation physics” (HIEIC), and has received funding from the European Union’s Horizon 2020 research and innovation program under grant agreement No. 824093.

\appendix

\section{The scalar propagator in LC coordinates in the discrete limit}
\label{app:scalar_prop}
	
	In this section we solve explicitly the $z^+$ and $p^-$ path integral given in \cref{eq:aux3}. In the continuous limit \cref{eq:aux3} can be written as follows:
	\begin{align}\label{eq:aux_1c}
		&\Delta_F(x,y) = \int_0^\infty dT \lim_{N \to \infty} 
		\int \prod_{l=1}^{N-1} d^4z_l
		\prod_{k=1}^N \frac{d^4 p_k}{(2\pi)^4}
		\breakp{0}{\times}
		\mathcal{P}
		\exp\Bigg\{ 
		i \sum_{n=1}^{N}
		\Bigg[ p_n^-(2 \frac{T}{N} p_n^+ - (z_n^+ - z_{n-1}^+)) -\frac{T}{N} ({\bf p}_n^2 + m_\epsilon^2)
		\breakp{1}{-} 
		\vec{p}_n \cdot (\vec{z}_n - \vec{z}_{n-1})  
		-g A^-(z_n^+, \vec{z}_n) (z_n^+ - z_{n-1}^+) -g \vec{A}(z_n^+, \vec{z}_n) \cdot (\vec{z}_n - \vec{z}_{n-1})
		\Bigg]
		\Bigg\}.
	\end{align}
	
	The $p^-$ and $z^-$ dependent part of this integral can therefore be solved as follows:
	\begin{align}\label{eq:aux1}
		&\int \prod_{l=1}^{N-1} dz_l^+ \prod_{n=1}^{N} \frac{dp_n^-}{2\pi} \mathcal{P} \exp\Bigg\{  
		i \sum_{n=1}^{N} \Bigg[
		p_n^-\left(\frac{2Tp_n^+}{N} - z_n^+ + z_{n-1}^+\right) -g A^-(z_n^+, \vec{z}_n) (z_n^+ - z_{n-1}^+) 
		\breakp{7}{-}
		g \vec{A}(z_n^+, \vec{z}_n) \cdot (\vec{z}_n - \vec{z}_{n-1})
		\Bigg]
		\Bigg\}
		\nonumber \\
		& = \int \prod_{l=1}^{N-1} dz_l^+ \prod_{k=1}^{N} \delta
		\left( \frac{2Tp_k^+}{N}- z_k^+ + z_{k-1}^+ 	\right) 
		\mathcal{P} \exp\Bigg\{  
		-i g \sum_{n=1}^{N} \Bigg[A^-(z_n^+, \vec{z}_n) (z_n^+ - z_{n-1}^+) 
		\breakp{7}{+}
		\vec{A}(z_n^+, \vec{z}_n) \cdot (\vec{z}_n - \vec{z}_{n-1})
		\Bigg]
		\Bigg\}
		\nonumber \\ & \hskip0cm = 
		\delta\left( 2 \avp T - (x^+-y^+) \right) 
		\breakp{2}{\times}
		\mathcal{P} \exp\left\{  
		-i g \sum_{n=1}^{N} \left[A^-(z_n^+, \vec{z}_n) \frac{2 T p_n^+}{N} +\vec{A}(z_n^+, \vec{z}_n) \cdot (\vec{z}_n - \vec{z}_{n-1})
		\right]
		\right\},
	\end{align}
	where we have defined
	\begin{align}\label{key}
		z_n^+ \equiv y^+ + 2 T \sum_{l=1}^n \frac{p_l^+}{N} , \qquad
		\avp = \sum_{n=1}^N \frac{p_n^+}{N}.
	\end{align}
	
	Plugging \cref{eq:aux1} into \cref{eq:aux_1c}, we obtain the discrete representation of \cref{eq:aux_1b}:
	\begin{align}\label{}
		\Delta_F(x,y) &= \int_0^\infty dT \lim_{N \to \infty} 
		\int \prod_{l=1}^{N-1} d^3\vec{z}_l
		\prod_{k=1}^N \frac{d^3 \vec{p}_k}{(2\pi)^3}
		\delta\left( 2 \avp T - (x^+-y^+) \right) 
		\nonumber \\ & \hskip-1cm \times
		\mathcal{P}
		\exp\Bigg\{ 
		- i \sum_{n=1}^{N}
		\Bigg[ \frac{{\bf p}_n^2 + m_\epsilon^2}{2 p_n^+} \frac{2p_n^+ T}{N}
		+ \vec{p}_n \cdot (\vec{z}_n - \vec{z}_{n-1})  
		\breakp{4}{+}
		g A^-(z_n^+, \vec{z}_n) \frac{2p_n^+ T}{N} + \vec{A}(z_n^+, \vec{z}_n) \cdot (\vec{z}_n - \vec{z}_{n-1})
		\Bigg]
		\Bigg\}.
	\end{align}

\section{Useful path integrals}
\label{sec:app_path_int}

	In this section, we compute the path integrals that are used throughout this manuscript. The approach for solving path-integrals adopted in this work, is based on the so-called time-slicing regularization \cite{bastianelli_vannieuwenhuizen_2006}. Path integrals are defined through their discrete representation and regulated by the number of slices $N$. We first solve them for fixed $N$, and once we obtain the final result, we take the $N\to\infty$ limit.
	
	We consider the D-dimensional Gaussian path integral with a "current term":
	\begin{align}\label{eq:app1}
		\int_{z(y^+)=y}^{z(x^+)=x} \mathcal{D}^D z(\tau^+) F[z] \int \mathcal{D}^D p(\tau^+) e^{-i \int_{y^+}^{x^+} d\tau^+ \left( a p^2 + p \cdot \dot{z} \right) },
	\end{align}
	where $F[z]$ is a generic functional of $z(\tau^+)$ and $a$ is a real constant. In the discrete representation we can write \cref{eq:app1} as follows:
	\begin{align}
		&\lim_{N \to \infty} \int \prod_{n=1}^{N-1} d^D z_n \prod_{k=1}^{N} \frac{d^Dp}{(2 \pi)^D} F[{z_i}] 
		\exp \Biggl\{
		-i \sum_{n=1}^{N} \frac{x^+-y^+}{N} \left[
		a p_n^2 +p_n \cdot (z_n-z_{n-1})
		\right]
		\Biggr\}
		\nonumber \\ & =
		\lim_{N \to \infty} \int \prod_{n=1}^{N-1} d^D z_n  F[\{z_i\}] \left(\frac{N}{4 \pi i a (x^+-y^+)}\right)^{ND/2}
		\exp \Biggl\{
		i \sum_{n=1}^{N} \frac{N}{4 a (x^+-y^+)}
		(z_n-z_{n-1})^2 
		\Biggr\},
	\end{align}
	where in the last step we have solved the analytical continuation of the Gaussian integral in $p^\mu$ and defined $z_0=x$ and $z_N=y$. Hence, in the continuous limit, we obtain the solution of \cref{eq:app1}:
	\begin{align}\label{eq:pathint_gauss}
		&\int_{z(y^+)=y}^{z(x^+)=x} \mathcal{D}^D z(\tau^+) \int \mathcal{D}^D p(\tau^+) e^{ -i \int_{y^+}^{x^+} d\tau^+ \left( a p^2 + p \cdot \dot{z} \right) } F[z]
		\breakp{4}{=}
		\int_{z(y^+)=y}^{z(x^+)=x} \mathcal{D}^D z(\tau^+) \exp\Bigg\{ i \int_{y^+}^{x^+} d\tau^+ \frac{\dot{z}^2}{4a} \Bigg\} F[z].
	\end{align}
	In \cref{eq:pathint_gauss}, we have defined the path integral measure as
	\begin{align}
		\mathcal{D}^D z(\tau^+) \equiv  \prod_{n=1}^{N-1} d^D z_n \left(\frac{N}{4 \pi i a (x^+-y^+)}\right)^{ND/2},
	\end{align}
	where this factor is introduced in the definition of the measure, after integration in the momentum coordinates, to regulate the Gaussian path integral.
	
	In the case in which $F[z]=1$, $D=2$ and $a=1/2p^+$ we have the following path integral we obtain the propagator of a free particle in a two-dimensional space. The result is well known \cite{Agostini:2022ctk}, and reads
	\begin{align}\label{eq:functional_cuadratica}
		\int_{{\bf z}(y^+)=y}^{{\bf z}(x^+)=x} \mathcal{D}^2 {\bf z}(\tau^+) \exp\Bigg\{ i \int_{y^+}^{x^+} d\tau^+ p^+ \frac{\dot{\bf z}^2}{2} \Bigg\} 
		=
		\frac{p^+}{2 \pi i (x^+-y^+)} e^{i \frac{p^+}{2(x^+-y^+)} ({\bf x}-{\bf y})^2}.
	\end{align}
	
	Now let us solve the following exponential path-integral:
	\begin{align}\label{eq:app2}
		\int_{z^-(y^+)=y^-}^{z^-(x^+)=x^-} \mathcal{D} z^-(\tau^+) e^{ i \int_{y^+}^{x^+} d\tau^+ z^-  \dot{p}^+ }.
	\end{align}
	 In the discrete representation this integral is given by
	\begin{align}
		\lim_{N \to \infty} \int \prod_{k=1}^{N-1} d z_k^-
		\exp \Biggl\{
		i \sum_{n=1}^{N-1} z_n^- (p_{n+1}^+-p_n^+) 
		\Biggr\} 
		&=
		\lim_{N \to \infty} \prod_{n=1}^{N-1} \deltabar(p_{n+1}^+-p_n^+) 
		\breakp{0}{=}
		\lim_{N \to \infty} \prod_{n=1}^{N-1} \deltabar(p_n^+-p_N^+).
	\end{align}
	
	Hence, the solution of \cref{eq:app2} is given by
	\begin{align}\label{eq:pathint_dirac}
		\int_{z^-(y^+)=y^-}^{z^-(x^+)=x^-} \mathcal{D} z^-(\tau^+) e^{ i \int_{y^+}^{x^+} d\tau^+ z^-  \dot{p}^+} = \deltabar \big[p(\tau^+) - p^+(x^+) \big],
	\end{align}
	where we have defined the functional Dirac delta as
	\begin{align}
		\deltabar \big[p(\tau^+) - k^+ \big] \equiv \lim_{N \to \infty} \prod_{n=1}^{N-1} \deltabar(p_n^+-k^+).
	\end{align}
	It is easy to realize that the momentum path-integral of any functional $F[p^+(\tau^+)]$ multiplied by the functional Dirac delta is given by:
	\begin{align}\label{eq:functional_dirac}
		\int \mathcal{D} p^+(\tau^+) F[p^+(\tau^+)] \deltabar \big[p(\tau^+) - p_f^+\big] = \int \frac{d p_f^+}{2\pi} F[p_f^+],
	\end{align}
	where we have introduced $p_f^+ \equiv p^+(x^+)$.

\section{Derivation of the propagator written in terms of the local momentum transfer}	
\label{app:local_momentum}

	In this section, we derive \cref{eq:medium_propagtor} starting from \cref{eq:aux_4}. The discrete representation of \cref{eq:aux_4} is given by
	\begin{align}\label{eq:app3}
		\Delta_{\rm m}(\vec{P},\vec{Q}) &=
		\int_{\vec{B},\vec{\Delta}}
		e^{i {\Delta} \cdot \hat{P}+i {B} \cdot \hat{Q}}
		\lim_{N_\Delta \to \infty}
		\int \prod_{k=1}^{N_\Delta} \frac{dp_k^+}{2 \pi}
		\frac{\Theta(\avp)}{2 \avp}
				\int \prod_{l=1}^{N_\Delta - 1} d^3 \vec{z}_l
		\breakp{0}{\times}
		\mathcal{P}_+
		\Bigg\{
		i \sum_{n=1}^{N_\Delta} \Bigg[
		-\frac{m_\epsilon^2}{2 \avp} \frac{\tau_f^+-\tau_i^+}{N_\Delta} 
		+ \frac{N_\Delta}{\tau_f^+-\tau_i^+} \frac{\avp}{2} ({\bf z}_n - {\bf z}_{n-1})^2 
		-p_n^+ (z_n^--z_{n-1}^-) 
		\breakp{2}{-}
		g A^-(z_n^+, \vec{z}_n) \frac{p_n^+}{\avp} \frac{\tau_f^+-\tau_i^+}{N_\Delta} + g {\bf A}(z_n^+, \vec{z}_n) \cdot ({\bf z}_n - {\bf z}_{n-1})
			\Bigg]
		\Bigg\},
	\end{align}
	where $\vec{z}_0 = \vec{B} - \vec{\Delta}/2$ and $\vec{z}_{N_\Delta} = \vec{B} + \vec{\Delta}/2$. Here, $N_\Delta$ is the number of discretization steps inside the medium and is written in terms of the total number of steps as follows:
	\begin{align}
		N_\Delta = \frac{\tau_f^+ - \tau_i^+ }{x^+ - y^+} N.
	\end{align}
	On the other hand, the discrete representation of the average longitudinal momentum is given by
	\begin{align}
		&\avp = \frac{\tau_i^+ - y^+}{x^+ - y^+} p_i^+ + \frac{\tau_f^+ - \tau_i^+ }{x^+ - y^+} \sum_{n=1}^{N_\Delta} \frac{p_n^+}{N_\Delta} + \frac{x^+ - \tau_f+}{x^+ - y^+} p_f^+
		\nonumber \\ & =
		\frac{(x^+ - y^+) - (\tau_f^+ - \tau_i^+)}{x^+ - y^+} P^+ + \frac{(x^+ + y^+) - (\tau_f^+ + \tau_i^+)}{x^+ - y^+} \frac{Q^+}{2}
		+ \frac{\tau_f^+ - \tau_i^+ }{x^+ - y^+} \sum_{n=1}^{N_\Delta} \frac{p_n^+}{N_\Delta}.
	\end{align}
	
	We can solve the $B^-$ and $\Delta^-$ integrals by writing the $z^-$ term of the kinetic part of the Lagrangian as
	\begin{align}
		-i \sum_{n=1}^{N_\Delta} p_n^+ (z_n^--z_{n-1}^-) &= -i \frac{p_{N_\Delta}^+ + p_1^+}{2} \Delta^- -i (p_{N_\Delta}^+ - p_1^+) B^- +i \sum_{n=1}^{N_\Delta-1} z_n^- (p_{n+1}^+ - p_n^+)
		\nonumber \\ & \hskip-2cm =
		-i \Delta^- \left( p_1^+ + \frac{1}{2} \sum_{n=1}^{N_\Delta-1} \frac{q_n^+}{N_\Delta} \right)  -i B^- \sum_{n=1}^{N_\Delta-1} \frac{q_n^+}{N_\Delta} +i \sum_{n=1}^{N_\Delta-1} z_n^- \frac{q_n^+}{N_\Delta},
	\end{align}
	where in the last step we have performed the change of variables
	\begin{align}
		\frac{q_n^+}{N_\Delta} = (p_{n+1}^+ - p_n^+), \qquad p_{n}^+ = p_1^+ + \sum_{l=1}^{n-1} \frac{q_n^+}{N_\Delta}.
	\end{align}
	
	Hence, assuming that $A^\mu(z_{N_\Delta}, \vec{z}{N\Delta}) = 0$, performing the change of variables $p_n^+ \to q_{n-1}^+$, and solving the $B^-$ and $\Delta^-$ integrals, we can write \cref{eq:app3} as follows:
	\begin{align}\label{eq:app4}
		\Delta_{\rm m}(\vec{P},\vec{Q}) &=
		e^{ i \hat{Q}^- B^+ + i \hat{P}^- \Delta^+}
		\int_{{\bf B}, {\bf \Delta}} e^{-i {\bf B} \cdot {\bf Q}-i {\bf \Delta} \cdot {\bf P}} 
		\lim_{N_\Delta \to \infty}
		\int \prod_{k=1}^{N_\Delta-1} \frac{dq_k^+}{2 \pi N_\Delta} 
		\int
		 \frac{dp_1^+}{2 \pi} 
		 \breakp{0}{\times}
		\deltabar\left( P^+ - p_1^+ - \frac{\langle q^+ \rangle}{2} \right) 
		\deltabar \left(Q^+ - \langle q^+ \rangle \right)
		\frac{\Theta(\avp)}{2 \avp}
		\int \prod_{l=1}^{N_\Delta - 1} d^3 \vec{z}_l
		\nonumber \\ & \hskip0cm \times
		\mathcal{P}_+
		\Bigg\{
		i \sum_{n=1}^{N_\Delta} \Bigg[
		-\frac{m_\epsilon^2}{2 \avp} \frac{\tau_f^+-\tau_i^+}{N_\Delta} 
		+ \frac{N_\Delta}{\tau_f^+-\tau_i^+} \frac{\avp}{2} ({\bf z}_n - {\bf z}_{n-1})^2 
		+ z_n^- \frac{q_n^+}{N_\Delta}
		\nonumber \\ & \hskip0cm
		- g A^-(z_n^+, \vec{z}_n) \frac{p_1^+ + \sum_{l=1}^{n-1} \frac{q_n^+}{N_\Delta}}{\avp} \frac{\tau_f^+-\tau_i^+}{N_\Delta} + g {\bf A}(z_n^+, \vec{z}_n) \cdot ({\bf z}_n - {\bf z}_{n-1})
		\Bigg]
		\Bigg\},
	\end{align}
	where $q_{N_\Delta}^+ = 0$. 
	
	The Dirac delta in the $p_1^+$ integral fixes the in-medium longitudinal momentum to:
	\begin{align}\label{eq:app5}
		p_n^+ = P^+ + \frac{1}{2}\sum_{l=1}^{n-1} \frac{q_l^+}{N_\Delta} - \frac{1}{2}\sum_{l=n}^{N_\Delta-1} \frac{q_l^+}{N_\Delta}.
	\end{align}
	Thus, by solving the $p_1^+$ integral in \cref{eq:app4}, we obtain the discrete representation of the in-medium propagator in terms of $q_n^+$:
	\begin{align}\label{eq:app6}
		\Delta_{\rm m}(\vec{P},\vec{Q}) &=
		e^{ i \hat{Q}^- B^+ + i \hat{P}^- \Delta^+}
		\int_{{\bf B}, {\bf \Delta}} e^{-i {\bf B} \cdot {\bf Q}-i {\bf \Delta} \cdot {\bf P}} 
		\lim_{N_\Delta \to \infty}
		\int \prod_{k=1}^{N_\Delta-1} \frac{dq_k^+}{2 \pi N_\Delta}
		\deltabar \left(Q^+ - \langle q^+ \rangle \right)
		\nonumber \\ & \hskip-2cm
		\frac{\Theta(\avp)}{2 \avp}
		\int \prod_{l=1}^{N_\Delta - 1} d^3 \vec{z}_l
		\
		\mathcal{P}_+
		\Bigg\{
		i \sum_{n=1}^{N_\Delta} \Bigg[
		-\frac{m_\epsilon^2}{2 \avp} \frac{\tau_f^+-\tau_i^+}{N_\Delta} 
		+ \frac{N_\Delta}{\tau_f^+-\tau_i^+} \frac{\avp}{2} ({\bf z}_n - {\bf z}_{n-1})^2 
		+ z_n^- \frac{q_n^+}{N_\Delta}
		\nonumber \\ & \hskip-2cm
		- g A^-(z_n^+, \vec{z}_n) \frac{1}{\avp} \left(P^+ + \frac{1}{2}\sum_{l=1}^{n-1} \frac{q_l^+}{N_\Delta} - \frac{1}{2}\sum_{l=n}^{N_\Delta-1} \frac{q_l^+}{N_\Delta}\right) \frac{\tau_f^+-\tau_i^+}{N_\Delta} 
		\nonumber \\ & \hskip6cm 
		+
		 g {\bf A}(z_n^+, \vec{z}_n) \cdot ({\bf z}_n - {\bf z}_{n-1})
		\Bigg]
		\Bigg\},
	\end{align}
	where the average momentum is given by
	\begin{align}\label{eq:app7}
		\avp &=
		P^+ + \frac{(x^+ + y^+) - (\tau_f^+ + \tau_i^+)}{x^+ - y^+} \frac{Q^+}{2}
		+\frac{1}{2} \frac{\tau_f^+ - \tau_i^+ }{x^+ - y^+} \sum_{n=1}^{N_\Delta} \left( \sum_{l=1}^{n-1} \frac{q_l^+}{N_\Delta} - \sum_{l=n}^{N_\Delta-1} \frac{q_l^+}{N_\Delta} \right).
	\end{align}
		
	Finally, taking the continuous limit of \cref{eq:app5,eq:app6,eq:app7}, we obtain the results of \cref{eq:momentum_medium,eq:medium_propagtor,eq:avg_momentum_q}, respectively.
		
	To finalize this section, we discuss the regularization of the $q^+$ path integral. We can read from \cref{eq:app6} that the $q^+$ path integral measure is regulated as
	\begin{align}
		\mathcal{D} q^+ = \lim_{N_\Delta \to \infty} \prod_{n=1}^{N_{\Delta}-1}\frac{dq_n^+}{2 \pi N_\Delta}.
	\end{align}
	This implies that the exponent path integral over the $v^-$ variable leads to the following result:
	\begin{align}\label{eq:pathint_dirac2}
		\int \mathcal{D} v^- e^{ i \int_{\tau_i^+}^{\tau_f^+} d\tau^+ v^- \frac{q^+}{\Delta \tau^+} } &= \lim_{N_{\Delta} \to \infty} \prod_{n=1}^{N_{\Delta}-1} dv_n^- \exp \Bigg\{ i \sum_{n=1}^{N_{\Delta}-1} v_n^- \frac{q_n^+}{N_{\Delta}} \Bigg\} 
		\breakp{0}{=}
		\lim_{N_{\Delta} \to \infty} \prod_{n=1}^{N_{\Delta}-1} \deltabar\left(\frac{q_n^+}{N_{\Delta}}\right) \equiv \deltabar[q^+(\tau^+)],
	\end{align}
	that is, the continuous limit of the functional Dirac delta is regulated by a factor $N_{\Delta}^{N_\Delta - 1}$ which cancels with the regulator of the $q^+$ path integral measure:
	\begin{align}\label{eq:pathint_dirac3}
		\int^{q^+(\tau_f^+) = 0} \mathcal{D}q^+(\tau^+) F[q^+] \deltabar[q^+(\tau^+)] &= 
		\lim_{N_{\Delta} \to \infty}\int \prod_{n=1}^{N_{\Delta}-1} \frac{dq_n^+}{2 \pi N_{\Delta}} \deltabar\left(\frac{q_n^+}{N_{\Delta}}\right) F(\{q_i^+\}) 
		\nonumber \\ & 
		=\lim_{N_{\Delta} \to \infty}\int \prod_{n=1}^{N_{\Delta}-1} \frac{dq_n^+}{2 \pi} \deltabar(q_n^+) F(\{q_i^+\})
		= F[0].
	\end{align}

	Moreover, the functional derivative at a point $\tau_1^+ \in [\tau_i^+, \tau_f^+]$ of the Dirac delta is defined in the discrete representation as follows:
	\begin{align}\label{app:pathint_dirac_der}
		\frac{\delta }{\delta q^+ (\tau_1^+)} \deltabar[q^+(\tau^+)] = \frac{\partial}{\partial q_{n_1}^+} \lim_{N_{\Delta} \to \infty} \prod_{n=1}^{N_{\Delta}-1} \deltabar\left(\frac{q_n^+}{N_{\Delta}}\right),
	\end{align}
	where
	\begin{align}
		n_1 = \frac{\tau_1^+ - \tau_i^+}{\tau_f^+ - \tau_i^+} N_\Delta.
	\end{align}

\section{The eikonal expansion at first order}
\label{sec:first_order_correction}
	
	In this section, we compute the first-order correction to the eikonal approximation using the eikonal expansion given in \cref{eq:eikonal_expansion}. The object analyzed in this section has recently been computed in \cite{Altinoluk_2022} using a different formalism and will serve as a double-check for our approach. 
	
	Before starting with our analysis, let us write schematically:
	\begin{align}
		\Delta_{\rm m}(\vec{P},\vec{Q}) = \Delta_{\rm m}^{(0)}(\vec{P},\vec{Q}) + \Delta_{\rm m}^{(1),-}(\vec{P},\vec{Q}) + \Delta_{\rm m}^{(1),\perp}(\vec{P},\vec{Q}) + \mathcal{O}\left(\frac{1}{(P^+)^2}\right),
	\end{align}
	where $\Delta_{\rm m}^{(0)}$ is the zeroth order, i.e., $m=0$, term in \cref{eq:eikonal_expansion}. $\Delta_{\rm m}^{(1),-}$ is the leading-order correction due to the $v^-$ expansion and gives a contribution of order $1/(P^+)^2$, which is subleading in our analysis. However, we study this term for illustrative purposes. On the other hand, the term $\Delta_{\rm m}^{(1),\perp}$ gives the leading order in $1/P^+$ in the eikonal expansion by computing corrections to the ${\bf z}$ classical trajectories and is the result of expanding \cref{eq:eikonal_expansion} up to order $m=2$.
	
	One can read from \cref{eq:eikonal_expansion} that the zeroth order of the expansion is simply given by
	\begin{align}\label{eq:correction_zero}
		\Delta_{\rm m}^{(0)}(\vec{P},\vec{Q}) &=
		e^{ i \hat{Q}^- B^+ + i \hat{P}^- \Delta^+-i \frac{{\bf P}^2+m_\epsilon^2}{2 P^+} \Delta^+}
		\frac{\Theta(P^+)}{2 P^+ }
		\int_{\vec{B}} e^{i \vec{B} \cdot \vec{Q}} 
		\
		U_{[z_f^+,z_i^+]}(\vec{B})
		\nonumber \\ & =
		\frac{\Theta(P^+)}{2 P^+ }
		\int_{\vec{B}} e^{i \vec{B} \cdot \vec{Q}} 
		\
		\left(1 + i \frac{{\bf P} \cdot {\bf Q}}{P^+}B^+ +i \frac{{\bf Q}^2}{8 P^+} \Delta^+\right)
		U_{[z_f^+,z_i^+]}(\vec{B}),
	\end{align}
	which is not but the result obtained in \cref{eq:eikonal_propagator} including the phases and with $B^- \ne 0$.

\subsection{Correction to the $z^-$ trajectory}
	
	We now analyze the corrections that arise due to the expansion around the fluctuations $v^-$. As we have pointed out in \cref{sec:eikonal_expansion}, this correction will result in a single derivative of the Dirac delta in the longitudinal momentum transfer $q_1^+ \equiv q^+(\tau_1^+)$. Thus, as we will see below, it is analogous to an expansion at leading order in $q_1^+/P^+$, since higher order corrections will be suppressed by the delta function. At this order, the insertion given in \cref{eq:insertion} can be written as
	\begin{equation}\label{key}
		\mathcal{I}_{\tau^+}\left[ v^-, 0, 0 \right] = v^- \partial^{+} \tilde{A}^-(z^+,0,{\bf B}) \dot{z}^+.
	\end{equation}
	
\begin{figure}[h!]
		\centering
		\includegraphics[scale=0.5]{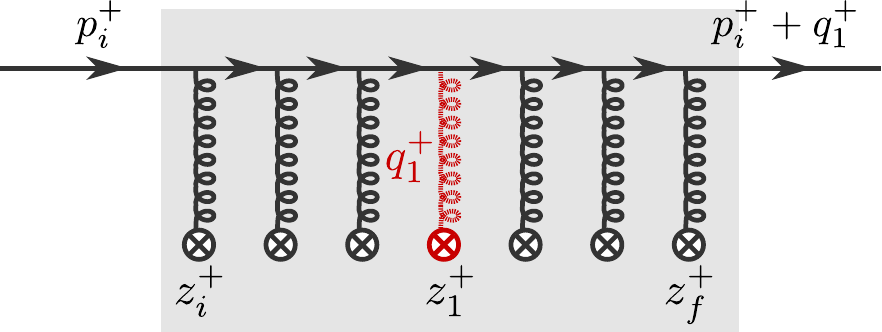}
		\caption{Diagrammatic interpretation of the leading order correction to the $v^-$ trajectory. The particle travels through the medium, defined in between $z_i^+$ and $z_f^+$, and scatters with multiple soft gluons, represented by black lines. At a longitudinal coordinate $z_1^+$, the particle interacts with a semi-hard gluon that gives a longitudinal "kick" of order $\partial^+ A^-$ and changes its longitudinal momentum. The red line represents the exchanged gluon responsible for the momentum transfer.}
		\label{fig:propagator_zm}
	\end{figure}

	On the other hand, we only need the first term of the summation in \cref{eq:eikonal_expansion}, so that the term in the in-medium propagator that contributes to the leading order correction is\footnote{From now on until the end of the calculation, we do not write the dependence of the eikonal Wilson line on $\vec{B}$ explicit in order to make the notation tidier.}
	\begin{align}\label{eq:aux16}
		&\Delta_{\rm m}^{(1),-}(\vec{P},\vec{Q}) = 
		e^{ i \hat{Q}^- B^+ + i \hat{P}^- \Delta^+}
		\int_{\vec{B}} e^{i \vec{B} \cdot \vec{Q}} 
		\int^{q^+(\tau_f^+) = 0} \mathcal{D}q^+
		\frac{\Theta(\avp)}{2 \avp }
		e^{-i \frac{{\bf P}^2 + m_\epsilon^2}{2 \avp} \Delta \tau^+} 
		\\ & \hskip2cm \times
		\int_{\tau_i^+}^{\tau_f^+} d\tau_1^+ U_{[z_f^+,z_1^+]} \dot{z}_1^+ \partial^{+} \tilde{A}_{z_1^+}^- U_{[z_1^+,z_i^+]} \frac{\delta \ }{i \delta q^+(\tau_1^+)} \deltabar[q^+(\tau^+)]
		\nonumber \\ & \hskip0cm =
		\int_{\vec{B}} e^{i \vec{B} \cdot \vec{Q}}
		\int dq_1^+
		(-i \delta'(q_1^+))
		\int_{\tau_i^+}^{\tau_f^+} d\tau_1^+ 
		\frac{\Theta(\avp)}{2 \avp }
		e^{-i \frac{{\bf P}^2 + m_\epsilon^2}{2 \avp} \Delta \tau^+} 
		\dot{z}_1^+
		U_{[z_f^+,z_1^+]} \partial^{+} \tilde{A}_{z_1^+}^- U_{[z_1^+,z_i^+]},
		\nonumber
	\end{align}
	where in the last step, we solved the path integral over $q^+$ by using the Dirac delta and dropped the non-eikonal phase, anticipating that the result after solving the $q_1^+$ integral would be of order $1/P^+$. This equation represents the leading-order correction to the longitudinal momentum transfer resulting from the interaction of the projectile with the medium. It describes the propagation of a particle that interacts with the medium by absorbing soft gluons, as illustrated in Figure \ref{fig:propagator_zm} with black lines. The particle absorbs a single semi-hard gluon at position $z_1^+$, leading to a longitudinal "kick" of order $\partial^+ A^-$, depicted by the red line in Figure \ref{fig:propagator_zm}. Consequently, the particle undergoes a change in its longitudinal momentum.
	
	Since the result in \cref{eq:aux16} is proportional to the derivative of the delta in $q_1^+$, it is enough to expand all the variables that depend on this quantity at leading order. Moreover, we also expand all the variables at order $Q^+/P^+$ since we are neglecting the $\mathcal{O}(1/(P^+)^2)$ correction. The quantities that depend on the longitudinal momentum transfer are the average longitudinal momentum, given by \cref{eq:avg_momentum_q}, and the LC proper time intervals, given by \cref{eq:geometric_relation}. Using \cref{eq:avg_momentum_q}, we can express the average momentum as
	\begin{align}
		\avp &=
		P^+ +
		\frac{Q^+}{2} 
		\frac{(x^+-\tau_f^+)+(y^+-\tau_i^+)}{x^+-y^+}
		+
		\frac{q_1^+}{2} 
		\frac{(\tau_f^+-\tau_1^+)+(\tau_i^+-\tau_1^+)}{x^+-y^+}
		.
	\end{align}
	This result can be written entirely in terms of $z_1^+ \equiv z^+(\tau_1^+)$ by using the relationships given in \cref{eq:geometric_rel2,eq:geometric_relation}. Doing that we obtain
	\begin{align}\label{eq:aux18}
		\avp &= P^+ \Bigg[
		1- \frac{Q^+}{2(x^+-y^+)} \left(\frac{x^+-z_f^+}{P^+ + Q^+/2} + \frac{y^+-z_i^+}{P^+ - Q^+/2} \right)
		\breakp{4}{-}
		\frac{q_1^+}{2(x^+-y^+)} \left(\frac{z_f^+-z_1^+}{P^+ + q_1^+/2} + \frac{z_i^+-z_1^+}{P^+ - q_1^+/2}\right)
		\Bigg]^{-1}
		\nonumber \\ & = 
		P^+ \left[ 1+\frac{Q^+}{P^+} \frac{B^+-\frac{x^+ + y^+}{2}}{x^+-y^+} +\frac{q_1^+}{P^+} \frac{z_1^+ - B^+}{x^+-y^+}+ \mathcal{O}\left( \frac{1}{(P^+)^2}\right) \right]^{-1}.
	\end{align}
	
	On the other hand, the term in the exponential of \cref{eq:aux16} can also be written in terms of $z_1^+$, using \cref{eq:geometric_relation}, as
	\begin{align}\label{eq:aux19}
		\frac{\Delta \tau^+}{\avp} = \frac{z_f^+-z_1^+}{P^+ + q_1^+/2} + \frac{z_1^+-z_i^+}{P^+ - q_1^+/2} = \frac{\Delta^+}{P^+} + \mathcal{O}\left( \frac{1}{(P^+)^2}\right).
	\end{align}
	
	Using \cref{eq:aux18}, the step function can be expanded as\footnote{At this point we only need to expand the $\Theta$ function in powers of $q_1^+$, since \cref{eq:aux16} is proportional to $\deltabar'(q_1^+$). However, we are anticipating that the final result \cref{eq:aux28}, after expanding around $B^-=0$, is going to be proportional to derivatives of $\deltabar(Q^+)$ and we are also expanding around $Q^+=0$.}
	\begin{align}\label{eq:aux20}
		\Theta(\avp) = \Theta(P^+) -\left[Q^+ \frac{B^+-\frac{x^+ + y^+}{2}}{x^+-y^+}+q_1^+ \frac{z_1^+ - B^+}{x^+-y^+}+ \cdots \right] \delta(P^+)+ \mathcal{O}(\delta'(P^+)).
	\end{align}
	However, as we stated in \cref{sec:scalar_prop_LC}, the scalar propagator studied in the present analysis neglects the contribution where the particle does not propagate in the longitudinal direction, i.e., $x^+=y^+$. This implies that the modes with $P^+=0$ do not contribute to the present analysis and are discarded. Thus, we can write $\Theta(\avp) = \Theta(P^+)$.
	
	Plugging \cref{eq:aux18,eq:aux19,eq:aux20} into \cref{eq:aux16} and solving the $q_1^+$ integral, we obtain
	\begin{align}
		&\Delta_{\rm m}^{(1),-}(\vec{P},\vec{Q}) = 
		\frac{\Theta(P^+)}{2P^+}
		e^{-i\frac{{\bf P}^2 + m_\epsilon^2}{2 P^+}\Delta^+}
		\frac{i}{P^+(x^+-y^+)}
		\int_{\vec{B}} e^{i \vec{B} \cdot \vec{Q}} 
		\nonumber \\ & \hskip1cm \times 
		\int_{\tau_i^+}^{\tau_f^+} d\tau_1^+ \dot{z}_1^+ (z_1^+-B^+)
		U_{[z_f^+,z_1^+]} \partial^{+} \tilde{A}_{z_1^+}^- U_{[z_1^+,z_i^+]}
		+
		\mathcal{O}\left( \frac{1}{(P^+)^2}\right)
		\nonumber \\ & \hskip0cm =
		\frac{\Theta(P^+)}{2P^+}
		\frac{i}{P^+(x^+-y^+)}
		\int_{\vec{B}} e^{i \vec{B} \cdot \vec{Q}} 
		\int_{z_i^+}^{z_f^+} dz_1^+ 
		(z_1^+-B^+)
		U_{[z_f^+,z_1^+]} \partial^{+} \tilde{A}_{z_1^+}^- U_{[z_1^+,z_i^+]}
		\breakp{1}{+}
		\mathcal{O}\left( \frac{1}{(P^+)^2}\right),
	\end{align}
	where in the last step we have performed the change of variables $\tau_1^+ \to z^+(\tau_1^+)$ which is trivial since the integrand is already multiplied by the Jacobian and dropped the phase which is subleading. This equation can be further simplified, and written in terms of derivatives of Wilson line, by using the following identities:	
	\begin{align}\label{eq:indentity1}
		\partial^\mu U_{[z_f^+,z_i^+]} &= \int_{z_i^+}^{z_f^+} dz^+ U_{[z_f^+,z^+]} \partial^{\mu} \tilde{A}_{z^+}^- U_{[z^+,z_i^+]},
		\\
		\label{eq:indentity2}
		\int_{z_f^+}^{z_f^+} dz^+ U_{[z_f^+,z^+]} \overleftrightarrow{\partial}^{\mu} U_{[z^+,z_i^+]} 
		&=-2 \int_{z_i^+}^{z_f^+} dz^+ (z^+-B^+) U_{[z_f^+,z^+]} \partial^{\mu} \tilde{A}_{z^+}^- U_{[z^+,z_i^+]},
	\end{align}
	where
	\begin{align}
		\overleftrightarrow{\partial}^{\mu} = \overrightarrow{\partial}^{\mu} - \overleftarrow{\partial}^{\mu}.
	\end{align}
	
	Thus, using \cref{eq:indentity2}, we can write the following correction to the in-medium propagator due to longitudinal scatterings of the projectile in the medium:
	\begin{align}\label{}
		\Delta_{\rm m}^{(1),-}(\vec{P},\vec{Q}) &=
		\frac{\Theta(P^+)}{2P^+}
		\frac{-i}{2P^+(x^+-y^+)}
		\int_{\vec{B}} e^{i \vec{B} \cdot \vec{Q}} 
		\int_{z_i^+}^{z_f^+} dz^+ U_{[z_f^+,z^+]} \overleftrightarrow{\partial}^{+} U_{[z^+,z_i^+]}.
	\end{align} 
	This term, although apparently it is $\mathcal{O}(1/P^+)$, is in fact a next-to-next-to-eikonal correction, i.e. $\sim\mathcal{O}(1/(P^+)^2)$. The reason for this dependence not being explicit is that we are working in a mixed representation where we express in terms of momentum, $\vec{p}$, and LC times, $x^+$. However, under a large boost in the right direction, the LC time is dilated as $z^+ \to e^{\omega} z^+$, so that the interval $(x^+-y^+)$ is enhanced. We can see that, in fact, $x^+-y^+$ scales as $P^+$ heuristically, by noting that $(x^+-y^+)$ is the conjugate of $P^-$, so that $(x^+-y^+) \sim 1/P^- \sim 2 P^+/P_\perp^2$.
	
	Therefore, we obtain that there is no next-to-eikonal correction coming from the expansion around $v^-=B^-$. So that,
	\begin{align}\label{eq:correction_minus}
		\Delta_{\rm m}^{(1),-}(\vec{P},\vec{Q}) &=
		\mathcal{O}\left(\frac{1}{(P^+)^2}\right).
	\end{align} 
	
\subsection{Correction to the ${\bf z}_\perp$ classical trajectory}
	
	We will now examine the corrections arising from the ${\bf z}$ trajectory. In this scenario, we only need to consider the corrections associated with the transverse coordinates. Hence, there is no longitudinal momentum transfer during the interaction, and we can set $\tau^+ = z^+$. Utilizing \cref{eq:eikonal_expansion}, we need to compute
	\begin{align}\label{eq:aux23}
		&\Delta_{\rm m}^{(1),\perp}(\vec{P},\vec{Q})
		= 
		\deltabar(Q^+)
		\frac{\Theta(P^+)}{2 P^+}
		e^{ i \hat{Q}^- B^+ + i \hat{P}^-\Delta^+}
		\int_{{\bf B}} e^{-i {\bf B} \cdot {\bf Q}}
		\sum_{m=1}^{2} 
		\Bigg(
		\prod_{n=m}^{1} \int_{z_i^+}^{z_{n+1}^+} dz_n^+
		\nonumber \\ & \hskip2cm \times
		U_{[z_{n+1}^+,z_{n}^+]}
		\mathcal{I}_{\tau_n^+} \left[0, \frac{\delta \ }{\delta {\bf J}},i \partial_{\bf P}\right] 
		\Bigg)
		U_{[z_{1}^+,z_{i}^+]}
		e^{-i \frac{{\bf P}^2+m_\epsilon^2}{2 P^+} \Delta^+} \frac{Z[{\bf J}]}{Z[0]} \Bigg|_{{\bf J}=0},
	\end{align}
	where in the term $m=1$ we have to expand the insertion to second order and in the term $m=2$ to first order.
	
	Thus, inserting the transverse expansion of the insertion, given in \cref{eq:insertion}, in \cref{eq:aux23} we can write
	\begin{align}\label{eq:aux24}
		&\Delta_{\rm m}^{(1),\perp}(\vec{P},\vec{Q}) = 
		\frac{\Theta(P^+)}{2 P^+}
		\int_{\vec{B}} e^{i \vec{B} \cdot \vec{Q}}  
		\Bigg\{
		\int_{z_i^+}^{z_f^+} dz_1^+ {U}_{[z_f^+,z_1^+]} 
		\Bigg[
		\frac{i\partial_{{\bf P}^i}}{\Delta^+} \left( (z_1^+-B^+) \partial_{{\bf B}^i} \tilde{A}_{z_1^+}^{-} - \tilde{\bf A}_{z_1^+}^{i}\right)
		\nonumber \\ & +
		\frac{1}{2} \left(\frac{i\partial_{{\bf P}^i} i\partial_{{\bf P}^j}}{(\Delta^+)^2} (z_1^+-B^+)^2 + {G}^{ij}(z_1^+,z_1^+) \right) \partial_{{\bf B}^i} \partial_{{\bf B}^j} \tilde{A}_{z_1^+}^{-}
		\nonumber \\ & \hskip0cm
		- 
		\left(\frac{i\partial_{{\bf P}^i} i\partial_{{\bf P}^j}}{(\Delta^+)^2} (z_1^+-B^+) + {G}^{ij}(z_1^+,z_1^+)^{\bullet} \right) \partial_{{\bf B}^i} \tilde{\bf A}_{z_1^+}^{j}
		\Bigg] {U}_{[z_1^+, z_i^+]}
		+
		\int_{z_i^+}^{z_f^+}dz_2^+ \int_{z_i^+}^{z_2^+} dz_1^+ {U}_{[z_f^+,z_2^+]}
		\nonumber \\ & \times
		\Bigg[
		\left(  
		\frac{i\partial_{{\bf P}^i} i\partial_{{\bf P}^j}}{(\Delta^+)^2} (z_2^+-B^+) (z_1^+-B^+) + {G}^{ij}(z_2^+,z_1^+)
		\right) 
		\partial_{{\bf B}^i} \tilde{A}_{z_2^+}^{-} {U}_{[z_2^+,z_1^+]} \partial_{{\bf B}^j} \tilde{A}_{z_1^+}^{-}
		 \\ & +
		\left(
		\frac{i\partial_{{\bf P}^i} i\partial_{{\bf P}^j}}{(\Delta^+)^2} + \prescript{\bullet}{}G (z_2^+,z_1^+)^{\bullet}
		\right) 
		\tilde{\bf A}_{z_2^+}^{i} {U}_{[z_2^+,z_1^+]} \tilde{\bf A}_{z_1^+}^{j}
		\nonumber \\ &
		-
		\left(
		\frac{i\partial_{{\bf P}^i} i\partial_{{\bf P}^j}}{(\Delta^+)^2} (z_2^+-B^+) +  {G}^{ij}(z_2^+,z_1^+)^{\bullet}
		\right)
		\partial_{{\bf B}^i} \tilde{A}_{z_2^+}^{-} {U}_{[z_2^+,z_1^+]} \tilde{\bf A}_{z_1^+}^{j}
		\nonumber \\ & \hskip0cm
		-
		\left(
		\frac{i\partial_{{\bf P}^i} i\partial_{{\bf P}^j}}{(\Delta^+)^2} (z_1^+-B^+) +  \prescript{\bullet}{}G (z_2^+,z_1^+)
		\right)
		\tilde{\bf A}_{z_2^+}^{i} {U}_{[z_2^+,z_1^+]} \partial_{{\bf B}^j} \tilde{A}_{z_1^+}^{-}
		\Bigg] {U}_{[z_1^+,z_i^+]}\Bigg\} e^{-i \frac{{\bf P}^2 + m_\epsilon^2}{2 P^+} \Delta^+}.
		\nonumber
	\end{align}
	This equation, although lengthy, is just composed of the 2-point function and its derivatives, defined in \cref{eq:2poinder0,eq:2poinder1,eq:2poinder2,eq:2poinder3}, as well as single and double derivatives of ${\bf P}^i$ with respect to the Gaussian. It can be simplified by using the expression of the 2-point function and performing the derivatives:
	\begin{align}\label{key}
		&\Delta_{\rm m}^{(1),\perp}(\vec{P},\vec{Q}) = 
		\frac{\Theta(P^+)}{2 P^+}
		\int_{\vec{B}} e^{i \vec{B} \cdot \vec{Q}} 
		\Bigg\{
		\int_{z_i^+}^{z_f^+} dz_1^+ {U}_{[z_f^+,z_1^+]}
		\Bigg[
		\frac{{\bf P}^i}{P^+} \left( (z_1^+-B^+) \partial_{{\bf B}^i} \tilde{A}_{z_1^+}^{-} - \tilde{\bf A}_{z_1^+}^{i}\right)
		\nonumber \\ & \hskip8cm
		+
		\frac{i \Delta^+}{8 P^+} \partial_{{\bf B}^i} \partial_{{\bf B}^i} \tilde{A}_{z_1^+}^{-}
		\Bigg] {U}_{[z_1^+, z_i^+]} 
		\nonumber \\ & \hskip0cm
		+
		\int_{z_i^+}^{z_f^+}dz_2^+ \int_{z_i^+}^{z_2^+} dz_1^+ {U}_{[z_f^+,z_2^+]}
		\Bigg[
		\frac{i}{4 P^+}
		\left(  
		\Delta^+-2(z_2^+-z_1^+)
		\right) 
		\partial_{{\bf B}^i} \tilde{A}_{z_2^+}^{-} {U}_{[z_2^+,z_1^+]} \partial_{{\bf B}^i} \tilde{A}_{z_1^+}^{-} 
		\nonumber \\ & \hskip4cm
		- \frac{i}{2 P^+}
		\left(
		\partial_{{\bf B}^i} \tilde{A}_{z_2^+}^{-} {U}_{[z_2^+,z_1^+]} \tilde{\bf A}_{z_1^+}^{i}
		-
		\tilde{\bf A}_{z_2^+}^{i} {U}_{[z_2^+,z_1^+]} \partial_{{\bf B}^i} \tilde{A}_{z_1^+}^{-}
		\right)
		\nonumber \\ & \hskip4cm
		+
		\frac{i}{P^+} \delta(z_2^+-z_1^+)
		\tilde{\bf A}_{z_2^+}^{i} {U}_{[z_2^+,z_1^+]} \tilde{\bf A}_{z_1^+}^{j}
		\Bigg] {U}_{[z_1^+,z_i^+]}
		\Bigg\},
	\end{align}
	where we have dropped the subleading phases at next-to-eikonal order. This equation can be further simplified by integrating by parts the second term in the sum that is proportional to $\partial_{\bf B}^2 A^-$. By doing so, we obtain:
	\begin{align}\label{eq:aux25}
		&\Delta_{\rm m}^{(1),\perp}(\vec{P},\vec{Q}) = 
		\frac{\Theta(P^+)}{2 P^+}
		\int_{\vec{B}} e^{i \vec{B} \cdot \vec{Q}} 
		\Bigg\{
		\int_{z_i^+}^{z_f^+} dz_1^+ {U}_{[z_f^+,z_1^+]}
		\Bigg[
		\frac{{\bf P}^i}{P^+} \left( (z_1^+-B^+) \partial_{{\bf B}^i} \tilde{A}_{z_1^+}^{-} - \tilde{\bf A}_{z_1^+}^{i}\right)
		\nonumber \\ & \hskip8cm
		-
		\frac{\Delta^+}{8 P^+} {\bf Q}^i \partial_{{\bf B}^i} \tilde{A}_{z_1^+}^{-}
		\Bigg] {U}_{[z_1^+, z_i^+]}
		\nonumber \\ & +
		\int_{z_i^+}^{z_f^+}dz_2^+ \int_{z_i^+}^{z_2^+} dz_1^+ {U}_{[z_f^+,z_2^+]}
		\Bigg[
		\frac{i}{2 P^+}
		\left(  
		z_1^+-z_2^+
		\right) 
		\partial_{{\bf B}^i} \tilde{A}_{z_2^+}^{-} {U}_{[z_2^+,z_1^+]} \partial_{{\bf B}^i} \tilde{A}_{z_1^+}^{-}
		\nonumber \\ & \hskip4cm
		- \frac{i}{2 P^+}
		\left(
		\partial_{{\bf B}^i} \tilde{A}_{z_2^+}^{-} {U}_{[z_2^+,z_1^+]} \tilde{\bf A}_{z_1^+}^{i}
		-
		\tilde{\bf A}_{z_2^+}^{i} {U}_{[z_2^+,z_1^+]} \partial_{{\bf B}^i} \tilde{A}_{z_1^+}^{-}
		\right)
		\nonumber \\ & \hskip4cm
		+
		\frac{i}{P^+} \delta(z_2^+-z_1^+)
		\tilde{\bf A}_{z_2^+}^{i} {U}_{[z_2^+,z_1^+]} \tilde{\bf A}_{z_2^+}^{j}
		\Bigg] {U}_{[z_1^+,z_i^+]}
		\Bigg\}.
	\end{align}
	
	Now, analogous to what we did for the $v^-$ correction, the next step is to express the derivatives of the gauge field in terms of the Wilson line. To do that, we use the identities given in \cref{eq:indentity1,eq:indentity2} as well as the following relation:
	\begin{align}\label{key}
		&\int_{z_i^+}^{z_f^+}dz_2^+ \int_{z_i^+}^{z_2^+} dz_1^+ {U}_{[z_f^+,z_2^+]} \left(  
		z_2^+-z_1^+
		\right) 
		\partial_{{\bf B}^i} \tilde{A}_{z_2^+}^{-} {U}_{[z_2^+,z_1^+]}(\vec{B}) \partial_{{\bf B}^i} \tilde{A}_{z_1^+}^{-} {U}_{[z_1^+,z_i^+]} 
		\nonumber \\ & \hskip5cm
		= 
		\int_{z_i^+}^{z_f^+} dz^+ {U}_{[z_f^+,z^+]} 
		\overleftarrow{\partial}_{{\bf B}^i} \overrightarrow{\partial}_{{\bf B}^i}
		{U}_{[z^+, z_i^+]}.
	\end{align}
	So that \cref{eq:aux25} can be written as
	\begin{align}\label{eq:aux27}
		&\Delta_{\rm m}^{(1),\perp}(\vec{P},\vec{Q}) = 
		\frac{\Theta(P^+)}{2 P^+}
		\int_{\vec{B}} e^{i \vec{B} \cdot \vec{Q}} 
		\Bigg\{
		- \frac{\Delta^+}{8 P^+} {\bf Q}^i \overrightarrow{\partial}_{{\bf B}^i}
		{U}_{[z_f^+,z_i^+]}
		\nonumber \\ & \hskip0cm +
		\int_{z_i^+}^{z_f^+} dz^+ {U}_{[z_f^+,z^+]}
		\Bigg[
		\frac{{\bf P}^i}{P^+} \left( -\frac{\overleftrightarrow{\partial}_{{\bf B}^i}}{2} - \tilde{\bf A}_{z_1^+}^{i}\right)
		\nonumber \\ & \hskip2cm
		- \frac{i}{2 P^+} 
		\left(
		\overleftarrow{\partial}_{{\bf B}^i} \overrightarrow{\partial}_{{\bf B}^i} - \tilde{\bf A}_{z^+}^{i} \overrightarrow{\partial}_{{\bf B}^i} + \overleftarrow{\partial}_{{\bf B}^i} \tilde{\bf A}_{z^+}^{i} - \tilde{\bf A}_{z^+}^{i} \tilde{\bf A}_{z^+}^{i}
		\right)
		\Bigg] {U}_{[z^+, z_i^+]}
		\Bigg\},
	\end{align}
	where the transverse derivatives only act on the terms inside the square brackets. Finally, analogous to \cite{Altinoluk:2022jkk}, we can write the final result in an explicit gauge covariant way by expressing the transverse components of the field in terms of the covariant derivatives. This can be done by noting that the transverse component of the gauge field can be written as:
	\begin{equation}\label{eq:aux26}
		\tilde{\bf A}^i = 
		\overrightarrow{D}_{{\bf B}^i} - \overrightarrow{\partial}_{{\bf B}^i} = 
		\overleftarrow{\partial}_{{\bf B}^i} - \overleftarrow{D}_{{\bf B}^i} =
		\frac{\overleftrightarrow{D}_{{\bf B}^i} - \overleftrightarrow{\partial}_{{\bf B}^i}}{2}.
	\end{equation}
	
	Inserting \cref{eq:aux26} into \cref{eq:aux27} and integrating by parts the first term, which is proportional to ${\bf Q}^i$, we obtain the result for the corrections of the scalar propagator due to fluctuations around the classical transverse trajectory expressed in a gauge covariant way:
	\begin{align}\label{eq:correction_trans}
		&\Delta_{\rm m}^{(1),\perp}(\vec{P},\vec{Q}) = 
		\frac{\Theta(P^+)}{2 P^+}
		\int_{\vec{B}} e^{i \vec{B} \cdot \vec{Q}} 
		\Bigg\{
		- i \frac{{\bf Q}^2 \Delta^+}{8 P^+}
		{U}_{[z_f^+,z_i^+]}
		\nonumber \\ & \hskip2cm
		-
		\int_{z_i^+}^{z_f^+} dz^+ {U}_{[z_f^+,z^+]}
		\Bigg[
		\frac{{\bf P}^i}{P^+} \overleftrightarrow{D}_{{\bf B}^i}
		+ \frac{i}{2 P^+} 
		\overleftarrow{D}_{{\bf B}^i}	\overrightarrow{D}_{{\bf B}^i}
		\Bigg] {U}_{[z^+, z_i^+]}
		\Bigg\}.
	\end{align}

\subsection{Summing all the corrections}
	
	Finally, we sum up all the $\mathcal{O}(1/P^+)$ corrections of the in-medium propagator given in \cref{eq:correction_zero,eq:correction_minus,eq:correction_trans}. The result gives
	\begin{align}\label{eq:correction_total}
		&\Delta_{\rm m}(\vec{P},\vec{Q}) = 
		\frac{\Theta(P^+)}{2 P^+}
		\int_{\vec{B}} e^{i \vec{B} \cdot \vec{Q}}
		\Bigg\{
		\left(1 + i \frac{{\bf P} \cdot {\bf Q}}{P^+}B^+\right){U}_{[z_f^+,z_i^+]}(\vec{B})
		\nonumber \\ &  \hskip0cm
		-
		\int_{z_i^+}^{z_f^+} dz^+ {U}_{[z_f^+,z^+]}(\vec{B})
		\Bigg[
		\frac{{\bf P}^i}{P^+} \overleftrightarrow{D}_{{\bf B}^i}
		+ \frac{i}{2 P^+} 
		\overleftarrow{D}_{{\bf B}^i}	\overrightarrow{D}_{{\bf B}^i}
		\Bigg] {U}_{[z^+, z_i^+]}(\vec{B})
		\Bigg\} + \mathcal{O}\left(\frac{1}{(P^+)^2}\right).
	\end{align}
	
	The last step is to expand the eikonal Wilson lines around $B^-=0$:
	\begin{align}
		{U}_{[z_f^+,z_i^+]}(\vec{B}) = \exp \left\{ B^- \partial^+ \right\} {U}_{[z_f^+,z_i^+]}(0,{\bf B}).
	\end{align}
	It is clear from \cref{eq:correction_total} that the zeroth order term in $B^-$ will give a Dirac delta $\deltabar(Q^+)$ that fixes the longitudinal momentum transfer to zero and higher order corrections will give higher order derivatives of the Dirac delta. The part of the retarded propagator, given in \cref{eq:propagator_momentum}, that depends on the longitudinal momentum transfer, $Q^+$, is the LC energy phases:
	\begin{align}\label{app:aux1}
		\hat{p}_f^-x^+=\frac{{\bf p}_f^2}{2P^+}\left(1 - \frac{Q^+}{2P^+} + \cdots\right)x^+, \qquad \hat{p}_i^-y^+=\frac{{\bf p}_i^2}{2P^+}\left(1 + \frac{Q^+}{2P^+} + \cdots\right)y^+.
	\end{align}
	Thus, the effect of the $n^{\rm th}$ derivative of the Dirac delta of $Q^+$, $\deltabar^{'n}(Q^+)$, is to drop a factor $\mathcal{O}(1/(P^+)^{n+1})x^+ \sim \mathcal{O}(1/(P^+)^{n})$ coming from the phase, after integration over $Q^+$. On the other hand, once we write the propagator in coordinate space, the integration over $Q^+$ comes through a Fourier transform that introduces the phase $Q^+ (x^- + y^-)/2$ so that, after integration over $Q^+$, the $n^{\rm th}$ derivative of the delta will also introduce a factor $\mathcal{O}((x^-+y^-)^n/2^n)$. This factor is not written in terms of powers of $1/P^+$ because, in this case, we are expressing the propagator in coordinate space. It is considered a non-eikonal correction because under a boost in the right direction, the $z^-$ component of the particle trajectory transforms as $z^- \to e^{-\omega} z^-$, making it subleading in the eikonal approximation. 
	
	So far, because of the aforementioned arguments, in order to get the next-to-eikonal corrections to the scalar propagator, it is enough to expand \cref{eq:correction_total} up to order $B^-$, while neglecting the terms of order $B^-/P^+$ which will lead to a next-to-next-to-eikonal correction. By doing this, we obtain:
	\begin{align}\label{eq:aux28}
		&\Delta_{\rm m}(\vec{P},\vec{Q}) = 
		\frac{\Theta(P^+)}{2 P^+}
		\int_{\vec{B}} e^{i \vec{B} \cdot \vec{Q}}
		\Bigg\{
		\left(1 + i \frac{{\bf P} \cdot {\bf Q}}{P^+}B^+\right){U}_{[z_f^+,z_i^+]}(0,{\bf B}) + B^- \partial^+ {U}_{[z_f^+,z_i^+]}(0,{\bf B})
		\nonumber \\ &  \hskip2cm
		-
		\int_{z_i^+}^{z_f^+} dz^+ {U}_{[z_f^+,z^+]}(0,{\bf B})
		\Bigg[
		\frac{{\bf P}^i}{P^+} \overleftrightarrow{D}_{{\bf B}^i}
		+ \frac{i}{2 P^+} 
		\overleftarrow{D}_{{\bf B}^i}	\overrightarrow{D}_{{\bf B}^i}
		\Bigg] {U}_{[z^+, z_i^+]}(0,{\bf B})
		\Bigg\}
		\nonumber \\ &  =
		\deltabar(Q^+)
		\frac{\Theta(P^+)}{2 P^+} 
		\int_{\bf{B}} e^{-i \bf{B} \cdot \bf{Q}}
		\Bigg\{
		\left(1 + i \frac{{\bf P} \cdot {\bf Q}}{P^+}B^+\right){U}_{[z_f^+,z_i^+]}(0,{\bf B})
		\nonumber \\ &  \hskip2cm
		-
		\int_{z_i^+}^{z_f^+} dz^+ {U}_{[z_f^+,z^+]}(0,{\bf B})
		\Bigg[
		\frac{{\bf P}^i}{P^+} \overleftrightarrow{D}_{{\bf B}^i}
		+ \frac{i}{2 P^+} 
		\overleftarrow{D}_{{\bf B}^i}	\overrightarrow{D}_{{\bf B}^i}
		\Bigg] {U}_{[z^+, z_i^+]}(0,{\bf B})
		\Bigg\}
		\nonumber \\ &  \hskip2cm
		-i\deltabar'(Q^+)
		\frac{\Theta(P^+)}{2 P^+}
		\int_{\bf{B}} e^{-i \bf{B} \cdot \bf{Q}} \partial^+ {U}_{[z_f^+,z_i^+]}(0,{\bf B})
		.
	\end{align}
	
	Introducing this result into \cref{eq:prop_momentum} we obtain the retarded scalar propagator in momentum space at next-to-eikonal order:
	\begin{align}\label{eq:aux29}
		&\tilde{\Delta}_R(x^+,\vec{p}_f;y^+,\vec{p}_i) =
		\Theta(x^+-y^+)
		e^{-i \hat{p}_f^- {x}^+ + i \hat{p}_i^- {y}^+} 
		\deltabar(p_f^+-p_i^+)
		\frac{\Theta(p_i^+)}{2 p_i^+} 
		\nonumber \\ &  \hskip0cm \times
		\int_{\bf{B}} e^{-i \bf{B} \cdot ({\bf p}_f-{\bf p}_i)}
		\Bigg\{
		\left(1 + i \frac{{\bf p}_f^2-{\bf p}_i^2}{2p_i^+}B^+\right){U}_{[z_f^+,z_i^+]}(0,{\bf B})
		\\ &  \hskip0cm
		-
		\int_{z_i^+}^{z_f^+} dz^+ {U}_{[z_f^+,z^+]}(0,{\bf B})
		\Bigg[
		\frac{{\bf p}_f^i+{\bf p}_i^i}{2 p_i^+} \overleftrightarrow{D}_{{\bf B}^i}
		+ \frac{i}{2 p_i^+} 
		\overleftarrow{D}_{{\bf B}^i}	\overrightarrow{D}_{{\bf B}^i}
		\Bigg] {U}_{[z^+, z_i^+]}(0,{\bf B})
		\Bigg\}
		\nonumber \\ &  \hskip0cm
		-i\deltabar'(p_f^+-p_i^+)
		\Theta(x^+-y^+)
		e^{-i \hat{p}_f^- {x}^+ + i \hat{p}_i^- {y}^+} 
		\frac{\Theta(p_i^+)\Theta(p_f^+)}{p_i^++p_f^+}
		\int_{\bf{B}} e^{-i \bf{B} \cdot ({\bf p}_f-{\bf p}_i)} \partial^+ {U}_{[z_f^+,z_i^+]}(0,{\bf B}),
		\nonumber
	\end{align}
	
	Performing the Fourier transform of \cref{eq:aux29}, we obtain the next-to-eikonal correction to the scalar propagator in coordinate space:
	\begin{align}\label{eq:final_result_coord}
		\Delta_R(x,y) &= \Theta(x^+-y^+) 
		\int_{P^+} \frac{\Theta(P^+)}{2 P^+} e^{-i P^+ (x^- - y^-)}
		\int_{{\bf p}_i,{\bf p}_f} e^{i{\bf p}_f \cdot {\bf x} - i {\bf p}_i \cdot {\bf y} -i \frac{{\bf p}_f^2+m^2}{2 P^+} {x}^+ + i \frac{{\bf p}_i^2+m^2}{2 P^+} {y}^+}
		\nonumber \\ &  \hskip0cm \times
		\int_{\bf{B}} e^{-i \bf{B} \cdot ({\bf p}_f-{\bf p}_i)}
		\Bigg\{
		\left(1 + i \frac{{\bf p}_f^2-{\bf p}_i^2}{2P^+}B^+\right){U}_{[z_f^+,z_i^+]}(0,{\bf B})
		\nonumber \\ &  \hskip0cm
		-
		\int_{z_i^+}^{z_f^+} dz^+ {U}_{[z_f^+,z^+]}(0,{\bf B})
		\Bigg[
		\frac{{\bf p}_f^i+{\bf p}_i^i}{2 P^+} \overleftrightarrow{D}_{{\bf B}^i}
		+ \frac{i}{2 P^+} 
		\overleftarrow{D}_{{\bf B}^i}	\overrightarrow{D}_{{\bf B}^i}
		\Bigg] {U}_{[z^+, z_i^+]}(0,{\bf B})
		\nonumber \\ &  \hskip0cm +
		\left[
		\frac{x^-+y^-}{2} - \frac{{\bf p}_f^2+m^2}{(2 P^+)^2}x^+ - \frac{{\bf p}_i^2+m^2}{(2 P^+)^2}y^+
		\right] \partial^+ {U}_{[z_f^+,z_i^+]}(0,{\bf B})
		\Bigg\}.
	\end{align}
	We note that although \cref{eq:final_result_coord} provides the correction to the scalar propagator due to fluctuations around the classical trajectory, it is not a new result, as it has already been computed recently in \cite{Altinoluk_2022}. However, the method used in \cite{Altinoluk_2022} was completely different, based on iterating the differential equations of the propagator and performing power counting in the boost parameter $e^\omega$ by boosting the background field instead of the particle. Therefore, \cref{eq:final_result_coord} serves as a cross-check for the eikonal expansion approach presented in this manuscript.

\section{The amputated retarded scalar propagator}
\label{app:LSZ}

	In this section, we derive the amputated retarded propagator using \cref{eq:aux9}.
	\Cref{eq:aux9} is particularly useful because it allows us to amputate the propagator legs straightforwardly using an LSZ-like reduction formula. In order to see that, we write \cref{eq:aux9} in momentum space:
	\begin{align}\label{eq:aux10}
		&\tilde{\Delta}_R(p_f,p_i) =
		\Theta(p_i^+) \Theta(p_f^+)
		\int_{x^+,y^+}
		e^{i (p_f^--\hat{p}_f^-) {x}^+ - i (p_i^--\hat{p}_i^-) {y}^+}
		\Theta(x^+-y^+) 
		\int_{\vec{z}_i,\vec{z}_f}
		e^{i \hat{p}_f \cdot {z}_f-i \hat{p}_i \cdot {z}_i}
		\nonumber \\ & \hskip0cm \times
		\int\mathcal{D}p_\Delta^+
		\frac{\Theta(\avp)}{2 \avp}
		\int_{\vec{z}(\tau^+_i)=\vec{z}_i}^{\vec{z}(\tau^+_f)=\vec{z}_f} \mathcal{D}^3\vec{z} 
		\
		e^{i \int_{\tau^+_i}^{\tau^+_f} d\tau^+ \left[-\frac{m_\epsilon^2}{2 \avp} + \frac{\avp}{2} \dot{\bf z}^2 -p^+ \dot{z}^- \right]}
		\mathcal{U}_{[\tau^+_f,\tau^+_i]}[z^+,\vec{z}].
	\end{align}
	
	Although this propagator is connected, in the sense that the vacuum diagrams have already been divided out, it is not amputated. In order to put its legs on-shell, we use a LSZ-like approach: we multiply the leg that we want to amputate by the inverse of the free propagator $(p^2-m^2)$ and we perform the on-shell limit $p^2 \to m^2$. Note that, in general, we should also divide by the factor $\sqrt{Z}$ arising from the renormalization of the scalar field. However, since the background field is classical and we are not taking into account self-interactions of the particle, this factor is just 1.
	
	Let us work out the amputation of the final leg. In this case, the LSZ reduction formula is given, in momentum space, by
	\begin{align}
		\tilde{\Delta}^{(c)}_R(p_f,p_i) = \lim_{p_f^2 \to m^2} (p_f^2-m^2) \tilde{\Delta}_R(p_f,p_i) = 
		2p_f^+
		\lim_{p_f^- \to \hat{p}_f^-} (p_f^--\hat{p}_f^-) \tilde{\Delta}_R(p_f,p_i),
	\end{align}
	where we denote the amputated propagator as $\tilde{\Delta}^{(c)}$. 
	Since \cref{eq:aux10} is of the form
	\begin{align}
		\tilde{\Delta}_R(p_f,p_i) = \int_{x^+} e^{i (p_f^--\hat{p}_f^-) {x}^+} f(x^+),
	\end{align}	
	we have that
	\begin{align}
		\tilde{\Delta}^{(c)}_R(p_f,p_i)
		&=
		2p_f^+
		\lim_{p_f^- \to \hat{p}_f^-} (p_f^--\hat{p}_f^-) \int_{x^+} e^{i (p_f^--\hat{p}_f^-) {x}^+} f(x^+)
		\nonumber \\ & 	= 
		2p_f^+ \lim_{p_f^- \to \hat{p}_f^-} \int_{x^+} \frac{\partial e^{i (p_f^--\hat{p}_f^-) {x}^+}}{i\partial x^+} f(x^+)
		\nonumber \\ & 	= 
		2 i p_f^+ \lim_{p_f^- \to \hat{p}_f^-} \int_{x^+} e^{i (p_f^--\hat{p}_f^-) {x}^+}  \frac{\partial f(x^+)}{\partial x^+}
		= 2 i p_f^+ \lim_{x^+ \to \infty} f(x^+),
	\end{align}
	where when we integrated by parts, in the third step, we have used the fact that the integral vanishes at $x^+ \to \infty$ due to the $i \epsilon$ prescription, hidden in the definition of $\hat{p}_f^-$. In the fourth step, we have performed the on-shell $p_f^- \to \hat{p}_f^-$ limit and used the fact that $f(x^+=-\infty)=0$ due to the step function. 
	
	We can do an analogous analysis for initial leg, $y^+$, of the propagator. Thus, the amputated propagator, i.e., the scattering amplitude for a scalar particle interacting with the medium, is given by 
	\begin{align}\label{eq:amputated_propagator}
		&\tilde{\Delta}^{(c)}_R(p_f,p_i) = 4 p_i^+ p_f^+ \lim_{x^+ \to \infty} \lim_{y^+ \to - \infty}
		\int_{\vec{z}_i,\vec{z}_f}
		e^{i \hat{p}_f \cdot {z}_f-i \hat{p}_i \cdot {z}_i}
		\int\mathcal{D}p_\Delta^+
		\frac{\Theta(\avp)}{2 \avp}
		\nonumber \\ & \hskip2cm \times
		\int_{\vec{z}(\tau^+_i)=\vec{z}_i}^{\vec{z}(\tau^+_f)=\vec{z}_f} \mathcal{D}^3\vec{z} 
		\
		e^{i \int_{\tau^+_i}^{\tau^+_f} d\tau^+ \left[-\frac{m_\epsilon^2}{2 \avp} + \frac{\avp}{2} \dot{\bf z}^2 -p^+ \dot{z}^- \right]}
		\mathcal{U}_{[\tau^+_f,\tau^+_i]}[z^+,\vec{z}].
	\end{align}
	We note that in \cref{eq:amputated_propagator}, although there is not a explicit dependence on the variables $x^+$ and $y^+$, the path integral depend on them through the definition of the average longitudinal momentum given in \cref{eq:inmed_avgp} as well as on the coordinates $\tau_i^+$ and $\tau_f^+$.

\input{pathint_propagator_JHEP.bbl}
\end{document}

%% file: pathint_propagator_JHEP.bbl
\providecommand{\href}[2]{#2}\begingroup\raggedright\endgroup

%% file: pathint_propagator_JHEP.bbl
\begin{thebibliography}{10}

\bibitem{GRIBOV19831}
L.~Gribov, E.~Levin and M.~Ryskin, \emph{Semihard processes in qcd},
  \href{https://doi.org/https://doi.org/10.1016/0370-1573(83)90022-4}{\emph{Physics
  Reports} {\bfseries 100} (1983) 1}.

\bibitem{MUELLER1986427}
A.~Mueller and J.~Qiu, \emph{Gluon recombination and shadowing at small values
  of x},
  \href{https://doi.org/https://doi.org/10.1016/0550-3213(86)90164-1}{\emph{Nuclear
  Physics B} {\bfseries 268} (1986) 427}.

\bibitem{McLerran_1994}
L.~McLerran and R.~Venugopalan, \emph{Computing quark and gluon distribution
  functions for very large nuclei},
  \href{https://doi.org/10.1103/physrevd.49.2233}{\emph{Physical Review D}
  {\bfseries 49} (1994) 2233}.

\bibitem{McLerran_1994b}
L.~McLerran and R.~Venugopalan, \emph{Gluon distribution functions for very
  large nuclei at small transverse momentum},
  \href{https://doi.org/10.1103/physrevd.49.3352}{\emph{Physical Review D}
  {\bfseries 49} (1994) 3352}.

\bibitem{McLerran_1994c}
L.~McLerran and R.~Venugopalan, \emph{Green's function in the color field of a
  large nucleus},
  \href{https://doi.org/10.1103/physrevd.50.2225}{\emph{Physical Review D}
  {\bfseries 50} (1994) 2225}.

\bibitem{kovchegov_levin_2012}
Y.V.~Kovchegov and E.~Levin, \emph{Quantum Chromodynamics at High Energy},
  Cambridge Monographs on Particle Physics, Nuclear Physics and Cosmology,
  Cambridge University Press (2012),
  \href{https://doi.org/10.1017/CBO9781139022187}{10.1017/CBO9781139022187}.

\bibitem{Blaizot_2017}
J.-P.~Blaizot, \emph{High gluon densities in heavy ion collisions},
  \href{https://doi.org/10.1088/1361-6633/aa5435}{\emph{Reports on Progress in
  Physics} {\bfseries 80} (2017) 032301}.

\bibitem{Baier:1996kr}
R.~Baier, Y.L.~Dokshitzer, A.H.~Mueller, S.~Peigne and D.~Schiff,
  \emph{{Radiative energy loss of high-energy quarks and gluons in a finite
  volume quark - gluon plasma}},
  \href{https://doi.org/10.1016/S0550-3213(96)00553-6}{\emph{Nucl. Phys. B}
  {\bfseries 483} (1997) 291}
  [\href{https://arxiv.org/abs/hep-ph/9607355}{{\ttfamily hep-ph/9607355}}].

\bibitem{Baier:1996sk}
R.~Baier, Y.L.~Dokshitzer, A.H.~Mueller, S.~Peigne and D.~Schiff,
  \emph{{Radiative energy loss and p(T) broadening of high-energy partons in
  nuclei}}, \href{https://doi.org/10.1016/S0550-3213(96)00581-0}{\emph{Nucl.
  Phys. B} {\bfseries 484} (1997) 265}
  [\href{https://arxiv.org/abs/hep-ph/9608322}{{\ttfamily hep-ph/9608322}}].

\bibitem{Baier:1998kq}
R.~Baier, Y.L.~Dokshitzer, A.H.~Mueller and D.~Schiff, \emph{{Medium induced
  radiative energy loss: Equivalence between the BDMPS and Zakharov
  formalisms}},
  \href{https://doi.org/10.1016/S0550-3213(98)00546-X}{\emph{Nucl. Phys. B}
  {\bfseries 531} (1998) 403}
  [\href{https://arxiv.org/abs/hep-ph/9804212}{{\ttfamily hep-ph/9804212}}].

\bibitem{Zakharov:1996fv}
B.G.~Zakharov, \emph{{Fully quantum treatment of the Landau-Pomeranchuk-Migdal
  effect in QED and QCD}}, \href{https://doi.org/10.1134/1.567126}{\emph{JETP
  Lett.} {\bfseries 63} (1996) 952}
  [\href{https://arxiv.org/abs/hep-ph/9607440}{{\ttfamily hep-ph/9607440}}].

\bibitem{Zakharov:1997uu}
B.G.~Zakharov, \emph{{Radiative energy loss of high-energy quarks in finite
  size nuclear matter and quark - gluon plasma}},
  \href{https://doi.org/10.1134/1.567389}{\emph{JETP Lett.} {\bfseries 65}
  (1997) 615} [\href{https://arxiv.org/abs/hep-ph/9704255}{{\ttfamily
  hep-ph/9704255}}].

\bibitem{Altinoluk_2014}
T.~Altinoluk, N.~Armesto, G.~Beuf, M.~Mart{\i}nez and C.A.~Salgado,
  \emph{Next-to-eikonal corrections in the {CGC}: gluon production and spin
  asymmetries in {pA} collisions},
  \href{https://doi.org/10.1007/jhep07(2014)068}{\emph{Journal of High Energy
  Physics} {\bfseries 2014} (2014) }.

\bibitem{Altinoluk_2016}
T.~Altinoluk, N.~Armesto, G.~Beuf and A.~Moscoso, \emph{Next-to-next-to-eikonal
  corrections in the {CGC}},
  \href{https://doi.org/10.1007/jhep01(2016)114}{\emph{Journal of High Energy
  Physics} {\bfseries 2016} (2016) }.

\bibitem{Altinoluk_2021}
T.~Altinoluk, G.~Beuf, A.~Czajka and A.~Tymowska, \emph{Quarks at
  next-to-eikonal accuracy in the {CGC}: Forward quark-nucleus scattering},
  \href{https://doi.org/10.1103/physrevd.104.014019}{\emph{Physical Review D}
  {\bfseries 104} (2021) }.

\bibitem{Altinoluk_2022}
T.~Altinoluk and G.~Beuf, \emph{Quark and scalar propagators at next-to-eikonal
  accuracy in the {CGC} through a dynamical background gluon field},
  \href{https://doi.org/10.1103/physrevd.105.074026}{\emph{Physical Review D}
  {\bfseries 105} (2022) }.

\bibitem{Altinoluk:2022jkk}
T.~Altinoluk, G.~Beuf, A.~Czajka and A.~Tymowska, \emph{{DIS dijet production
  at next-to-eikonal accuracy in the CGC}},
  \href{https://doi.org/10.1103/PhysRevD.107.074016}{\emph{Phys. Rev. D}
  {\bfseries 107} (2023) 074016}
  [\href{https://arxiv.org/abs/2212.10484}{{\ttfamily 2212.10484}}].

\bibitem{Agostini_2019}
P.~Agostini, T.~Altinoluk and N.~Armesto, \emph{Non-eikonal corrections to
  multi-particle production in the color glass condensate},
  \href{https://doi.org/10.1140/epjc/s10052-019-7097-5}{\emph{The European
  Physical Journal C} {\bfseries 79} (2019) }.

\bibitem{Agostini:2022ctk}
P.~Agostini, T.~Altinoluk, N.~Armesto, F.~Dominguez and J.G.~Milhano,
  \emph{{Multiparticle production in proton\textendash{}nucleus collisions
  beyond eikonal accuracy}},
  \href{https://doi.org/10.1140/epjc/s10052-022-10962-1}{\emph{Eur. Phys. J. C}
  {\bfseries 82} (2022) 1001}
  [\href{https://arxiv.org/abs/2207.10472}{{\ttfamily 2207.10472}}].

\bibitem{Agostini:2019hkj}
P.~Agostini, T.~Altinoluk and N.~Armesto, \emph{{Effect of non-eikonal
  corrections on azimuthal asymmetries in the Color Glass Condensate}},
  \href{https://doi.org/10.1140/epjc/s10052-019-7315-1}{\emph{Eur. Phys. J. C}
  {\bfseries 79} (2019) 790}
  [\href{https://arxiv.org/abs/1907.03668}{{\ttfamily 1907.03668}}].

\bibitem{Agostini:2022oge}
P.~Agostini, T.~Altinoluk and N.~Armesto, \emph{{Finite width effects on the
  azimuthal asymmetry in proton-nucleus collisions in the Color Glass
  Condensate}},
  \href{https://doi.org/10.1016/j.physletb.2023.137892}{\emph{Phys. Lett. B}
  {\bfseries 840} (2023) 137892}
  [\href{https://arxiv.org/abs/2212.03633}{{\ttfamily 2212.03633}}].

\bibitem{EuropeanMuon:1987isl}
{\scshape European Muon} collaboration, \emph{{A Measurement of the Spin
  Asymmetry and Determination of the Structure Function g(1) in Deep Inelastic
  Muon-Proton Scattering}},
  \href{https://doi.org/10.1016/0370-2693(88)91523-7}{\emph{Phys. Lett. B}
  {\bfseries 206} (1988) 364}.

\bibitem{Chirilli_2021}
G.A.~Chirilli, \emph{High-energy operator product expansion at sub-eikonal
  level}, \href{https://doi.org/10.1007/jhep06(2021)096}{\emph{Journal of High
  Energy Physics} {\bfseries 2021} (2021) }.

\bibitem{Hatta:2016aoc}
Y.~Hatta, Y.~Nakagawa, F.~Yuan, Y.~Zhao and B.~Xiao, \emph{{Gluon orbital
  angular momentum at small-$x$}},
  \href{https://doi.org/10.1103/PhysRevD.95.114032}{\emph{Phys. Rev. D}
  {\bfseries 95} (2017) 114032}
  [\href{https://arxiv.org/abs/1612.02445}{{\ttfamily 1612.02445}}].

\bibitem{Chirilli:2018kkw}
G.A.~Chirilli, \emph{{Sub-eikonal corrections to scattering amplitudes at high
  energy}}, \href{https://doi.org/10.1007/JHEP01(2019)118}{\emph{JHEP}
  {\bfseries 01} (2019) 118}
  [\href{https://arxiv.org/abs/1807.11435}{{\ttfamily 1807.11435}}].

\bibitem{Jalilian-Marian:2018iui}
J.~Jalilian-Marian, \emph{{Quark jets scattering from a gluon field: from
  saturation to high $p_t$}},
  \href{https://doi.org/10.1103/PhysRevD.99.014043}{\emph{Phys. Rev. D}
  {\bfseries 99} (2019) 014043}
  [\href{https://arxiv.org/abs/1809.04625}{{\ttfamily 1809.04625}}].

\bibitem{Jalilian-Marian:2019kaf}
J.~Jalilian-Marian, \emph{{Rapidity loss, spin, and angular asymmetries in the
  scattering of a quark from the color field of a proton or nucleus}},
  \href{https://doi.org/10.1103/PhysRevD.102.014008}{\emph{Phys. Rev. D}
  {\bfseries 102} (2020) 014008}
  [\href{https://arxiv.org/abs/1912.08878}{{\ttfamily 1912.08878}}].

\bibitem{Altinoluk:2023qfr}
T.~Altinoluk, N.~Armesto and G.~Beuf, \emph{{Probing quark transverse momentum
  distributions in the Color Glass Condensate: quark-gluon dijets in Deep
  Inelastic Scattering at next-to-eikonal accuracy}},
  \href{https://arxiv.org/abs/2303.12691}{{\ttfamily 2303.12691}}.

\bibitem{Bhattacharya:2022vvo}
S.~Bhattacharya, R.~Boussarie and Y.~Hatta, \emph{{Signature of the Gluon
  Orbital Angular Momentum}},
  \href{https://doi.org/10.1103/PhysRevLett.128.182002}{\emph{Phys. Rev. Lett.}
  {\bfseries 128} (2022) 182002}
  [\href{https://arxiv.org/abs/2201.08709}{{\ttfamily 2201.08709}}].

\bibitem{Boussarie:2020vzf}
R.~Boussarie and Y.~Mehtar-Tani, \emph{{Gauge invariance of transverse momentum
  dependent distributions at small $x$}},
  \href{https://doi.org/10.1103/PhysRevD.103.094012}{\emph{Phys. Rev. D}
  {\bfseries 103} (2021) 094012}
  [\href{https://arxiv.org/abs/2001.06449}{{\ttfamily 2001.06449}}].

\bibitem{Boussarie:2020fpb}
R.~Boussarie and Y.~Mehtar-Tani, \emph{{A novel formulation of the unintegrated
  gluon distribution for DIS}},
  \href{https://doi.org/10.1016/j.physletb.2022.137125}{\emph{Phys. Lett. B}
  {\bfseries 831} (2022) 137125}
  [\href{https://arxiv.org/abs/2006.14569}{{\ttfamily 2006.14569}}].

\bibitem{Li:2020uhl}
M.~Li, X.~Zhao, P.~Maris, G.~Chen, Y.~Li, K.~Tuchin et~al.,
  \emph{{Ultrarelativistic quark-nucleus scattering in a light-front
  Hamiltonian approach}},
  \href{https://doi.org/10.1103/PhysRevD.101.076016}{\emph{Phys. Rev. D}
  {\bfseries 101} (2020) 076016}
  [\href{https://arxiv.org/abs/2002.09757}{{\ttfamily 2002.09757}}].

\bibitem{Li:2021zaw}
M.~Li, T.~Lappi and X.~Zhao, \emph{{Scattering and gluon emission in a color
  field: A light-front Hamiltonian approach}},
  \href{https://doi.org/10.1103/PhysRevD.104.056014}{\emph{Phys. Rev. D}
  {\bfseries 104} (2021) 056014}
  [\href{https://arxiv.org/abs/2107.02225}{{\ttfamily 2107.02225}}].

\bibitem{Kovchegov_2016}
Y.V.~Kovchegov, D.~Pitonyak and M.D.~Sievert, \emph{Helicity evolution at small
  x}, \href{https://doi.org/10.1007/jhep01(2016)072}{\emph{Journal of High
  Energy Physics} {\bfseries 2016} (2016) }.

\bibitem{Kovchegov:2016weo}
Y.V.~Kovchegov, D.~Pitonyak and M.D.~Sievert, \emph{{Small-$x$ asymptotics of
  the quark helicity distribution}},
  \href{https://doi.org/10.1103/PhysRevLett.118.052001}{\emph{Phys. Rev. Lett.}
  {\bfseries 118} (2017) 052001}
  [\href{https://arxiv.org/abs/1610.06188}{{\ttfamily 1610.06188}}].

\bibitem{Kovchegov:2016zex}
Y.V.~Kovchegov, D.~Pitonyak and M.D.~Sievert, \emph{{Helicity Evolution at
  Small $x$: Flavor Singlet and Non-Singlet Observables}},
  \href{https://doi.org/10.1103/PhysRevD.95.014033}{\emph{Phys. Rev. D}
  {\bfseries 95} (2017) 014033}
  [\href{https://arxiv.org/abs/1610.06197}{{\ttfamily 1610.06197}}].

\bibitem{Kovchegov:2017jxc}
Y.V.~Kovchegov, D.~Pitonyak and M.D.~Sievert, \emph{{Small-$x$ Asymptotics of
  the Quark Helicity Distribution: Analytic Results}},
  \href{https://doi.org/10.1016/j.physletb.2017.06.032}{\emph{Phys. Lett. B}
  {\bfseries 772} (2017) 136}
  [\href{https://arxiv.org/abs/1703.05809}{{\ttfamily 1703.05809}}].

\bibitem{Kovchegov:2017lsr}
Y.V.~Kovchegov, D.~Pitonyak and M.D.~Sievert, \emph{{Small-$x$ Asymptotics of
  the Gluon Helicity Distribution}},
  \href{https://doi.org/10.1007/JHEP10(2017)198}{\emph{JHEP} {\bfseries 10}
  (2017) 198} [\href{https://arxiv.org/abs/1706.04236}{{\ttfamily
  1706.04236}}].

\bibitem{Kovchegov:2018znm}
Y.V.~Kovchegov and M.D.~Sievert, \emph{{Small-$x$ Helicity Evolution: an
  Operator Treatment}},
  \href{https://doi.org/10.1103/PhysRevD.99.054032}{\emph{Phys. Rev. D}
  {\bfseries 99} (2019) 054032}
  [\href{https://arxiv.org/abs/1808.09010}{{\ttfamily 1808.09010}}].

\bibitem{Kovchegov:2018zeq}
Y.V.~Kovchegov and M.D.~Sievert, \emph{{Valence Quark Transversity at Small
  $x$}}, \href{https://doi.org/10.1103/PhysRevD.99.054033}{\emph{Phys. Rev. D}
  {\bfseries 99} (2019) 054033}
  [\href{https://arxiv.org/abs/1808.10354}{{\ttfamily 1808.10354}}].

\bibitem{Kovchegov:2020hgb}
Y.V.~Kovchegov and Y.~Tawabutr, \emph{{Helicity at Small $x$: Oscillations
  Generated by Bringing Back the Quarks}},
  \href{https://doi.org/10.1007/JHEP08(2020)014}{\emph{JHEP} {\bfseries 08}
  (2020) 014} [\href{https://arxiv.org/abs/2005.07285}{{\ttfamily
  2005.07285}}].

\bibitem{Borden:2023ugd}
J.~Borden and Y.V.~Kovchegov, \emph{{Analytic Solution for the Revised Helicity
  Evolution at Small $x$ and Large $N_c$: New Resummed Gluon-Gluon Polarized
  Anomalous Dimension and Intercept}},
  \href{https://arxiv.org/abs/2304.06161}{{\ttfamily 2304.06161}}.

\bibitem{Cougoulic:2019aja}
F.~Cougoulic and Y.V.~Kovchegov, \emph{{Helicity-dependent generalization of
  the JIMWLK evolution}},
  \href{https://doi.org/10.1103/PhysRevD.100.114020}{\emph{Phys. Rev. D}
  {\bfseries 100} (2019) 114020}
  [\href{https://arxiv.org/abs/1910.04268}{{\ttfamily 1910.04268}}].

\bibitem{Cougoulic:2020tbc}
F.~Cougoulic and Y.V.~Kovchegov, \emph{{Helicity-dependent extension of the
  McLerran\textendash{}Venugopalan model}},
  \href{https://doi.org/10.1016/j.nuclphysa.2020.122051}{\emph{Nucl. Phys. A}
  {\bfseries 1004} (2020) 122051}
  [\href{https://arxiv.org/abs/2005.14688}{{\ttfamily 2005.14688}}].

\bibitem{Cougoulic:2022gbk}
F.~Cougoulic, Y.V.~Kovchegov, A.~Tarasov and Y.~Tawabutr, \emph{{Quark and
  gluon helicity evolution at small x: revised and updated}},
  \href{https://doi.org/10.1007/JHEP07(2022)095}{\emph{JHEP} {\bfseries 07}
  (2022) 095} [\href{https://arxiv.org/abs/2204.11898}{{\ttfamily
  2204.11898}}].

\bibitem{Li:2023tlw}
M.~Li, \emph{{Small $x$ Physics Beyond Eikonal Approximation: an Effective
  Hamiltonian Approach}},  \href{https://arxiv.org/abs/2304.12842}{{\ttfamily
  2304.12842}}.

\bibitem{Strassler:1992zr}
M.J.~Strassler, \emph{{Field theory without Feynman diagrams: One loop
  effective actions}},
  \href{https://doi.org/10.1016/0550-3213(92)90098-V}{\emph{Nucl. Phys. B}
  {\bfseries 385} (1992) 145}
  [\href{https://arxiv.org/abs/hep-ph/9205205}{{\ttfamily hep-ph/9205205}}].

\bibitem{Schubert:2001he}
C.~Schubert, \emph{{Perturbative quantum field theory in the string inspired
  formalism}}, \href{https://doi.org/10.1016/S0370-1573(01)00013-8}{\emph{Phys.
  Rept.} {\bfseries 355} (2001) 73}
  [\href{https://arxiv.org/abs/hep-th/0101036}{{\ttfamily hep-th/0101036}}].

\bibitem{Tarasov:2020cwl}
A.~Tarasov and R.~Venugopalan, \emph{{Role of the chiral anomaly in polarized
  deeply inelastic scattering: Finding the triangle graph inside the box
  diagram in Bjorken and Regge asymptotics}},
  \href{https://doi.org/10.1103/PhysRevD.102.114022}{\emph{Phys. Rev. D}
  {\bfseries 102} (2020) 114022}
  [\href{https://arxiv.org/abs/2008.08104}{{\ttfamily 2008.08104}}].

\bibitem{Tarasov:2021yll}
A.~Tarasov and R.~Venugopalan, \emph{{Role of the chiral anomaly in polarized
  deeply inelastic scattering. II. Topological screening and transitions from
  emergent axionlike dynamics}},
  \href{https://doi.org/10.1103/PhysRevD.105.014020}{\emph{Phys. Rev. D}
  {\bfseries 105} (2022) 014020}
  [\href{https://arxiv.org/abs/2109.10370}{{\ttfamily 2109.10370}}].

\bibitem{Zapp:2012nw}
K.C.~Zapp and U.A.~Wiedemann, \emph{{Coherent Radiative Parton Energy Loss
  beyond the BDMPS-Z Limit}},
  \href{https://doi.org/10.1140/epjc/s10052-012-2028-8}{\emph{Eur. Phys. J. C}
  {\bfseries 72} (2012) 2028}
  [\href{https://arxiv.org/abs/1202.1192}{{\ttfamily 1202.1192}}].

\bibitem{Feal:2018bru}
X.~Feal and R.A.~Vazquez, \emph{{Transverse spectrum of bremsstrahlung in
  finite condensed media}},
  \href{https://doi.org/10.1103/PhysRevD.99.016002}{\emph{Phys. Rev. D}
  {\bfseries 99} (2019) 016002}
  [\href{https://arxiv.org/abs/1810.02645}{{\ttfamily 1810.02645}}].

\bibitem{Feal:2018sml}
X.~Feal and R.~Vazquez, \emph{{Intensity of gluon bremsstrahlung in a finite
  plasma}}, \href{https://doi.org/10.1103/PhysRevD.98.074029}{\emph{Phys. Rev.
  D} {\bfseries 98} (2018) 074029}
  [\href{https://arxiv.org/abs/1811.01591}{{\ttfamily 1811.01591}}].

\bibitem{Caron-Huot:2010qjx}
S.~Caron-Huot and C.~Gale, \emph{{Finite-size effects on the radiative energy
  loss of a fast parton in hot and dense strongly interacting matter}},
  \href{https://doi.org/10.1103/PhysRevC.82.064902}{\emph{Phys. Rev. C}
  {\bfseries 82} (2010) 064902}
  [\href{https://arxiv.org/abs/1006.2379}{{\ttfamily 1006.2379}}].

\bibitem{Andres:2020vxs}
C.~Andres, L.~Apolin\'ario and F.~Dominguez, \emph{{Medium-induced gluon
  radiation with full resummation of multiple scatterings for realistic
  parton-medium interactions}},
  \href{https://doi.org/10.1007/JHEP07(2020)114}{\emph{JHEP} {\bfseries 07}
  (2020) 114} [\href{https://arxiv.org/abs/2002.01517}{{\ttfamily
  2002.01517}}].

\bibitem{Sadofyev:2021ohn}
A.V.~Sadofyev, M.D.~Sievert and I.~Vitev, \emph{{Ab~initio coupling of jets to
  collective flow in the opacity expansion approach}},
  \href{https://doi.org/10.1103/PhysRevD.104.094044}{\emph{Phys. Rev. D}
  {\bfseries 104} (2021) 094044}
  [\href{https://arxiv.org/abs/2104.09513}{{\ttfamily 2104.09513}}].

\bibitem{Andres:2022ndd}
C.~Andres, F.~Dominguez, A.V.~Sadofyev and C.A.~Salgado, \emph{{Jet broadening
  in flowing matter: Resummation}},
  \href{https://doi.org/10.1103/PhysRevD.106.074023}{\emph{Phys. Rev. D}
  {\bfseries 106} (2022) 074023}
  [\href{https://arxiv.org/abs/2207.07141}{{\ttfamily 2207.07141}}].

\bibitem{Laenen:2008gt}
E.~Laenen, G.~Stavenga and C.D.~White, \emph{{Path integral approach to eikonal
  and next-to-eikonal exponentiation}},
  \href{https://doi.org/10.1088/1126-6708/2009/03/054}{\emph{JHEP} {\bfseries
  03} (2009) 054} [\href{https://arxiv.org/abs/0811.2067}{{\ttfamily
  0811.2067}}].

\bibitem{Fabbrichesi:1993kz}
M.~Fabbrichesi, R.~Pettorino, G.~Veneziano and G.A.~Vilkovisky,
  \emph{{Planckian energy scattering and surface terms in the gravitational
  action}}, \href{https://doi.org/10.1016/0550-3213(94)90361-1}{\emph{Nucl.
  Phys. B} {\bfseries 419} (1994) 147}
  [\href{https://arxiv.org/abs/hep-th/9309037}{{\ttfamily hep-th/9309037}}].

\bibitem{Brink:1976sc}
L.~Brink, P.~Di~Vecchia and P.S.~Howe, \emph{{A Locally Supersymmetric and
  Reparametrization Invariant Action for the Spinning String}},
  \href{https://doi.org/10.1016/0370-2693(76)90445-7}{\emph{Phys. Lett. B}
  {\bfseries 65} (1976) 471}.

\bibitem{Fradkin:1991ci}
E.S.~Fradkin and D.M.~Gitman, \emph{{Path integral representation for the
  relativistic particle propagators and BFV quantization}},
  \href{https://doi.org/10.1103/PhysRevD.44.3230}{\emph{Phys. Rev. D}
  {\bfseries 44} (1991) 3230}.

\bibitem{Affleck:1981bma}
I.K.~Affleck, O.~Alvarez and N.S.~Manton, \emph{{Pair Production at Strong
  Coupling in Weak External Fields}},
  \href{https://doi.org/10.1016/0550-3213(82)90455-2}{\emph{Nucl. Phys. B}
  {\bfseries 197} (1982) 509}.

\bibitem{Feal:2022iyn}
X.~Feal, A.~Tarasov and R.~Venugopalan, \emph{{QED as a many-body theory of
  worldlines: General formalism and infrared structure}},
  \href{https://doi.org/10.1103/PhysRevD.106.056009}{\emph{Phys. Rev. D}
  {\bfseries 106} (2022) 056009}
  [\href{https://arxiv.org/abs/2206.04188}{{\ttfamily 2206.04188}}].

\bibitem{Feal:2022ufw}
X.~Feal, A.~Tarasov and R.~Venugopalan, \emph{{QED as a many-body theory of
  worldlines. II. All-order S-matrix formalism}},
  \href{https://doi.org/10.1103/PhysRevD.107.096021}{\emph{Phys. Rev. D}
  {\bfseries 107} (2023) 096021}
  [\href{https://arxiv.org/abs/2211.15712}{{\ttfamily 2211.15712}}].

\bibitem{Mehtar-Tani:2006vpj}
Y.~Mehtar-Tani, \emph{{Relating the description of gluon production in pA
  collisions and parton energy loss in AA collisions}},
  \href{https://doi.org/10.1103/PhysRevC.75.034908}{\emph{Phys. Rev. C}
  {\bfseries 75} (2007) 034908}
  [\href{https://arxiv.org/abs/hep-ph/0606236}{{\ttfamily hep-ph/0606236}}].

\bibitem{bastianelli_vannieuwenhuizen_2006}
F.~Bastianelli and P.~van Nieuwenhuizen, \emph{Path Integrals and Anomalies in
  Curved Space}, Cambridge Monographs on Mathematical Physics, Cambridge
  University Press (2006),
  \href{https://doi.org/10.1017/CBO9780511535031}{10.1017/CBO9780511535031}.

\end{thebibliography}
